\documentstyle[12pt]{article}

\font\blackboard=msbm10 at 12pt
\font\blackboards=msbm7
\font\blackboardss=msbm5
\newfam\black
\textfont\black=\blackboard
\scriptfont\black=\blackboards
\scriptscriptfont\black=\blackboardss

\newcommand{\junk}[1]{}

\newcommand{\ba}{\begin{array}}
\newcommand{\ea}{\end{array}}
\newcommand{\be}{\begin{equation}}
\newcommand{\ee}{\end{equation}}
\newcommand{\bea}{\begin{eqnarray}}
\newcommand{\eea}{\end{eqnarray}}
\newcommand{\beas}{\begin{eqnarray*}}
\newcommand{\eeas}{\end{eqnarray*}}

\def\identity{{\rlap{1} \hskip 1.6pt \hbox{1}}}

\def\laplace{{\kern1pt\vbox{\hrule height 1.2pt\hbox{\vrule width
1.2pt\hskip
  3pt\vbox{\vskip 6pt}\hskip 3pt\vrule width 0.6pt}\hrule height
  0.6pt}
  \kern1pt}}
\def\scriptlap{{\kern1pt\vbox{\hrule height 0.8pt\hbox{\vrule width
  0.8pt
  \hskip2pt\vbox{\vskip 4pt}\hskip 2pt\vrule width 0.4pt}\hrule height
  0.4pt}
  \kern1pt}}
\def\slash#1{{\rlap{$#1$} \thinspace /}}
\def\roughly#1{\raise.3ex\hbox{$#1$\kern-.75em\lower1ex\hbox{$\sim$}}}

\def\str{{\rm STr} \,}
\def\sym{{\rm Sym} \,}

\def\tr{{\rm Tr} \,}

\newcommand{\NP}{{\em Nucl.\ Phys.\ }}

\newcommand{\PL}{{\em Phys.\ Lett.\ }}
\newcommand{\PR}{{\em Phys.\ Rev.\ }}

\newcommand{\PRL}{{\em Phys.\ Rev.\ Lett.\ }}

\newcommand{\gone}[1]{}
\begin{document}
\pagestyle{plain}
\setcounter{page}{1}

\baselineskip16pt

\begin{titlepage}

\begin{flushright}
PUPT-1828\\
MIT-CTP-2810\\
hep-th/9812239
\end{flushright}
\vspace{8 mm}

\begin{center}

{\Large \bf Supergravity currents and linearized interactions\\
for Matrix Theory configurations\\
with fermionic backgrounds\\}

\end{center}

\vspace{7 mm}

\begin{center}

Washington Taylor IV$^a$ and Mark Van Raamsdonk$^b$

\vspace{3mm}
${}^a${\small \sl Center for Theoretical Physics} \\
{\small \sl MIT, Bldg. 6-306} \\
{\small \sl Cambridge, MA 02139, U.S.A.} \\
{\small \tt wati@mit.edu}\\

\vspace{3mm}
${}^b${\small \sl Department of Physics} \\
{\small \sl Joseph Henry Laboratories} \\
{\small \sl Princeton University} \\
{\small \sl Princeton, New Jersey 08544, U.S.A.} \\
{\small \tt mav@princeton.edu}
\end{center}

\vspace{8 mm}

\begin{abstract} 
The leading terms in the long-range interaction potential between an
arbitrary pair of matrix theory objects are calculated at one-loop
order.  This result generalizes previous calculations by including
arbitrary fermionic background field configurations.  The interaction
potential at orders $1/r^7$ and $1/r^8$ is shown to correspond
precisely with the leading terms expected from linearized supergravity
interactions between arbitrary objects in M-theory.  General
expressions for the stress tensor, membrane current and 5-brane
current of an arbitrary matrix configuration are derived, including
fermionic contributions.  Supergravity effects which are correctly
reproduced include membrane/5-brane interactions, 0-brane/6-brane
interactions, supercurrent/supercurrent interactions and the spin
contributions to moments of the supergravity currents.  The matrix
theory description of the supergravity stress tensor, membrane current
and 5-brane current are used to propose an explicit formulation of
matrix theory in an arbitrary background metric and 3-form field.
\end{abstract}

\vspace{1cm}
\begin{flushleft}
December 1998
\end{flushleft}
\end{titlepage}
\newpage

\section{Introduction}

Over a decade ago, a simple model of
supersymmetric matrix quantum mechanics 
\cite{Claudson-Halpern,Flume,brr}
was
proposed as a regulated form of the quantum supermembrane theory in 11
dimensions \cite{Goldstone-Hoppe,bst,dhn}.  Interest in this model
declined when it was found to have a continuous spectrum \cite{dln},
which seemed incompatible with a first-quantized interpretation of the
theory in 11D.  In 1996, interest in this  ``Matrix Theory'' was rekindled
when Banks, Fischler, Shenker and Susskind (BFSS)
proposed that it gives a
complete description of light-front M-theory in the large $N$ limit
\cite{BFSS}.  They pointed out that the continuous spectrum is
natural since the theory should be understood as a second-quantized
theory in 11 dimensions.  BFSS argued that M-theory interactions,
including those of classical supergravity, should arise as quantum
effects in matrix theory.  They gave as one piece of evidence for
their conjecture the result that the leading long-range interaction
between a pair of gravitons in 11D supergravity can be reproduced by a
one-loop calculation in matrix theory \cite{DKPS}.

In the two years following the BFSS conjecture, it has become clear
that matrix theory encodes a remarkable amount of the structure of
M-theory and 11D supergravity (for reviews, see
\cite{banks-review,Susskind-review,Bilal-review,WT-Trieste}).  Matrix
theory configurations corresponding to supergravitons
\cite{BFSS,Sethi-Stern}, membranes \cite{dhn,BFSS} and fivebranes
\cite{grt,bss} have been identified.  The interactions between objects
in Matrix theory have been found to agree with supergravity
in a variety of situations.  Two-graviton interactions
have been shown to agree with supergravity up to terms of the
form $v^6/r^{14}$ \cite{bbpt,pss,pss2}.  The spin-dependent
interactions between two gravitons have been checked to leading order
in $1/r$
for each power of the fermion fields encoding the graviton spin by
performing one-loop matrix theory calculations
\cite{Kraus-spin,mss-2,McArthur,bhp,psw}.  It has been shown that the
first nonlinear gravitational correction to 3-graviton scattering is
correctly reproduced in a two-loop matrix theory calculation
\cite{Dine-Rajaraman,ffi,Mark-Wati-2,Okawa-Yoneya,deg2}.  A subset of
terms in the $N$-graviton interaction potential have been reproduced
by an $(N -1)$-loop matrix theory calculation, although there is an
indication that some 3-loop matrix theory terms may disagree with
4-graviton interactions in supergravity \cite{deg2}.

For more general matrix theory configurations, progress has been made
towards identifying the structure responsible for the agreement
between matrix theory and supergravity at leading orders.  It was
shown in \cite{Dan-Wati-2} that the supergravity potential between an
arbitrary pair of bosonic M-theory objects arising from the exchange
of quanta with zero longitudinal momentum is exactly reproduced by
terms in the one-loop matrix theory potential.  This result was
achieved through the identification of matrix theory quantities
corresponding to the supergravity stress-energy tensor,
membrane current and fivebrane current.  Linearized interactions
between multipole moments of the supergravity currents were shown to
be reproduced correctly by an infinite series of terms in the
effective matrix theory potential of the form $F^4 X^l/r^{7 + l}$.

In this work, we extend the results of \cite{Dan-Wati-2} to a pair of
completely general configurations, including non-vanishing fermionic
fields in our matrix theory background.  As in \cite{Dan-Wati-2}, we
do not require that the matrix theory configurations preserve any
supersymmetry.  We demonstrate in complete generality that the
on-shell matrix theory potential vanishes below $1/r^7$ and that at
orders $1/r^7$ and $1/r^8$ the matrix theory calculation reproduces
exactly the leading-order interactions expected from supergravity
arising from the exchange of gravitons, gravitinos, and three form
quanta with zero longitudinal momentum.  In addition to the
interactions considered in \cite{Dan-Wati-2}, these include a
supercurrent-supercurrent interaction mediated by gravitino exchange,
as well as membrane-fivebrane and zerobrane-sixbrane interactions
(previously considered in \cite{Berkooz-Douglas,bdflrs}).  We give
explicit matrix expressions for the integrated stress tensor, membrane
current and 5-brane current of general matrix theory objects,
including the fermionic contributions to these currents.  We identify
the fermionic contribution to the first moments of the currents,
generalizing the previously known result for the spin contribution to
graviton angular momentum \cite{Kraus-spin}.  We are also able to
identify matrix expressions for the fermionic components of the
supercurrent as well as a 6-brane current and its first moments.

One motivation for the work described in this paper is the goal of
developing a systematic formalism for describing general $N$-body
interactions in matrix theory in terms of structures such as the
stress tensor and membrane current which have natural interpretations
in supergravity.  The nonlinear structure of gravity can be probed by
studying 3-body processes in matrix theory.  As was mentioned in
\cite{Mark-Wati-2}, however, knowledge of the full two-body
interaction between objects with nonzero fermionic degrees of freedom
is a key step towards carrying out a general calculation of 3-body
interactions even between three purely bosonic configurations.

Another application of the results in this paper is to the
formulation of matrix theory in a general metric and 3-form
background.  If the matrix theory conjecture is correct, then the
effective action for matrix theory in a wide class of backgrounds can
be determined by including extra background blocks in the matrices and
integrating out the off-diagonal fields connecting to the background.
Armed with a precise knowledge of the form of the matrix theory stress
current, membrane and 5-brane currents, we propose a  concrete
algorithm for constructing the matrix theory action in a general
background, and we explicitly give the leading terms for this action
in the presence of a weak background
field.

It has been proposed that the full set of matrix theory interaction
terms which correspond to classical 2-body interactions in gravity
should take the form of a nonabelian Born-Infeld theory
\cite{Chepelev-Tseytlin2,Esko-Per2,Balasubramanian-gl}.  Assuming this
to be true, our result for the $1/r^7$ term in the effective 2-body
potential should be related to a supersymmetric nonabelian Born-Infeld
action.  We compare our results to previous results on supersymmetric
Born-Infeld theory \cite{brs,Metsaev-Rahmanov} and find that this
correspondence seems to hold.

In Section \ref{sec:action} we calculate to order $1/r^7$ the complete
one-loop matrix theory effective potential for a background
corresponding to general widely separated systems.  We also calculate
the potential to order $1/r^8$ including all terms which are at most
quadratic in the fermion backgrounds.  As the calculations are
somewhat involved, we present only a summary of some salient features
in Section \ref{sec:action}, leaving a more detailed account to
Appendix A.  The results for the effective potential are summarized in
Section \ref{sec:action-summary}.
    
In Section \ref{sec:interactions-currents} we describe the
supergravity interactions at orders $1/r^7$ and $1/r^8$ which arise
from the exchange of gravitons, gravitinos and 3-form quanta.  We show
that these interactions correspond precisely with the matrix theory
potential computed in Section \ref{sec:action} when the proper identification
is made of the matrix theory stress tensor, membrane current, 5-brane
current, 6-brane current and fermionic components of the
supercurrent.  The expressions we find for the supergravity currents
are summarized in subsection \ref{sec:currents-summary}.
     
Section \ref{sec:applications} contains a discussion of our results, a
comparison to other related results in the literature, and some
suggestions for how the results of this paper might be applied.  In
particular, we discuss the connection to higher-order terms in the
nonabelian Born-Infeld theory, we briefly discuss the application to
the general 3-body problem, we mention possible generalizations to
higher-dimensional theories, and we give a concrete suggestion for a
formulation of matrix theory in a general background using the matrix
forms we have found for the supergravity currents.

\section{Calculation of the effective action}
\label{sec:action}

We wish to study the Matrix Theory interaction between an arbitrary
pair of isolated systems, whose centers of mass are separated at time
$t$ by a distance $r$ which is large compared to the sizes of the two
individual systems.  In particular, we wish to calculate the matrix
theory effective action around a background which is block-diagonal in
both bosonic and fermionic matrices.  

There are a number of approaches which can be used to perform such a
calculation.  The simplest approach is to use the quasi-static
approximation to calculate the ground state energy of the harmonic
oscillator modes corresponding to the off-diagonal degrees of freedom
in the theory.  This approach was used in \cite{Dan-Wati} to compute
the leading term in the interaction potential between a pair of purely
bosonic matrix theory systems.  Unfortunately, in the presence of
fermionic backgrounds this approach cannot be applied in a
straightforward fashion.  The terms in the tree-level action which are
quadratic in the off-diagonal fluctuations include products of bosonic
and fermionic fluctuations, complicating the problem significantly.
Furthermore, as we shall discuss in more detail, unlike the case of a
purely bosonic background, in the case of a completely general
background there are non-vanishing terms involving fermions which
appear in the effective action below order $1/r^7$.  These terms
vanish when we enforce the matrix theory equations of motion to
calculate the effective potential between two physical systems; the
fact that such terms appear already at order $1/r^3$, however, makes
the validity of the quasi-static approximation in the general case
rather questionable.

Another approach one might consider using to calculate the effective
potential for general backgrounds would be to find a supersymmetric
completion of the bosonic potential of the form $F^4/r^7$ computed in
\cite{Dan-Wati}.  Unfortunately, however, the
supersymmetry transformations under which the tree level action is
invariant are corrected by subleading terms \cite{pss}.  Determining
precisely how the supersymmetry transformations are modified seems to
be just as difficult a problem as doing the full calculation of the
effective action.

In this paper we chose to do the calculation using the most
conceptually straightforward approach: an explicit summation over all
one-loop diagrams in the chosen background.  This approach leads to a
rather lengthy but finite calculation.  In this section we describe the
tools used to do the calculation and give brief descriptions of each
of the pieces of the calculations.  More details are included in
Appendix A for the reader interested in following the calculations in
detail or carrying out analogous calculations themselves.  The reader
primarily interested in the results may wish to skip from Section
\ref{sec:setup} directly to the summary of results for the effective
action in section \ref{sec:action-summary}

\subsection{Action and propagators}
\label{sec:setup}

We begin with the matrix theory action \footnote{We use Euclidian
conventions, taking $t \rightarrow i\tau$, $A \rightarrow
-i{\bf X}_0$. Throughout the work, indices $i,j,k,...$ run from 1 to 9 while
indices $a,b,c,...$ run from 0 to 9.}
\begin{eqnarray*}
S &=& -{1 \over 2 R} \int d\tau \tr \biggl\lbrace -D_{\tau} {\bf X}_i D_{\tau} 
{\bf X}_i
+ \frac{1}{2} [{\bf X}_i,{\bf X}_j] [{\bf X}_i,{\bf X}_j]
- \left(D_a^{\rm cl} {\bf X}_a^{{\rm qu}}\right)^2 \\
&& \qquad \qquad \qquad +  \Theta_{\alpha} D_{\tau} \Theta_{\alpha} - 
\Theta_{\alpha} 
\gamma^i_{\alpha \beta}
[{\bf X}_i,\Theta_{\beta}] \biggr\rbrace
\end{eqnarray*}
with a covariant background-field gauge fixing term
\[
D_a^{\rm cl} {\bf X}_a^{{\rm qu}} \equiv D_\tau {\bf X}_0 + 
i[B_i , {\bf X}_i]
\]
plus the corresponding ghost action not written here.
We expand the bosonic and fermionic matrices in terms
of background and fluctuation degrees of freedom
\begin{equation}
{\bf X}_a = \left[\begin{array}{cc}
\hat{r}_a + \hat{X}_a + \hat{Z}_a   & Y_a\\
\ Y_a^{\dagger} & \tilde{r}_a +\tilde{X}_a + \tilde{Z}_a  
\end{array}\right],
\;\;\;\;\; \;\;\;\;\;
\Theta_\alpha = \left[\begin{array}{cc}
\hat{\theta}_\alpha + \hat{\eta}_\alpha   & \chi_{\alpha}\\
\chi_{\alpha}^{\dagger} & \tilde{\theta}_\alpha + \tilde{\eta}_\alpha  
\end{array}\right] \; .
\label{fields}
\end{equation} 
where $r, X$ and $\theta$ describe background fields and $Y, Z, \chi$
and $\eta$ are the fluctuating fields.
The two systems are described respectively by hatted matrices of
size $\hat{N} \times \hat{N}$ and tilded matrices of size $\tilde{N}
\times \tilde{N}$.  We take $\hat{r}_0=0$, $\tilde{r}_0=0$, while
$\hat{r}_i$ and $\tilde{r}_i$ are constants, interpreted as the
centers of mass of the two systems at $t=0$ ($\tr(X_i)$ is taken to
vanish at $t=0$ for both systems).

We would like to compute the effective action obtained by integrating out the  
off-diagonal fields $Y$ and $\chi$ to one loop. For this purpose, we need only 
keep the terms quadratic in these fields, and we find that the relevant terms 
in the action (including the term quadratic in off-diagonal ghost fields) are:
\begin{eqnarray}
S_{quad} &=& -{1 \over R} \int d \tau  \left[
 Y_a^{\dagger} \left(
(\partial_{\tau}^2 - r^2 - 2r \cdot K - K^2 + i \dot{K_0} + 
2iK_0 \partial_{\tau})\delta_{ab} +2iF_{ab}\right)
Y_b \right. \nonumber \\
 && \hspace{1 in} + \chi^{\dagger} (\partial_{\tau} - \slash{r} - 
\slash{K})\chi 
 + \chi^{\dagger} \gamma^a L Y_a + Y^{\dagger}_a L \gamma^a \chi 
\label{squad}\\
\noalign{\vskip 0.2 cm}
 &&\left.
 \hspace{1 in} + C^* (\partial_{\tau}^2 - r^2 - 2r \cdot K - K^2 + i \dot{K_0} 
+ 
2iK_0 \partial_{\tau})C
 \right] 
\nonumber
\end{eqnarray}
Here, we are treating $Y_a$ and $\chi_\alpha$ as $(\hat{N}
\tilde{N})$-component vectors acted on by matrices
\begin{equation}
K_i \equiv \hat{X}_i \otimes \identity_{\tilde{N} \times \tilde{N}} 
-  \identity_{\hat{N} \times \hat{N}} \otimes \tilde{X}_i^T\; ,
\label{splits}
\end{equation}
\[
L_i \equiv \hat{\theta}_i \otimes \identity_{\tilde{N} \times \tilde{N}} 
-  \identity_{\hat{N} \times \hat{N}} \otimes \tilde{\theta}_i^T
\]
and
\begin{eqnarray*}
F_{0i} & = &  \partial_\tau K^i + i[K_0, K^i]\\
F_{ij} & = &  i[K^i, K^j]
\end{eqnarray*}
We have defined 
\[
\gamma^0 = -i
\]
(though this does not satisfy the usual anticommutation relations with
the other gamma matrices).  The only contribution from the quadratic
ghost term here will be a set of terms containing purely bosonic
fields which will cancel terms from the boson and fermion loops.
 
For our calculation, we shall treat all terms involving the matrices
$K, L$ and $F$ as vertices, 
using only the terms
\[
S_{prop} = -{1 \over R} \int d \tau (
 Y_a^{\dagger} (\partial_{\tau}^2 - r^2 )Y_a  + \chi^{\dagger} 
(\partial_{\tau} 
- \slash{r} )\chi ) 
\]
to determine the propagators. These are given simply by:
\[
<Y_a^{kl}(\tau)Y_b^{\dagger \, mn}(\sigma)> = \delta_{ab} \delta^{lm} 
\delta^{kn} \int{dk \over 2\pi} {e^{ik(\tau - \sigma)} \over k^2 +r^2} \equiv 
\delta_{ab} \delta^{lm} 
\delta^{kn} \Delta(\tau - \sigma) 
\] 
 for the bosonic propagator, and 
\[
 <\chi_{\alpha}^{kl}(\tau)\chi_{\beta}^{\dagger\, mn}(\sigma)> =  \delta^{lm} 
\delta^{kn} \int{dk \over 2\pi} {e^{ik(\tau - \sigma)} 
(\slash{r} + ik)_{\alpha \beta}\over k^2 +r^2} \equiv \delta^{lm} \delta^{kn} 
(\slash{r} + \partial_{\tau})_{\alpha \beta} 
\Delta(\tau-\sigma) 
\]
for the fermionic propagator. We thus have 
a fermion-fermion vertex,
\begin{equation}
\label{fermvert}
-\chi^{\dagger}\slash{K}\chi
\end{equation}
two mixed vertices,
\begin{equation}
\label{fermbos}
\chi^{\dagger} \gamma^a L Y_a \; ,
\end{equation}
\begin{equation}
\label{bosferm}
Y^{\dagger}_a L \gamma^a \chi.
\end{equation}
and boson-boson vertices given by
\begin{equation}
\label{bos1}
iY_a^{\dagger} \dot{K_0} Y_b \; ,
\end{equation}
\begin{equation}
\label{bos2}
2iY_a^{\dagger} K_0 \partial_{\tau} Y_a \; ,
\end{equation}
and
\begin{equation}
\label{bos3}
- Y_a^{\dagger} M_{ab} Y_b
\end{equation}
where we define:
\[
M_{ab} = 2r \cdot K +K^2 - 2iF_{ab}
\]

\subsection{Tools}

Before proceeding with the calculation, we make a few observations which help 
to 
simplify the calculation considerably.

\subsubsection{Gauge invariance}

First, we may take advantage of the 0+1 dimensional 
gauge symmetry, noting that by our choice of background field gauge, the 
effective action we calculate should be invariant under a $0+1$ dimensional 
gauge transformation. In particular, the result must contain only covariant 
derivatives, with $\partial_\tau$ and $K_0$ appearing in the combination 
$\partial_\tau X + i[K_0, X]$. Thus, we could set $K_0$ = 0, calculating terms 
with all numbers of derivatives and restoring $K_0$'s in the end by replacing 
derivatives with covariant derivatives. Alternatively, we may restrict the 
calculation to non-derivative terms, keeping a non-zero $K_0$, and deduce the 
derivative terms from the $K_0$ terms. We will use both approaches, 
depending on which part of the calculation we are interested in. One approach 
may also be used as a check of the other.

\subsubsection{0+0 dimensional calculation}

At this point, we note that the action (\ref{squad}) above is closely
related to the $0+0$ dimensional action
\begin{eqnarray}
S_{0+0} &=& -{1 \over R}  (
 Y_a^{\dagger} (-r^2)\delta_{ab} -M_{ab})Y_b \nonumber\\
 && \hspace{1 in} - \chi^{\dagger} (\slash{r} + \slash{K})\chi 
 + \chi^{\dagger} \gamma^a L Y_a + Y^{\dagger}_a L \gamma^a \chi )
\label{szero}
\end{eqnarray} which arises from the dimensional reduction of the theory to 0
dimensions.  This is the quadratic part in fluctuations of the
(0+0)-dimensional action which was used by Ishibashi, Kawai, Kitazawa
and Tsuchiya (IKKT) to conjecture a matrix formulation of IIB string
theory \cite{IKKT}.  If we restrict to considering terms involving no
derivatives or factors of $K_0$, we find that the $(0+0)$- and
$(0+1)$-dimensional calculations are exactly analogous, except that
the $(0+0)$-dimensional propagators are simply
\[
 <Y_a^{kl}(\tau)Y_b^{\dagger \, mn}(\sigma)> = \delta_{ab} \delta^{lm} 
\delta^{kn} {1 \over r^2} ,
 \]
 and 
\[
 <\chi_{\alpha}^{kl}(\tau)\chi_{\beta}^{\dagger\, mn}(\sigma)> =  \delta^{lm} 
\delta^{kn} {\slash{r}_{\alpha \beta} \over r^2} 
\]   
The form of the leading $1/r^8$ term in the bosonic 2-body potential
in the $(0 + 0)$-dimensional theory
was computed by IKKT in \cite{IKKT} and is the same up to a constant
as the $1/r^7$ 2-body term calculated for matrix theory in \cite{Dan-Wati}.
As we shall see below, the answer from the simpler $0+0$ dimensional 
calculation will often aid us in computing the full $0+1$ dimensional result.  

\subsubsection{$r \cdot K$ and $\slash{r}$ terms}

A further simplification arises from the fact that in the action 
(\ref{squad}), 
$r_i$ and 
$K_i$ appear only in the combination $(r_i + K_i)$ (a trivial consequence of 
their definitions). Thus, a 
transformation 
\begin{equation}
\label{rtrans1}
r_i \rightarrow r_i + \Lambda_i
\end{equation}
is exactly equivalent to a transformation 
\begin{equation}
\label{Ktrans1} 
K_i \rightarrow K_i + \Lambda_i.
\end{equation}
As described in detail in the body of the calculation, this property will 
allow 
us to ignore all terms containing $r \cdot K$ and certain terms with an $r^i$ 
coupled to a gamma matrix during the calculation, and to deduce these at the 
end 
using a transformation (\ref{Ktrans1}) on the remaining terms.

\subsubsection{D=10 Notation}

It turns out that a certain subset of terms in the effective action at each 
order are the dimensional reduction of $D=10$ Lorentz and gauge invariant 
terms, while many of the remaining terms are simply related to these 
by insertions of operators such as $\vec{K}^2$, $r \cdot K$ and $(D_0)^2$ 
which are not invariant under the $D=10$ symmetries but which do have the 
$D=0+1$ dimensional gauge invariance and $SO(9)$ rotational invariance of the 
Matrix Theory action. 
 
   Thus, we will find that the fermionic terms in the effective action may 
often be written most naturally and concisely using $D=10$ gamma matrices 
defined by
\begin{equation}
\Gamma^i = \left[
\ba{cc}
0 & \gamma^i \\
\gamma^i & 0
\ea
\right] \; , \; \; 
\Gamma^0 =\left[
\ba{cc}
0 & \gamma^0 \\
-\gamma^0 & 0
\ea
\right]
\label{eq:first-gamma}
\end{equation}
and a 32 component spinor $L$ whose first sixteen components are taken
to vanish and whose remaining components are the 16 component spinor
considered previously.  Note that these gamma matrices correctly
satisfy the relations defining the $D=10$ Clifford algebra,
\[
\{\Gamma^a, \Gamma^b\} = 2 \delta_{ab}.
\] 

  Given that the original action has a $D=10$ covariant form, one might wonder 
why not all terms in the one loop effective action may be written in this way. 
However, in computing the one loop potential, we integrate over only a single 
momentum rather than a ten-component momentum vector. Further, in writing our 
result, we make an expansion in the transverse seperation $r$. Neither of these 
respect the ten dimensional structure that we started with, so we should not 
expect the full one loop effective action to be the dimensional reduction of 
the $D=10$ result. 
  
\subsection{Summary of the calculation} 

We divide the calculation up by the number of fermionic fields
appearing in the result. The leading term with $2n$ fermions comes
from choosing $n$ each of the vertices (\ref{fermbos}) and
(\ref{bosferm}). Such a term will have $n$ bosonic and $n$ fermionic
propagators, and thus appear at order $(1/r)^{3n-1}$. Thus, up to order
$1/r^7$, where we expect the leading-order interactions, we will have
terms with zero, two, and four fermions.
 
\subsubsection{bosonic terms}

The purely bosonic terms at one loop were previously calculated for a
general background in \cite{Dan-Wati}, using the quasistatic
approximation in which the background fields (and their derivatives)
appearing in the action (\ref{squad}) were assumed to be
time-independent.  Since we are no longer restricting to this
approximation, it is interesting to check whether any higher
derivative terms appear up to $1/r^7$. In our calculation, these extra
derivative terms come from keeping higher order terms in the Taylor
expansions of the background fields about a particular time, rather
than assuming that the background fields are time independent.

Calculating with $K_0 = 0$, the non-vanishing vertices containing only bosonic 
background fields are (\ref{fermvert}), (\ref{bos3}), and the ghost vertex. 
The 
only diagrams we may build with these are loops with a single one of these 
vertex types repeated arbitrarily many times. These three types of 
contributions 
are calculated explicitly in Section \ref{sec:bosterms} of the appendix, and 
it 
is found that  
all terms with explicit derivatives cancel up to and including order $1/r^7$, 
leaving exactly the action 
\begin{equation}
\label{V7}
\Gamma^{\rm bos}_{1/r^7} =  \int d\tau {15 \over 16r^7}  
\str(F_{ab} 
F_{bc} F_{cd} F_{da} - {1 \over 
4} F_{ab} F_{ab} F_{cd} F_{cd})
\end{equation}
calculated in \cite{Dan-Wati}. Here, $\str$ denotes a symmetrized
trace in which we average over all possible orderings of the matrices
in the trace (treating any commutators such as $F$ as a unit).

Thus, to this order the result of our full calculation agrees with the
quasistatic calculation. However, as we shall see below, we expect the
appearance of higher-derivative terms at order $1/r^9$ which would not
appear in a quasistatic approximation. These are proportional to the
average of all possible insertions of $D_0^2$ (covariant derivative
squared) into the terms (\ref{V7}) of the $1/r^7$
action. Interestingly, the cancellation up to order $1/r^7$ is off
shell, that is, it does not require use of the equations of motion. We
will see that this is not the case for terms with fermionic fields.
   
We will also be interested in terms of order $1/r^8$. At this order,
we find two purely bosonic contributions. The first, studied in
\cite{Mark-Wati,Dan-Wati-2},
is proportional to an insertion of $r \cdot K$ into the $1/r^7$
action,
\[
-{105 \over 16}  \str(F_{ab} F_{bc} F_{cd} F_{da}(r \cdot K) - {1 \over 
4} F_{ab} F_{ab} F_{cd} F_{cd}(r \cdot K))
\]
The second contribution contains a nine index 
totally antisymmetric tensor coming from the trace of nine gamma matrices, and 
is given by
\[
{35r_s\over 256r^9}\str(F_{ij}F_{kl}F_{mn}\dot{K_p}K_q)\epsilon^{ijklmnpqs}
\]
We will discuss the physical interpretation of these two terms below.

\subsubsection{two-fermion terms}

We now move on to terms in the effective action with two factors of
the background fermion fields $L$. These terms arise from loops with
a single insertion of each of the two boson-fermion vertices
(\ref{fermbos}) and (\ref{bosferm}) to give two factors of $L$. In
addition, we may have an arbitrary number of boson-boson vertices and
fermion-fermion vertices.

The details of the calculation, performed using a non-vanishing $K_0$ and 
ignoring derivative terms, are found in Section \ref{sec:twoferm} of the 
appendix. In 
contrast to the case of the purely bosonic terms, we find that the off-shell 
action does not vanish below $1/r^7$, and in fact there are non-vanishing 
contributions to the effective action starting at $1/r^3$. However, if we are 
interested in the effective potential between two physical systems, we must 
require that the background fields satisfy the Matrix Theory equations of 
motion, and in this case, the non-vanishing contributions begin at order 
$1/r^7$, in agreement with supergravity.

   We now give our result for the two-fermion terms in the one loop matrix 
theory effective action to order $1/r^7$. As mentioned above, we have a subset 
of terms which are the dimensional reduction of a $D=10$ Lorentz and gauge 
invariant action. These are given in $D=10$ notation (omitting the leading 
$1/r^n$) by
\begin{eqnarray}
\Gamma^{\rm cov}_{1/r^3} &=&  \tr(\bar{L} \slash{D} L)\nonumber\\
\Gamma^{\rm cov}_{1/r^5} &=& 
         {3 \over 8}i \; \tr(\bar{L}\; \Gamma^b \; D_a F_{ab}\, L)\nonumber\\
         & & +{3 \over 16}i \; \tr(\bar{L}\; \Gamma^{[ab]} \; F_{ab}\, 
\slash{D} 
L)  
         -{3 \over 16}i \; \tr(\bar{L} \; \Gamma^{[ab]} \; \slash{D} L \, 
F_{ab})\nonumber\\ 
\Gamma^{\rm cov}_{1/r^7} 
&=& -{5 \over 64} \tr(\bar{L}\; \Gamma^e \Gamma^a \Gamma^b \Gamma^d \; F_{ab}
                                        \,  \slash{D} L\, F_{de} )
\label{eq:boff}\\ 
& & - {5 \over 16} \tr(\bar{L} \; \Gamma^a \Gamma^c \; F_{ab}\, 
F_{bc}\,\slash{D} 
L 
) 
- {5 \over 16} \tr(\bar{L} \; \Gamma^c \Gamma^a \; \slash{D} L \,F_{ab} 
\,F_{bc} 
)\nonumber\\
& & + {5 \over 32} \tr(\bar{L} \; \Gamma^d \Gamma^b \Gamma^c \;  D_a
         F_{ab} \, L \,F_{cd} )\nonumber\\
& & - {5 \over 16} \tr(\bar{L}\; \Gamma^c \; D_a F_{ab}\, L\, F_{cd} )      
- {5 \over 16} \tr(\bar{L} \; \Gamma^c \; F_{cb}\, D_a F_{ab}\, L)\nonumber\\ 
& & + {15 \over 16} \str(\bar{L} \; \Gamma^b \Gamma^c \Gamma^d \; F_{ab}\, 
F_{cd}\, 
D_a L) \nonumber
\end{eqnarray}     
All other terms up to order $1/r^7$ come from insertions of $r \cdot K$, 
$\vec{K}^2$, or $D_0^2$ into the covariant terms. To write these, we define an 
operation 
$\sym (\tr(A_1...A_n);B_1,...,B_n)$ to be the average of all possible 
different 
insertions of the operators $B_i$ between elements of the trace. For 
example, 
\begin{eqnarray*}
\lefteqn{\sym ( \tr(\bar{L} \; \slash{D} L);\vec{K}^2, D_0^2) =}\nonumber \\
   && {1 \over 6} \left[\tr(D_0^2 \bar{L}\;\vec{K}^2\; \slash{D} L) + 
\tr(\bar{L}\;D_0^2\vec{K}^2\; \slash{D} L) + \tr(\bar{L}\;\vec{K}^2\; 
D_0^2\slash{D} L) \right.\\
   & &\hspace{0.4in} \left.
+\tr(D_0^2\bar{L}\; \slash{D} L \; \vec{K}^2) + \tr(\bar{L}\; D_0^2\slash{D} 
L \; \vec{K}^2) + \tr(\bar{L}\; \slash{D} L \;D_0^2\vec{K}^2) \right]
\end{eqnarray*}
Note that any commutators in the trace are treated as single element. We can 
now write the full set of two-fermion terms in the $D=0+1$ off-shell action up 
to order $1/r^7$ as (again omitting factors of $1/r^n$)
\begin{eqnarray}
\Gamma_{1/r^3} &=& \Gamma^{\rm cov}_{1/r^3} \nonumber\\
\Gamma_{1/r^4} &=& -3\;\sym (\Gamma^{\rm cov}_{1/r^3};r \cdot K)\nonumber\\    
\Gamma_{1/r^5} &=& \Gamma^{\rm cov}_{1/r^5}
 + {1 \over 4} \sym (\Gamma^{\rm cov}_{1/r^3}; D_0^2)
             -{3 \over 2} \sym (\Gamma^{\rm cov}_{1/r^3};\vec{K}^2) 
              +{15 \over 2}\sym (\Gamma^{\rm cov}_{1/r^3};r \cdot K, r \cdot 
K)\nonumber\\ 
\Gamma_{1/r^6} &=& -5 \;\sym (\Gamma^{\rm cov}_{1/r^5}; r \cdot K)
-{15 \over 8}  \sym (\Gamma^{\rm cov}_{1/r^3}; D_0^2,r \cdot K)
\nonumber\\ 
             & &
             +{15 \over 2} \sym (\Gamma^{\rm cov}_{1/r^3};\vec{K}^2,r
             \cdot K) 
 -{35 \over 2} \sym (\Gamma^{\rm cov}_{1/r^3}; r \cdot K, r \cdot K,r
             \cdot K)\label{7off}\\    
\Gamma_{1/r^7} &=& \Gamma^{\rm cov}_{1/r^7}
      + {5 \over 8}\sym (\Gamma^{\rm cov}_{1/r^5}; D_0^2 )
       - {5 \over 2}\sym (\Gamma^{\rm cov}_{1/r^5}; \vec{K}^2) +
          {35 \over 2}\sym (\Gamma^{\rm cov}_{1/r^5}; r \cdot K, r \cdot 
K)\nonumber\\ 
       & & + {15 \over 8} \sym (\Gamma^{\rm cov}_{1/r^3}; \vec{K}^2, 
\vec{K}^2)
        -{15 \over 16} \sym (\Gamma^{\rm cov}_{1/r^3}; D_0^2, \vec{K}^2)
        +{1 \over 16} \sym (\Gamma^{\rm cov}_{1/r^3}; D_0^2,D_0^2)\nonumber\\
       & & -{105 \over 4} \sym (\Gamma^{\rm cov}_{1/r^3}; \vec{K}^2, r \cdot 
K, r 
\cdot K)
        +{105 \over 48} \sym (\Gamma^{\rm cov}_{1/r^3}; D_0^2, r \cdot K, r 
\cdot 
K)\nonumber\\
        & & +{315 \over 8} \sym (\Gamma^{\rm cov}_{1/r^3}; r \cdot K, r \cdot 
K,r 
\cdot K, r \cdot K)\nonumber
\end{eqnarray}

We will see that things simplify greatly when we restrict to a background that 
satisfies the matrix theory equations of motion. In our $D=10$ notation, these 
equations of motion read   
\begin{equation}
\label{fermeom}
\slash{D} L = 0
\end{equation}
and
\begin{equation}
\label{boseom}
D_aF_{ab} = i\bar{L}\Gamma^bL
\end{equation}
\junk{
These are the classical matrix theory equations of motion.  A complete
analysis would include quantum corrections to these equations.
Quantum effects will cause two types of corrections for the 2-body
system we are considering.  On the one hand there will be quantum
corrections to the equations of motion arising from precisely the
leading 2-body potential which we are computing.  These effects will
be suppressed by a factor of the order of $1/r^7$ and therefore can be
ignored in our computation.  On the other hand, there will also be
quantum corrections to the equations of motion for each body from
gravitational self-interactions.  These will correspond to
higher-order terms in the supergravity theory which will modify the
linearized supergravity interactions we will derive in the next
section.  A complete understanding of these higher-order terms could
be achieved by breaking each of the two bodies into multiple pieces
and considering the leading terms in the $N$-body interaction between
these pieces.
In order to compare matrix theory with linearized supergravity,
however, we can ignore all quantum corrections to (\ref{fermeom}) and
(\ref{boseom}).}

Examining the covariant terms above, we see that apart from the final
term in the $1/r^7$ action, all terms listed contain either $\slash{D}
L$ or $D_aF_{ab}$, and so these terms either cancel directly or may be
canceled by the four-fermion terms which we will soon calculate.

In the applications below, we will also be interested in certain terms at $1/ 
r^8$. We find that apart from terms containing $r \cdot K$, the two 
fermion terms at $1/r^8$ are given by the $D=10$ gauge invariant expression 
\begin{eqnarray}
\Gamma_{1/r^8} &=& -{35 \over 256r^9}r_g \str(\bar{L}\, F_{ab} F_{cd} F_{ef} 
\Gamma^{[abcdefg]} L) -{105 \over 32r^9}r_e \str(\bar{L}\, F_{ab} F_{bc} 
F_{cd} 
\Gamma^{[ade]} L)\nonumber\\
 & & -{105 \over 128r^9}r_e \str(\bar{L}\, F_{ab} F_{ab} F_{cd} \Gamma^{[cde]} 
L)
\label{cov8}
\end{eqnarray}
Again, we'll see the physical interpretation of these terms below.

\subsubsection{four-fermion terms}
\label{sec:4ferms}

 The remaining terms in the one-loop Matrix theory effective action up to 
order $1/r^7$  contain four fermions. These arise from a loop with two each of 
the vertices  (\ref{fermbos}) and (\ref{bosferm})  plus insertions of the 
boson-boson vertices and the fermion-fermion vertex.  

   The calculation of four-fermion terms, performed setting $K_0$ to 0 and 
calculating derivative terms explicitly, appears in Section \ref{sec:fourferm} 
of the 
appendix. The calculation relies heavily on the use of Fierz identities for 
$16 \times 16$ symmetric gamma matrices, and a discussion of these as they 
apply to non-abelian four-fermion terms is given in Appendix B.

We find first a set of terms (\ref{df5}), (\ref{dfterms}), (\ref{df6})
listed in the appendix which are obtained simply by replacing
$D_aF_{ab}$ with $D_aF_{ab} - i\bar{L}\Gamma^bL$ each time it appears
in the two-fermion terms. These terms serve to cancel all two-fermion
terms containing $D_aF_{ab}$ in the on-shell effective action.
   
The remaining four-fermion terms which do not contain $\slash{D} L$ all
appear at $1/r^7$, so we indeed have complete cancellation of all
terms below $1/r^7$ in the on-shell effective potential, in agreement
with supergravity. These remaining four-fermion terms are given by

\begin{eqnarray}
\Gamma^{4L}_{1/r^7} 
&=& {5 \over 32} \left( \tr(\bar{L} \Gamma^{a} D_bL \; \bar{L} \Gamma^{a} 
D_bL) 
+ 
\tr(D_b\bar{L} \Gamma^{a} L \; D_b\bar{L} \Gamma^{a} L) + 2 \tr(\bar{L} 
\Gamma^{a} 
L \; D_b\bar{L} \Gamma^{a} D_bL) \right) \nonumber\\ 
& & + {5i \over 32} \left( \tr(\bar{L} \Gamma^{c} F_{ab} L \; \bar{L} 
\Gamma^{[cab]} 
L)+\tr(\bar{L} \Gamma^{[cab]} L F_{ab} \bar{L} \Gamma^{c} L)+\tr(\bar{L} 
\Gamma^{c} \bar{L} F_{ab} L \Gamma^{[cab]} L)   \right)
\nonumber\\   
& & + {5 \over 128}\left( \tr(L \gamma^{[ki]} [\dot{K_i} , L] \; L \gamma^k L) 
+ 
\tr(L 
\gamma^{[ki]} L [\dot{K_i} , L] \gamma^k L) \right)
\nonumber\\
& & - {5 \over 128} \left( \tr(L \gamma^{[kl]} \dot{L} \; 
\dot{L} \gamma^{[lk]} L) + 2\tr(L \gamma^{k} \dot{L} \; \dot{L} \gamma^{k} L) 
+ 
6\tr(L  \dot{L} \; \dot{L}  L) \right) \label{eq:4L}\\ 
& & + { 5 \over 256} \left( \tr(L \gamma^{[kli]} K_i \dot{L} \; L 
\gamma^{[lk]} 
L)+6\tr(L \slash{K} \dot{L} \; L L)+2\tr(L K_i \dot{L} \; L \gamma^{i} 
L)\nonumber 
 \right. \\
& & \left. \hspace{0.5in} - \tr(\dot{L} \gamma^{[kli]} K_i L \; L 
\gamma^{[lk]} 
L)-6\tr(\dot{L} \slash{K} L \; L L)-2\tr(\dot{L} K_i L \; L \gamma^{i} L) 
\right)\nonumber\\
& & + { 15 \over 256} \left( \tr(L \gamma^{[kli]} K_i L \; \dot{L} 
\gamma^{[lk]} 
L)+6\tr(L \slash{K} L \; \dot{L} L)+2\tr(L K_i L \; \dot{L} \gamma^{i} 
L)\nonumber 
 \right. \\
& & \left. \hspace{0.5in}- \tr(L \gamma^{[kli]} K_i L \; L \gamma^{[lk]} 
\dot{L})-6\tr(L 
\slash{K} L \; L \dot{L})-2\tr(L K_i L \; L \gamma^{i} \dot{L})
\right)
\nonumber\\ 
& & + { 5 \over 64} \left( \tr(L \gamma^{[kli]} K_i L \; L \gamma^{[lkj]} K_j  
L)+2\tr(L \gamma^{[ki]} K_i L \; L \gamma^{[kj]} K_j  L)
\nonumber \right. \\
& & \left. \hspace{0.5in}
+6\tr(L \gamma^{i} K_i L 
\; L \gamma^{j} K_j  L) +2\tr(L \gamma^{j} K_i L \; L \gamma^{i} K_j
L)
\nonumber \right. \\
& & \left. \hspace{0.5in}-2\tr(L 
\gamma^{i} K_j L \; L \gamma^{i} K_j  L)+2\tr(L K_i L \; L K_i L)
\right) \nonumber
\end{eqnarray}

Here, we note that some of the terms have a $D=10$ form, while others do not 
seem to combine into $D=10$ covariant expressions. In particular, it appears 
that there is no way to rewrite the above expression in a form with 
all $K$'s appearing in commutators, even taking into account all possible 
Fierz identities. 

\subsection{Summary of results for effective potential}
\label{sec:action-summary}

We have now calculated the complete one-loop matrix theory effective action to 
order $1/r^7$ for a background corresponding to two widely separated systems. 
The off-shell action is given by expressions (\ref{eq:boff}), (\ref{7off}), 
(\ref{eq:4L}), (\ref{df5}), (\ref{dfterms}) and (\ref{df6}) and has 
non-vanishing contributions beginning at order $1/r^3$. However, as we saw 
above, enforcing the Matrix theory equations of motion (\ref{fermeom}) and 
(\ref{boseom}) we find complete cancellation of all terms at orders less than 
$1/r^7$, in agreement with supergravity. The on-shell effective action (which 
we may interpret as the negative of an effective potential for the system) is 
given 
by
\begin{eqnarray}
\label{on7}
\Gamma_{1/r^7} &=&  {15 \over 16r^7}  \str(F_{ab} F_{bc} F_{cd} F_{da} - {1 
\over 
4} F_{ab} F_{ab} F_{cd} F_{cd})\\ 
& & + {15 \over 16r^7} \str(\bar{L} \; \Gamma^b \Gamma^c \Gamma^d \; F_{ab}\, 
F_{cd}\, D_a L) 
 + \Gamma^{4L}\nonumber
\end{eqnarray}
where $\Gamma^{4L}$ is given in (\ref{eq:4L}) (replacing, if we like, factors 
of 
$\dot{L}$ 
by $[\slash{K},L]$).
At $1/r^8$, we have terms proportional to the insertion of $r \cdot K$
into the
symmetrized trace of the $1/r^7$ terms plus additional terms.
The full expression to quadratic order in in the fermion backgrounds
is given by
\begin{eqnarray}
\label{on8}
\Gamma_{1/r^8} &=& -7\;\sym (\Gamma_{1/r^7}; r \cdot K) +
{35r_s\over 256r^9} \str(F_{ij}F_{kl}F_{mn}\dot{K_p}K_q) 
\epsilon^{ijklmnpqs}\\
& & -{35 \over 256r^9}r_g \str(\bar{L}\, F_{ab} F_{cd} F_{ef} 
\Gamma^{[abcdefg]} L) -{105 \over 32r^9}r_e \str(\bar{L}\, F_{ab} F_{bc} 
F_{cd} 
\Gamma^{[ade]} L)\nonumber\\
 & & -{105 \over 128r^9}r_e \str(\bar{L}\, F_{ab} F_{ab} F_{cd} \Gamma^{[cde]} 
L)\nonumber
 \end{eqnarray}
There are also four-fermion and six-fermion terms which contribute to
$\Gamma_{1/r^8}$ which we have not calculated.  In the next section,
we will compare the terms in (\ref{on7}) and (\ref{on8}) to
leading-order supergravity interactions, and will find a complete
interpretation for all of these terms in terms of physical quantities
familiar from supergravity.

\section{Comparison to supergravity}
\label{sec:interactions-currents}

In order to compare our Matrix theory result with supergravity, we
would like to understand from a supergravity point of view the
interactions that exist between two arbitrary widely separated systems
at leading orders in the inverse separation distance. 
The leading terms in the long-range effective potential between two
separated systems are described in supergravity by the linearized
theory.  The terms in the linearized theory arising from graviton and
3-form exchange between arbitrary bosonic systems were analyzed in
\cite{Dan-Wati-2} and shown to agree with matrix theory for processes
with no longitudinal momentum transfer.  In this
section we generalize that discussion to include fermionic
backgrounds.  This gives rise to additional contributions from the
fermion backgrounds to the gravitational, ``electric'' and
``magnetic'' interactions mediated by the graviton and 3-form quantum,
as well as new interactions arising from gravitino exchange.
Extending the analysis out to order $1/r^8$ we also find new
interactions in the bosonic sector describing ``dyonic''
membrane-5-brane and 0-brane-6-brane interactions.

All the interactions in the linearized theory can be written in a
current-current form corresponding to an instantaneous potential
in light-front time
between the two separated systems proportional to the product of
moments of the stress tensor, membrane current, 5-brane current,
6-brane current and
fermionic components of a supercurrent.  To identify the matrix theory
and supergravity interaction potentials the only structure needed is
the detailed form of these currents in matrix theory language,
expressed as traces of products of the backgrounds for the two
systems.  A synopsis of our results for these currents is given in
section \ref{sec:currents-summary}.

\subsection{Linearized supergravity interactions}
\label{sec:linear-supergravity}

In this subsection we discuss the general structure of the linearized
supergravity interactions we expect to reproduce in matrix theory.
The detailed forms of the specific interactions are discussed and
compared to matrix theory in the following subsections.
The propagating fields of eleven dimensional supergravity are a graviton, a
three form gauge field, and a gravitino.  The classical linearized
supergravity theory arises from considering all tree-level processes
in which a single quantum is exchanged between two classical sources.
For an eleven-dimensional spacetime
with one compact direction, the propagators for all the fields go like 
$1/r^7$,
so at orders below $1/r^{14}$ all classical supergravity
interactions can be described by the linearized theory.

Arbitrary sources can be coupled linearly to the supergravity fields by adding
extra terms to the action of the form
\begin{equation}
\label{sources}
S = S_{{\rm SUGRA}} (h, A, \psi)
 + \int d^{11}x(h_{IJ}T^{IJ} + A_{IJK}J^{IJK} +  
A^D_{IJKLMN} M^{IJKLMN} +
\bar{\psi}_{\alpha I} S_\alpha^I + \bar{S}_\alpha^I \psi_{\alpha I})
\end{equation}

where $T^{IJ}, J^{IJK}, M^{IJKLMN}$ and $S_\alpha^I$ are the
stress-energy tensor, membrane current, 5-brane current and fermionic
supercurrent components of the source.  The field $A^D$ has a 7-form
field strength dual\footnote{The true duality relation is somewhat
more complicated, however we ignore the additional terms in the
linearized theory
since they
are products of more than one field.} to that of the 3-form field $A$
\begin{equation}
F^D = dA^D ={}^* F={}^* (dA).
\label{eq:duality-constraint}
\end{equation}
It is difficult to formulate a consistent quantum theory which
contains both $A$ and $A^D$ and which
couples
both to membranes and 5-branes since (\ref{eq:duality-constraint})
is difficult to impose at the quantum level.  We are only interested
in the classical theory here, however, so we may impose
(\ref{eq:duality-constraint}) as a classical condition.
This still leads to some complications since there is not always a
single-valued solution for $A^D$ of (\ref{eq:duality-constraint}) for
a given field $A$.  We discuss this issue further in \ref{sec:memfive}.

The classical interactions in the linearized theory can be determined
by simply taking the quadratic part of the supergravity action and
solving explicitly for the propagating fields in the presence of the
given background.  If we have a background containing two
well-separated objects,
we may treat one object as a source, solve for the fields produced by
this source, and use (\ref{sources}) to find the effective action of
the second object which is treated as a probe in the fields produced
by the first object.

Another approach to explicitly describing the linearized theory, which
is slightly more transparent in the light-front formalism, is to
follow the standard field theory prescription for writing the
interactions in terms of the propagators of the gravity fields.
Keeping only the quadratic terms in the supergravity action, we have
interactions arising from graviton, 3-form and gravitino exchange.  To
study interactions between two isolated systems, we assume that each
of the sources in (\ref{sources}) may be decomposed into the sum of
two terms whose supports are separated by some large distance, for
example $T^{IJ}$ = $\hat{T}^{IJ} + \tilde{T}^{IJ}$. The leading-order
interactions are then given by all diagrams of the form 
\begin{equation}
\label{diag}
\widehat{\times} \! \! \! \!- \! \!\! -\!\! \! -\!\! \!-\!\! \!-
\!\!\! - \!\!\! \widetilde{\times} 
\end{equation}
coupling a hatted source to a tilded source.  The following
non-vanishing terms appear in the effective potential (up to
overall
coefficients):
\begin{eqnarray}
V_{\rm gravity} &=& \int d^{11}x \int d^{11}y \; \hat{T}^{IJ}(x) \langle
h_{IJ}(x)  
h_{KL}(y) \rangle \tilde{T}^{KL}(y)\label{vgra}\\
V_{\rm electric} &=& \int d^{11}x \int d^{11}y \; \hat{J}^{IJK}(x)
\langle A_{IJK}(x)  
A_{LMN}(y) \rangle \tilde{J}^{LMN}(y)\label{vel}\\
V_{\rm magnetic} &=& \int d^{11}x \int d^{11}y \; \hat{M}^{IJKLMN}(x) \langle 
A^D_{IJKLMN}(x) A^D_{PQRSTU}(y) \rangle \tilde{M}^{PQRSTU}(y)\label{vmag}\\
V_{\rm super} &=& \int d^{11}x \int d^{11}y \;
\bar{\hat{S}}_\alpha^{I}(x) \langle  
\psi_{\alpha I}(x) \bar{\psi}_{\beta J}(y) \rangle \tilde{S}_\beta^{J}(y) 
\label{vsup}\\
& & + \int d^{11}x \int d^{11}y \; \bar{\tilde{S}}_\alpha^{I}(x) \langle 
\psi_{\alpha 
I}(x) \bar{\psi}_{\beta J}(y) \rangle \hat{S}_\beta^{J}(y)\nonumber 
\end{eqnarray}
In addition, there are membrane-fivebrane interactions of the form
$\hat{J} \tilde{M}$ and $\hat{M} \tilde{J}$ proportional to a
propagator $\langle A (x) A^D (y)\rangle$.  These terms, however,
cannot be completely described by a potential.  These terms are
discussed separately in section (\ref{sec:memfive}) below.

To compare with matrix theory, we must take supergravity on a
spacetime with a lightlike direction compactified, and also consider
only processes involving the exchange of quanta with zero longitudinal
momentum. The appropriate Green's function and propagators for the
bosonic fields are
discussed and calculated in  \cite{Dan-Wati-2}.  Denoting the light-front time
coordinate by $x^+$, the compact direction by $x^-$ , and the
remaining spatial directions by $\vec{x}$, the appropriate propagators
are all proportional to
\[
\delta(x^+ - y^+){1 \over |\vec{x} - \vec{y}|^7}
\]
Note that this is independent of $x^-$, a consequence of the fact that
we have restricted to the exchange of quanta with zero momentum in
$x^-$, which therefore have wavefunctions spread evenly over this
direction. Also notable is the fact that such a propagator gives rise
to interactions which are instantaneous with respect to light-front
time (see, for example \cite{Hellerman-Polchinski}).

We now show schematically how we can relate these supergravity interactions 
to the matrix theory potential we have calculated. Examining the 
interactions (\ref{vgra}-\ref{vsup}) above, we see that all have the form 
\begin{equation}
\int d^{11}x \int d^{11}y \; \hat{C}(x) \langle \phi(x) \phi(y) \rangle 
\tilde{C}(y)
\label{eq:general-interaction}
\end{equation}
where $C$ is a general source and $\phi$ is a propagating field.  Note
that in general, a gauge-fixing choice may be necessary to explicitly
calculate the propagator and to determine how the components of the
tensor currents are contracted.
Since the propagator is independent of $x^-$ and $y^-$, the integrals over 
these variables act only on the background currents. We may rewrite the 
interaction as an explicit series in $1/r$ by replacing the currents $C(x)$ by 
equivalent distributions
\begin{equation}
\label{dist}
\int dx^- C(x^+, x^-, \vec{x}) = C(x^+)\delta(\vec{x} - \vec{x}_0) - 
C^{(i)} (x^+)\partial_i \delta(\vec{x} - \vec{x}_0)
 +{1 \over 2} C^{(ij)} (x^+) \partial_i 
\partial_j \delta(\vec{x} - \vec{x}_0) + \cdots
\end{equation}
where we define the spatial moments of the current $C(x)$ about the point 
$\vec{x}_0$ by 
\[
C^{(i_1...i_n)}(x^+) = \int dx^- d\vec{x}\; C(x^+, x^-, \vec{x}_0 + \vec{x})\; 
x^{i_1} \cdots x^{i_n}
\]
We choose the points $\vec{x}_0$ and $\vec{y}_0$ to be the centres of mass of 
the two objects at some initial time $t = x^+ = 0$, so that $\vec{x}_0 - 
\vec{y}_0$ may be identified with the vector $\vec{r}$ defined in our matrix 
theory background. With these definitions, the supergravity
interaction (\ref{eq:general-interaction})
may be 
rewritten as (up to an overall proportionality constant)
\begin{eqnarray}
\lefteqn{\int d\tau \sum_{n=0}^{\infty} \sum_{k=0}^n {(-1)^k \over 
k!(n-k)!}\hat{C}^{(i_1...i_k)}\tilde{C}^{(j_{k+1}...j_n)}\partial_{i_1}
\ldots
\partial_{j_n}({1 \over r^7})=}\nonumber \\
&&\hspace{1in}
 \int d\tau \{{1 \over r^7} \hat{C}\tilde{C} + 7{r^i \over r^9} ( 
\hat{C}^{(i)}\tilde{C} + \hat{C} \tilde{C}^{(i)}) + \cdots \}
\label{super}
\end{eqnarray}
Hence the supergravity interactions we are interested in may be written order 
by order in $1/r$ with terms at each order coupling some moment of the current 
for the hatted system with a moment of the current for the tilded system.
      
Let us compare this discussion to the general form of the
corresponding potential in matrix theory.  In Matrix theory, recalling
the definitions (\ref{splits}), we see that each term in the one-loop
matrix theory effective action is a single trace of a product of
matrices which are the tensor product of an $\hat{N} \times \hat{N}$
matrix and a $\tilde{N} \times \tilde{N}$ matrix.  Thus, the trace in
a given term splits into the product of two traces as 
\begin{equation}
\label{decomp}
\tr\left(
(\hat{A}_1 \otimes \tilde{A}_1) \cdots  (\hat{A}_n \otimes
\tilde{A}_n)\right) = 
\tr(\hat{A}_1 \cdots \hat{A}_n) \tr(\tilde{A}_1 \cdots \tilde{A}_n)
\end{equation}
Exactly as in the supergravity expression (\ref{super}), the one-loop matrix 
theory action may be written as a series in $1/r$ with each term coupling a 
single hatted quantity to a tilded quantity. This suggests that for each 
moment 
of a given supergravity current, we may identify a corresponding matrix theory 
quantity given by a single trace of matrix theory variables.

    We will now examine the explicit forms of the supergravity and matrix 
theory potentials and show that this correspondence is precise, identifying 
explicit quantities corresponding to moments of the supergravity currents and 
showing that the supergravity interactions are exactly reproduced in the one 
loop matrix theory action.

\subsection{Gravitational and 3-form interactions
at order $1/r^7$}   
      
We first consider the interactions (\ref{vgra}), (\ref{vel}), and 
(\ref{vmag}) mediated by the 
exchange of bosonic particles. 
In \cite{Dan-Wati-2} these terms were analyzed for purely bosonic
backgrounds.  It was shown that the leading contribution to the long-range
potential of both linearized supergravity and matrix
theory  could be expressed in the form
\begin{eqnarray}
V_{\rm sugra} &=& V_{\rm gravity} + V_{\rm electric} + V_{\rm 
magnetic} \label{eq:linear-supergravity}\\
V_{\rm gravity} &=& - {15 R^2 \over 4 r^7} \left(\hat{T}^{IJ} \tilde{T}_{IJ} - 
{1 \over 9} \hat{T}^I{}_I
\tilde{T}^J{}_J\right)\label{vg2} \\
V_{\rm electric} &=& - {45 R^2 \over r^7} \hat{J}^{IJK} 
\tilde{J}_{IJK}\label{ve2}\\
V_{\rm magnetic} &=& - {3 R^2 \over 2 r^7} \hat{M}^{IJKLMN}
\tilde{M}_{IJKLMN}\label{vm2}. 
\end{eqnarray}
where in the supergravity theory the tensors $T^{IJ},J^{IJK}$
and $M^{IJKLMN}$ are identified with the integrated stress
tensor, membrane current and 5-brane current of the two separated
systems.  The identification with matrix theory was made by
identifying matrix forms for each of these tensors written as traces
of functions of the bosonic fields describing the background.

In the more general situation we are considering here with background
fermions, we can perform a similar analysis.  We can decompose the
leading terms in the matrix theory effective potential (\ref{on7}) in
terms of a sum of products of traces of the bosonic and fermionic
degrees of freedom of the two systems as in (\ref{decomp}), where we
only retain terms with bosonic indices contracted between the hatted
and tilde'd variables.  We find that in general this part of the
effective potential (\ref{on7}) can be expressed in the form
(\ref{eq:linear-supergravity}), where in the presence of fermionic
backgrounds the matrix theory expressions for the integrated 
currents are defined by (returning to a Minkowski 
formalism)
\begin{eqnarray}
T^{++} &=& {1 \over R}\str\left(\identity\right)\nonumber\\
T^{+i} &=& {1 \over R}\str\left(\dot{X_i}\right)\nonumber\\
T^{+-} &=& {1 \over R}\str\left({1 \over 2} \dot{X_i} \dot{X_i} + {1 \over 4} 
F_{ij}  
F_{ij} + {1 \over 2} \theta\gamma^i[X^i,\theta]\right)\nonumber\\
T^{ij} &=& {1 \over R}\str\left( \dot{X_i} \dot{X_j} +  F_{ik}  F_{kj} - {1 
\over 4} 
\theta\gamma^i[X_j,\theta] - {1 \over 4} 
\theta\gamma^j[X_i,\theta]\right)\nonumber\\
T^{-i} &=& {1 \over R} \str\left({1 \over 2}\dot{X_i}\dot{X_j}\dot{X_j} + {1 
\over 
4} \dot{X_i} F_{jk} F_{jk} + F_{ij} F_{jk} \dot{X_k}\right) \nonumber\\ & & - 
{1 \over 
4R} 
\str\left(\theta_\alpha 
\dot{X_k}[X_m,\theta_\beta]\right)\{\gamma^k\delta_{im} 
+\gamma^i\delta_{mk} -2\gamma^m\delta_{ki} \}_{\alpha \beta}\nonumber\\ & & - 
{1 
\over 8R} 
\str\left(\theta_{\alpha} F_{kl}[X_m,\theta_{\beta}]\right)\{ \gamma^{[iklm]} 
+ 
2 \gamma^{[lm]} \delta_{ki} + 4\delta_{ki}\delta_{lm} \}_{\alpha 
\beta}\nonumber\\  & & + 
{i \over 8R} \tr(\theta \gamma^{[ki]} \theta \; \theta \gamma^k 
\theta)\nonumber\\
T_f^{--} &=& {1 \over 4R} \str\left(F_{ab}F_{bc}F_{cd}F_{da} - {1 \over 
4}F_{ab} 
F_{ab} F_{cd} F_{cd}  + {\theta} \Gamma^a \Gamma^b \Gamma^c F_{ab} F_{cd} 
D_a\theta + {\cal O} ({\theta^4})\right)\nonumber\\
J^{+ij} &=& {1 \over 6R} \str\left(F_{ij}\right) \label{eq:currents}\\
J^{+-i} &=& {1 \over 6R} \str\left( F_{ij} \dot{X_j} - {1 \over 2} 
\theta[X_i,\theta] 
+ {1 
\over 4} \theta \gamma^{[ki]} [X_k, \theta]\right)\nonumber\\
J^{ijk} &=& {1 \over 6R} \str\left( \dot{X_i} F_{jk} +  \dot{X_j}
F_{ki}  +
\dot{X_k} F_{ij} -{1 \over 4} \theta 
\gamma^{[ijkl]}[X_l,\theta]\right)\nonumber\\
J^{-ij} &=& {1 \over 6R} \str\left(+\dot{X_i} \dot{X_k} F_{kj} - 
\dot{X_j}\dot{X_k} 
F_{ki} - {1 \over 2} \dot{X_k}\dot{X_k} F_{ij} + {1 \over 4}F_{ij} F_{kl} 
F_{kl} 
+ F_{ik} F_{kl} F_{lj}\right)\nonumber\\
& & +{1 \over 24R} \str\left(\theta_\alpha 
\dot{X_k}[X_m,\theta_\beta]\right)\{\gamma^{[kijm]} + 
\gamma^{[jm]} \delta_{ki} - \gamma^{[im]} \delta_{kj} + 2 \delta_{jm} 
\delta_{ki} - 2 \delta_{im} \delta_{kj}\}_{\alpha \beta}\nonumber\\
& & + {1 \over 8} \str\left(\theta_{\alpha} 
F_{kl}[X_m,\theta_{\beta}]\right)\{\gamma^{[jkl]} 
\delta_{mi} - \gamma^{[ikl]} \delta_{mj} + 2 \gamma^{[lij]} \delta_{km} + 2 
\gamma^l \delta_{jk} \delta_{im} - 2 \gamma^l \delta_{ik} 
\delta_{jm}\nonumber\\
& & \hspace{1in} + 2 \gamma^j \delta_{il} \delta_{km} - 2 \gamma^i \delta_{jl} 
\delta_{km}\}_{\alpha \beta}\nonumber\\ & & + {i \over 48R} \str\left(\theta 
\gamma^{[kij]} \theta \; \theta 
\gamma^k \theta - \theta \gamma^{[ij]} \theta \; \theta 
\theta\right)\nonumber\\
M^{+-ijkl} &=& {1 \over 12R} \str\left(F_{ij}F_{kl} +F_{ik}F_{lj} + 
F_{il}F_{jk} + 
\theta \gamma^{[jkl}[X^{i]},\theta]\right)\nonumber
\end{eqnarray}
The hatted and tilded currents $\hat{T}^{IJ}, \ldots$ are given by
using the appropriate variables $\hat{X}, \hat{\theta}$ or $\tilde{X},
\tilde{\theta}$ in these expressions.  Time derivatives are taken with
respect to Minkowski time $t$.  
In these expressions we have used the definitions $F_{0i} = \dot{X}^i,
F_{ij} = i[X^i, X^j]$.
The order $\theta^4$ term in $T^{--}$ is given by (\ref{eq:4L}) where
$K$ and
$L$ are replaced by $X$ and $\theta$.

A few comments may be helpful regarding the derivation of the currents
(\ref{eq:currents}).  The purely bosonic contributions to these
tensors are those given in \cite{Dan-Wati-2}.  To determine the
fermionic contributions from (\ref{on7}) it is helpful to write $C =
C_{bos} + C_{f}$ (for $C$ = $T$, $J$ or $M$, suppressing tensor
indices in this schematic discussion), where $C_{bos}$ is the set of
purely bosonic terms in $C$ and $C_f$ contains all other terms.  If
the matrix theory potential contains the interactions (\ref{vg2}),
(\ref{ve2}), and (\ref{vm2}), then it must contain terms corresponding
to $\hat{C}_{bos} \tilde{C}_{f}$, $\hat{C}_{f} \tilde{C}_{bos}$, and
$\hat{C}_{f} \tilde{C}_{f}$ in addition to the purely bosonic terms
$\hat{C}_{bos} \tilde{C}_{bos}$.

The terms $\hat{C}_{bos} \tilde{C}_{f}$ and $\hat{C}_{f}
\tilde{C}_{bos}$, in particular, each have one trace with purely
bosonic fields and another trace which contains fermion fields. If we
isolate such terms in the matrix theory effective potential, we can
simply read off all the components of $T_f$, $J_f$, and $M_f$. For
example, to determine $T_f^{-i}$ we collect all terms in the leading
order potential (\ref{on7}) of the form
$\tr(-i\dot{\hat{X}_i})\tr(\tilde{A}_i)$ and identify $\tilde{T}_f^{-i}$
as the sum of all the $\tr(\tilde{A}_i)$ terms appearing (with an overall
coefficient determined by (\ref{vg2})).  Note that most of the bosonic
currents have more than one term, so that this presciption is only
consistent if each bosonic term in a given current couples to exactly
the same set of terms containing fermions.  The Matrix theory
potential we have calculated passes this non-trivial consistency check
with the currents defined as above.

Note that,
on first glance, it appears that
the last two terms in (\ref{eq:4L}) should yield terms not
accounted for in the preceding analysis, where $\tr(X_i X_j)$ couples
to a four-fermion expression and $\tr(X_i)$ couples to a four-fermion
expression.  Both of these four-fermion expressions vanish, however,
by Fierz identities.

Note also that only the component $M^{+-ijkl}$ of the 5-brane current
has been identified.  Only this component appears in the supergravity
interaction because 5-branes which are not wrapped around the
longitudinal ($x^-$) direction decouple from the theory so that
$M^{ijklmn} = M^{+ijklm}= 0$.  This decoupling follows from the fact
that 5-branes act as Dirichlet boundaries for membranes, and so do not
appear as interacting transverse objects in the light-front formalism
\cite{bss}.  The remaining component $M^{-ijklm}$ can be determined
from current conservation \cite{mvr}, as we will discuss shortly.

 To complete the demonstration that $V_{\rm gravity}$, $V_{\rm electric}$, and 
$V_{\rm magnetic}$ are completely reproduced in Matrix theory at leading 
order, we 
should show that the terms $\hat{T}_{f} \tilde{T}_{f}$, and the corresponding 
terms for $J$ and $M$ also appear in the matrix theory potential. These must 
appear in the decomposition of the matrix theory potential as terms with two 
fermions in each trace (since $T$, $J$, and $M$ may have no free fermionic 
indices) and therefore come from the four-fermion terms (\ref{eq:4L}). 

We note here that any decomposition of a four-fermion term to two traces with 
two $L$ matrices in 
each trace may be written using a Fierz identity in a form where both the 
traces have no free fermionic indices. We must therefore consider all possible 
decompositions of the four-fermion terms into pairs of traces of fermion 
bilinears and show that these agree with the expressions for $\hat{T}_{f} 
\tilde{T}_{f}$. Examining the list of moments, we see that all such terms take 
the form
\begin{equation}
\label{desterms}
Tr(\hat{L} \hat{\gamma} [\hat{K},\hat{L}])Tr(\hat{L} \tilde{\gamma} 
[\tilde{K}, 
\tilde{L}])
\end{equation}
in which $\hat{\gamma}$ and $\tilde{\gamma}$ are antisymmetric products of the 
same number of gamma matrices. 

In the decomposition of (\ref{eq:4L}) to pairs of traces with two fermions in 
each trace, 
we indeed find terms of this form, though there is also the possibility of 
terms 
in which both factors of $K$ appear in a single trace, the other trace being 
either
\[
\tr(L \gamma^{ij} L)
\]
or
\[
\tr(L \gamma^{ijk} L).
\]
These two expressions are identified below as part of the higher moments
$J^{+ij(k)}$ and $T^{+i(j)}$, and should not affect the interactions
at $1/r^7$, so we expect that all such terms will vanish, leaving only
the desired terms (\ref{desterms}). The demonstration of this, and the
verification that terms of the form (\ref{desterms}) appear with the
correct coefficients for their interpretation as $\hat{T}_{f}
\tilde{T}_{f}$, will not be completed here, as it would involve a
large amount of algebra and is not important for our main
applications.

There are various checks that may be performed on the expressions we
have found.  For example, those tensors with a $+$ index correspond to
the light-front time component of a conserved current integrated over
all space, and therefore correspond to conserved charges of the
theory. In particular, $T^{++}$, $T^{+i}$, and $T^{+-}$ are
longitudinal momentum, transverse momentum and energy; $J^{+ij}$ and
$J^{+-i}$ are the transverse and longitudinal membrane (string) charges, and
$M^{+-ijkl}$ is the longitudinal fivebrane charge. These charges all
appear in the fully extended eleven-dimensional supersymmetry algebra,
and matrix theory expressions for these charges were previously
identified in \cite{bss} by taking commutators of the matrix theory
supersymmetry generators. Our expressions agree completely with that
work, providing a check for all components involving a $+$.
The remaining components of the currents can be checked by ensuring
that  they correspond to the zeroth moments of components of
conserved currents in
11 dimensions.  To perform this check it is necessary to have
expressions for the first moments of the currents, which appear in the
interaction potential at order $1/r^8$, to which we now turn.

\subsection{Gravitational and 3-form interactions
at order $1/r^8$}   

We now consider the terms in the matrix theory potential at order
$1/r^8$ which correspond to bosonic interactions in the linearized
theory.  These terms arise from interactions involving first moments
of the supergravity currents, as in (\ref{super}).  For purely bosonic
backgrounds, these terms were studied in \cite{Mark-Wati,Dan-Wati-2},
where matrix theory expressions for the first moments of the bosonic
currents were calculated and shown to give precise agreement between
matrix theory and supergravity for terms at this order.  As an example
of the terms which appear at order $1/r^8$, from expressions
(\ref{vg2}) and (\ref{super}) we see that the potential due to the
exchange of a graviton with zero longitudinal momentum contains terms
of the form
\[
V^{1/r^8}_{\rm gravity} = - {105 R^2 r_i\over 4 r^9} \left(\hat{T}_{IJ} 
\tilde{T}^{IJ(i)}+\hat{T}^{IJ(i)} \tilde{T}_{IJ} - {1 \over 9} 
(\hat{T}^{I(i)}{}_I \tilde{T}^J{}_J +\hat{T}^I{}_I
\tilde{T}^{J(i)}{}_J )
\right) 
\\ 
\]
Writing the $1/r^8$ matrix theory potential (\ref{on8}) as the product
of a hatted trace and a tilded trace, and comparing it to this
expression as we did for the $1/r^7$ potential, we may now read off
the expressions for the first moments $\tilde{T}^{IJ (i)}$.

We can perform this analysis for all of the first moments of the
bosonic supergravity currents.  For each current, we find that the
first moment in the $i$ direction has a set of terms arising from
simply inserting a matrix $X_i$ into the symmetrized trace expression
for the zeroeth moment. Contributions of this type to first and higher
moments of the currents were studied in \cite{Mark-Wati,Dan-Wati-2},
and come from the terms in the Matrix theory potential with an
insertion of $r \cdot K$ (the first term in (\ref{on8})).  There are
additional contributions to the first moments from terms where the
moment index $i$ appears on a gamma matrix.  It is a fairly
straightforward exercise to derive the additional contributions to
each of the first moments which are quadratic in fermions from
(\ref{on8}).  It is interesting to note that the last two terms in
(\ref{on8}) decompose completely into interactions with first moments
of the type we are interested in, while the second and third terms have no
such interpretation.  We will return to these other terms in the next
subsection.  Here, we list some of the $\theta^2$ contributions to the
higher moments.  We give explicit expressions only for the set of
terms we will need to check the results of the previous subsection.
\begin{eqnarray*}
T_{\rm add}^{+i(j)} &=& {1 \over 8R} \str\left(\theta \gamma^{[ij]} 
\theta\right)\\
T_{\rm add}^{+-(i)} &=& {1 \over 16R} \str\left(-i\theta F_{kl} \gamma^{[kli]} 
\theta + 
2i\theta \dot{X_l} 
\gamma^{[li]} \theta\right)\\
J_{\rm add}^{+ij(k)} &=& {i \over 48R} \str\left(\theta \gamma^{[ijk]} 
\theta\right)\\
J_{\rm add}^{+-i(j)} &=& {1 \over 48R} \str\left(i\theta \dot{X_k} 
\gamma^{[kij]} \theta 
+ i \theta 
F_{ik} 
\gamma^{[kj]}\theta\right)\\
M_{\rm add}^{+-ijkl(m)} &=& -{i \over 16R} \str\left(\theta 
F^{[jk}\gamma^{il]m}\theta\right)
\end{eqnarray*}  
These first moments have no contribution quartic in the
fermion backgrounds.  
Note that the expression given here for $T^{i(j)}$ is responsible for
a ``spin'' contribution to the angular momentum
\[
{\cal J}^{ij} = T^{+ i (j)} -T^{+ j (i)}.
\]
This generalizes the calculation in \cite{Kraus-spin} where the
analogous computation was done in the abelian case of $1 \times 1$ matrices.
   
We can now use these expressions for the first moments of the
currents to check that the expressions (\ref{eq:currents}) for the
zeroth moments of the currents are correct.  In \cite{mvr} it was
shown that conservation of the supergravity
stress-energy tensor, membrane current, and fivebrane current implies
relations among the matrix theory quantities identified with spatial
moments of these currents. In particular, for a current $C^I$,
agreement with the supergravity conservation relations implies that
\[
\partial_t C^{+(i_1\cdots i_n)} = C^{i_1(i_2\cdots i_n)} + \cdots + C^{i_n(i_1 
\cdots i_{n-1})}
\]
In \cite{mvr}, these relations as they apply to $T$, $J$, and $M$ were
shown to hold for a bosonic background as exact matrix theory
identities at finite $N$. Since we expect this to remain true for a
general background, these relations should provide checks on the
components of the currents we have determined with fermionic backgrounds.

In particular, we find for the case $n=1$ that the relations for the 
stress-energy tensor, membrane current, and fivebrane current are
\begin{eqnarray}
 \partial_t T^{++(i)} &=& T^{+i}\nonumber\\ 
 \partial_t T^{+i(j)} &=& T^{ij}\nonumber\\ 
 \partial_t T^{+-(i)} &=& T^{-i}\nonumber\\
 \partial_t J^{+ij(k)} &=& J^{ijk}\label{cons}  \\ 
 \partial_t J^{+-i(j)} &=& J^{-ij}\nonumber\\
 \partial_t M^{+-ijkl(m)} &=& M^{-ijklm}\nonumber\\
 \partial_t M^{+ijklm(n)} &=& M^{ijklmn}\nonumber
\end{eqnarray}
Thus, all components with a spatial index may be checked by comparing
with the time derivative of the first spatial moment of the conserved
charges previously checked.  Note that none of the moments on the left
hand sides of these relations contain four-fermion terms, so that we
have exact expressions for both sides of each of these equations.  The
first five of the relations in (\ref{cons}) can be verified with some
straightforward but tedious algebra.  Note that the term $M^{-ijklm}$,
which appears in the penultimate expression, could not be identified
from the interaction potential, since it couples to the transverse
fivebrane charge $M^{+ijklm}$, whose matrix theory expression appears
to vanish. The conservation relation therefore serves to determine
this component, as was done for its bosonic part in \cite{mvr}.  We
find that
\[
M^{-ijklm} = { 5 \over 4R} \str\left( \dot{X}_{[i}F_{jk}F_{lm]}  
-{1 \over 3}\theta\dot{X}^{[i}\gamma^{jkl}[X^{m]},\theta] - {1 \over 6} \theta 
F^{[ij}\gamma^{klm]}\gamma^i [X^i,\theta]\right).
\]
The final
expression in (\ref{cons}) is consistent with the statement that in
the light-front theory
$M^{ijklmn} = M^{+ijklm}= 0$ which, as mentioned above, arises from
the fact that transverse M5-branes decouple in the light-front theory.

We have now performed an explicit check on all components of the
stress energy tensor, membrane current, and fivebrane current given in
(\ref{eq:currents}) except for $T^{--}$.  However, $T^{--}$ is exactly
the expression for the complete Matrix theory action at $1/r^7$, with
$K$ and $L$ replaced by $X$ and $\theta$ (since
$T^{++}$ is a constant). Since this is the expression from which we
derived all of the other components, it must be correct up to terms
which would not contribute to any of the other components.  Such terms
are four-fermion terms with all factors of $K_i$ appearing in
commutators with $L$s.  Thus, these are the only terms in the currents
we have calculated for which we do not have an independent check.
In principle these terms could be checked by verifying that they
decompose correctly into the product $\hat{C}_f \tilde{C}_f$ of
two-fermion terms we have explicitly computed.  We have not performed
this check in detail.

\subsection{Membrane-fivebrane and ``zerobrane-sixbrane'' interactions}
\label{sec:memfive}

We have now given a complete interpretation in terms of supergravity
interactions of all terms which appear in a decomposition of the
leading-order interaction potential (\ref{on7}) into currents with
purely bosonic indices.  We have also interpreted a subset of the
analogous terms in the potential (\ref{on8}) at order $1/r^8$.  We
have shown that the $r \cdot K$ terms and the last two terms in
(\ref{on8}) have supergravity interpretations and correspond to first
multipole corrections to (\ref{vgra}-\ref{vmag}), coupling the first
spatial moment of each current to the zeroeth moment. The remaining
terms in (\ref{on8}) have not yet been given an interpretation in terms of
supergravity.  We can rewrite these terms using equation (\ref{anti})
from the appendix as \begin{equation}
\label{EM8}
{35 \;r_i\over 256 \;r^9}
 \epsilon^{ijklmnpqr}\str\left(F_{jk}F_{lm}F_{np}\dot{K_q}K_r 
- L \, \dot{K}_j F_{kl} F_{mn} \gamma^{[pqr]} L +{i \over 2} L \,
 F_{jk} F_{lm}  
F_{np} \gamma^{[qr]} L\right) 
\end{equation}
We now consider all possible ways of decomposing this into a hatted
trace times a tilded trace. Guided by the definitions of the moments
above, we find that the decomposition gives a set of terms which can
be identified as  
\begin{eqnarray}
V_{{\rm 2-5}} & = & 
\frac{63 R^2}{32} 
{r^i \over r^9}\epsilon^{ijklmnpqr} \left( \hat{J}^{+jk}\tilde{M}^{-lmnpq(r)}+ 
\hat{J}^{+jk(l)}\tilde{M}^{-mnpqr} + 
({\;\widehat{} \leftrightarrow \widetilde{} \; }) \right)\nonumber\\
& &
+ \frac{105 R^2}{32} 
{r^i \over r^9}\epsilon^{ijklmnpqr} \left(
\hat{J}^{jkl}\tilde{M}^{+-mnpq(r)}  +
\hat{J}^{jkl(m)}\tilde{M}^{+-npqr} +  ({\;\widehat{} \leftrightarrow
\widetilde{} \; })
\right) \label{eq:2-5}
\\
 & = &  \frac{7 R^2}{64} \frac{r^I}{r^9} 
\epsilon_{IJKLMNPQRST} 
 \left( \hat{J}^{JKL}\tilde{M}^{MNPQSR(T)}+ 
 \hat{J}^{JKL(M)}\tilde{M}^{NPQSRT}+ 
({\;\widehat{} \leftrightarrow \widetilde{} \; }) \right) \nonumber
\end{eqnarray}
There are additional terms in the decomposition of (\ref{EM8})
which do not have a simple interpretation in terms of the moments of
the currents we have already discussed.
These terms can be written in the form
\begin{eqnarray}
V_{{\rm 0-6}} & = & 
\frac{5 R^2}{256} 
{r^i \over r^9}\epsilon^{ijklmnpqr} 
\left( \hat{T}^{++}\tilde{S}^{jklmnpq(r)}+ 
\hat{T}^{++(j)}\tilde{S}^{klmnpqr} + 
({\;\widehat{} \leftrightarrow \widetilde{} \; }) \right)\nonumber\\
& & +
\frac{35 R^2}{256} 
{r^i \over r^9}\epsilon^{ijklmnpqr} 
\left( \hat{T}^{+j}\tilde{S}^{+klmnpq(r)}+ 
\hat{T}^{+j(k)}\tilde{S}^{+lmnpqr} + 
({\;\widehat{} \leftrightarrow \widetilde{} \; }) \right) \label{eq:0-6}\\
 & = &  \frac{5 R^2}{256} \frac{r^a}{r^9} 
\epsilon_{abcdefghij} 
 \left( \hat{T}^{+b}\tilde{S}^{cdefghi(j)}+ 
 \hat{T}^{+b(c)}\tilde{S}^{defghij}+ 
({\;\widehat{} \leftrightarrow \widetilde{} \; }) \right) \nonumber
\end{eqnarray}
where we have defined the currents
\begin{eqnarray}
S^{+ijklmn} & = & \frac{1}{R}  \str\left(F_{[ij} F_{kl} F_{mn]}\right) 
\nonumber\\
S^{+ijklmn(p)} & = & \frac{1}{R}  \str\left(F_{[ij} F_{kl} F_{mn]} X_{p} 
-\theta F_{[kl}F_{mn}\gamma_{pqr]}  \theta\right) \label{eq:6-current}\\
S^{ijklmnp} & = & 
\frac{7}{R}  \str\left(F_{[ij} F_{kl} F_{mn} \dot{X}_{p]}\right) \nonumber\\
S^{ijklmnp(q)} & = & 
\frac{7}{R}  \str\left(F_{[ij} F_{kl} F_{mn} \dot{X}_{p]} X_{q} 
- \theta \, \dot{X}_{[j} F_{kl} F_{mn} \gamma_{pqr]} \theta +{i \over 2} 
\theta 
\,
 F_{[jk} F_{lm}  
F_{np} \gamma_{qr]} \theta
\right) \nonumber
\end{eqnarray}

We will now discuss the supergravity interpretation of the terms
(\ref{eq:2-5}) and (\ref{eq:0-6}).  We begin with the potential
$V_{\rm 2-5}$ in (\ref{eq:2-5}).  It is clear that this corresponds to a
membrane-5-brane interaction potential.  Indeed, we expect to see such
interactions since both the membrane and the 5-brane couple to the 11D
supergravity 3-form field $A_{IJK}$.  It is fairly straightforward to
see that, up to  an overall multiplicative constant, (\ref{eq:2-5}) is
precisely the right form for this interaction, as we now discuss in
more detail.

Matrix theory interactions between a membrane and 5-brane were first
discussed by Berkooz and Douglas in \cite{Berkooz-Douglas}, where a
5-brane was put in as a static background for a dynamical matrix
membrane.  They showed that  the fermion 0-modes corresponding to
strings stretching between the membrane and 5-brane experience a correct
Berry's phase shift when the membrane is carried around the 5-brane.

A configuration containing a membrane and a 5-brane which are both of
infinite extent is analogous to a configuration in 4D electromagnetism
containing both an electrically charged particle and a magnetic
monopole.  In principle, one might hope to describe an effective
action for the electrically charged particle in the long-range field
produced by the monopole.  The vector potential $A^\mu$ around a
monopole is not single-valued, however, and has a Dirac singularity.
As a result, it is not possible to describe  an interaction between
electrically and magnetically charged particles in classical
electromagnetism in terms of an effective potential.
Similarly, it is not possible to construct an effective potential
describing M-theory or matrix theory interactions between infinite
membranes and 5-branes, so a study of fermion 0-modes is necessary to
understand the interactions between such objects.

In this work, however, we are concerned with long-range interactions
between finite-sized objects in matrix theory.  Such objects cannot
have net membrane or 5-brane charge.  This can be seen directly from
(\ref{eq:currents}), where the membrane charge $J^{+ ij}$ and the
5-brane charge $M^{+ -ijkl}$ vanish for finite $N$.  Finite $N$ matrix
theory configurations can, however, have nonzero moments of the
membrane and 5-brane charges.  Simple examples are given by the
membrane with $S^2$ geometry and the L5-brane with $S^4$ geometry
studied in \cite{Dan-Wati,clt}.  Unlike the interaction between an
infinite membrane and an infinite 5-brane, an interaction between
either brane and any configuration with higher moments of the other
brane charge {\em can} be described by an effective potential.  This
is analogous to an interaction between an electron and a particle with a
magnetic dipole moment.  The gauge field $A_\mu$ around the dipole 
is single valued, so that the interaction can be written in
the form $\int A^\mu j_\mu$ where $j_{\mu}$ is the 4-current of the
electron.

The ``electric'' interactions between
a pair of matrix theory configurations with nonzero moments of
membrane charge are described by higher-order generalizations of
(\ref{ve2}).  Although the first nonvanishing term of this form in the
interaction potential between a pair of finite $N$ configurations
appears at order $1/r^9$, the expression (\ref{ve2}) which formally
appears in the potential at order $1/r^7$ is useful because it implies
the precise forms of all the higher-moment electric interactions, as
discussed in \cite{Dan-Wati-2}.  This is precisely analogous to the
story in the case of membrane-5-brane interactions.  The leading
interaction (\ref{eq:2-5}) which formally appears in the long-range
potential at order $1/r^7$
vanishes for a pair of finite-size configurations.  Nonetheless, it
correctly encodes all the higher-moment membrane-5-brane interactions
which do not vanish.

The fact that (\ref{eq:2-5}) has precisely the right form for a
membrane-5-brane interaction potential can be seen by a simple analogy
with the electron-magnetic dipole case mentioned above.  It may be
helpful to review this situation explicitly.  The long-range
field of a magnetic monopole can be encoded in a dual vector potential
\[
A_0^{D} = -\frac{g}{r} 
\]
where $g$ is the magnetic charge.  This is related to a dual field
strength $F^D_{\mu \nu} = \partial_\mu A^D_\nu -\partial_\nu A^D_\mu$ with
nonzero components
\[
F^D_{0i} = -\frac{gr^i}{r^3} 
\]
which is related through $F ={}^*F^D$ to the usual field strength of
electromagnetism.  This field has nonzero components
\[
F_{ij} = -\frac{g\epsilon_{ijk}r^k}{r^3} 
\]
and cannot be written as the field strength of a single-valued
potential $A_\mu$.  On the other hand, the field strength around a
magnetic dipole is found by taking the derivative of the above
expressions in the direction of the dipole $d^i$ where $m^i = gd^i$.
This gives
\begin{equation}
F_{ij} = - \frac{\epsilon_{ijk}m^k}{r^3}  +\frac{3 \epsilon_{ijl} r^l
(r\cdot m)}{r^5}  = \partial_iA_j-\partial_jA_i
\end{equation}
where the vector potential $A_\mu$ has nonzero components
\begin{equation}
A_i = \frac{\epsilon_{ijk}d^j r^k}{r^3} 
\end{equation}

The generalization to the membrane-5-brane situation is
straightforward.  The long-range field around
a longitudinal 5-brane is described by a dual potential (dropping
overall constants)
\[
A^D_{+ -ijkl} = \int d^4 \xi \frac{M^{+ -ijkl} (\xi)}{r^7} 
\]
given by an integral over the local L5-brane charge.  A finite volume
configuration has vanishing net 5-brane charge but may have a
nontrivial first moment $M^{+ -ijkl (m)}$ which gives rise to a
long-range field strength.  $F ={}^*F^D$ which is
the curvature of a well-defined 3-form field
\[
A_{ijk} = \epsilon_{ijklmnpqs} \frac{M^{+ -lmnp (q)}r^s}{r^{9}} 
\]
If we consider a probe membrane in this long-range field, we find an
effective potential described by the first term in the second line of
(\ref{eq:2-5}).  The other terms in (\ref{eq:2-5}) correspond in a
similar fashion to the action for a probe 5-brane moving in the
long-range field produced by a configuration with a nonzero first
moment of the membrane current, and to terms related to these two by
Galilean invariance.

Thus, we see that (\ref{eq:2-5}) precisely encodes the form of
interaction we expect to see at leading order between a pair of finite
size objects in light-front supergravity.  To connect this situation
with the general discussion in section \ref{sec:linear-supergravity}
in which classical supergravity interactions were described in terms
of two-point functions of the propagating fields, we can consider a
term coupling the membrane current and 5-brane current which appears
in the effective action analogous to the expressions
(\ref{vgra}-\ref{vmag}) for the interactions previously
considered. 
The sources in (\ref{sources})
give rise to terms
in the effective action
\begin{eqnarray}
V_{\rm EM} &=& \int d^{11}x \int d^{11}y \hat{J}^{IJK}(x) \langle 
A_{IJK}(x)A^D_{PQRSTU}(y) \rangle \tilde{M}^{PQRSTU}(y)
+({\;\widehat{} \leftrightarrow \widetilde{} \; })
\nonumber
\label{EMI}
\end{eqnarray} 
It is not clear that the expectation values written here make
sense in a quantum theory, since they require a definition of the six form
field as an operator written in terms of creation and annihilation
operators for the basic supergravity fields.  
We are interested in terms with at least one higher moment, however,
which give rise to propagators such as
\begin{equation}
\langle 
A_{IJK}(x)\partial_N A^D_{PQRSTU}(y) \rangle 
\sim \langle A_{IJK} (x) \epsilon^{GHLMNPQRSTU} \partial_G A_{HLM} (y)\rangle
\end{equation}
which can be understood in terms of the original 3-form propagator.
This analysis again leads to a term of the form (\ref{eq:2-5}) in the
supergravity potential.


We now return to the terms (\ref{eq:0-6}) which describe long-range
interactions between gravitons and a new set of currents defined in
(\ref{eq:6-current}).  In the language of type IIA string theory,
where we have a system of $N$ 0-branes, the current $S^{+ijklmn}$
gives the 6-brane charge of the 0-brane configuration described by the
matrices $X^i$.  This is simply the T-dual of the statement that $\int
F \wedge F \wedge F$ on a 6-brane counts 0-brane charge
\cite{WT-Trieste}.  Given this interpretation, it is clear from
current conservation and 10-dimensional symmetry that the remaining
components of $S$ correspond to other integrated components of the IIA
6-brane current.  In (\ref{eq:0-6}) we see a coupling between the
6-brane current and a vector $T^{+ a}$ in the 10D IIA theory
corresponding to the integrated 0-brane current.  Of course, in type IIA 
string
theory we expect to see precisely such a coupling.  The 0-brane and
6-brane couple electrically and magnetically to the R-R 1-form field
in IIA.  The interaction between 0-branes and an infinite 6-brane is
analogous to the interaction between an infinite membrane and 5-brane.
Such an interaction cannot be described by an effective potential, but
can be seen from the fermion 0-modes in the system, as in the case of
the membrane-5-brane system.  This leading-order interaction between
0-branes and an infinite 6-brane was studied in
\cite{Lifschytz-46,Vijay-Finn2,hlw,Vakkuri-Kraus,Pierre-06,bdflrs}.

The 6-brane charge of any set of matrices with finite size $N$ must
vanish, just as the membrane and L5-brane charge vanish.  We therefore
expect that the interactions between moments of the 6-brane current of
one finite-size configuration and the 0-brane current of a second
configuration should be described by a potential which has precisely
the form (\ref{eq:0-6}), based on considerations precisely analogous
to those above in the membrane-5-brane case.  This 0-6 interaction,
however, is completely described in terms of 10-dimensional currents.
Unlike all the other interactions we have found it cannot be expressed
in terms of a simple interaction potential with 11-dimensional
invariance.  This is a consequence of the fact that the IIA 6-brane
has a more subtle connection with the 11-dimensional M-theory than the
0-, 2- and 4-branes, which correspond to extended objects in M-theory
of dimension 0, 2 and 5.  The infinite flat 6-brane as a IIA
background can be interpreted in 11D supergravity as a Kaluza-Klein
monopole background metric.  The interpretation of a finite size IIA
object with local 6-brane charge is somewhat less clear.  Presumably
it corresponds to a more complicated background metric in 11
dimensions.  Thus, we have a physical interpretation of the
interactions in (\ref{eq:0-6}) in 10 dimensions which is somewhat more
transparent than the 11-dimensional interpretation, in which the
matrix theory configuration with 6-brane charge seems to correspond to
a nontrivial background geometry in M-theory.

\subsection{Supercurrent interactions}

In the previous subsections, we have considered all terms in the
decomposition of the matrix theory action at orders $1/r^7$ and
$1/r^8$ into two traces with no free fermionic indices and shown that these
terms correspond precisely with the leading and subleading order
supergravity interactions due to the exchange of a graviton or three
form quantum.  The only remaining terms in the decomposition of the
matrix theory potential contain a single free fermionic index in each
trace.  (We argued above that terms with two fermionic indices in each
trace could be rewritten using a Fierz identity into a form with no
free fermionic indices in either trace.)  We would like to identify
these terms with interactions of the form (\ref{vsup}) arising from
the exchange of a gravitino.  It may seem unusual to consider an
interaction between two fixed sources in which a fermionic gravitino
is exchanged, as this would seem to imply a change of the statistics
of the two objects in the course of the interaction.  This can be
understood for classical objects, however in terms of a probe-source
picture where one object probes the long-range gravitino field
$\psi_J$ produced by a distant source, giving an interaction potential
of the form $\int \bar{\psi}_J S^J$ where $S^J$ is the fermionic part
of the supercurrent of the probe.  Interactions of this form have also
been discussed recently in \cite{bktw}.
   
To see the form of the leading-order supergravity interaction due to 
gravitino exchange, we must explicitly determine the gravitino propagator. 
In the supergravity action, the quadratic terms are 
\[
\Gamma_{\Psi\Psi} = - {1 \over 2}\bar{\Psi}_I \Gamma^{IJK} \partial_K \Psi_J
\]
Supergravity is invariant under local supersymmetry transformations
which affect the gravitino field, so as for the case of a local gauge
symmetry, we must add gauge fixing terms to the action in order to
quantize the theory \cite{Nieuwenhuizen}.  These affect the form of
the propagator, but as for the case of graviton and three-form
exchange, there appears to be a natural choice which makes the
correspondance with Matrix theory most clear. In this case, we choose
a gauge-fixing term
\[
\Gamma_{\rm fix} = {9 \over 8} \bar{\Psi}_I \Gamma^I \slash{\partial} \Gamma^J 
\Psi_J \;.
\]
This choice is analogous to Feynman gauge in electromagnetism, ensuring that 
the 
momentum space propagator contains only terms with a single $k$ in the 
numerator. 

  The propagator $\Delta_{IJ}^{\alpha \beta}$ is then determined by 
  
\[  
(- {1 \over 2}  \Gamma^{IJL} \partial_L  + {9 \over 8} \Gamma^I 
\slash{\partial} 
\Gamma^J) \Delta_{JK}(x-y) = \delta^I_K \delta(x - y) \; .
\]
As with the graviton and three form propagators, we restrict to the $x^-$ 
independent term corresponding to exchange of gravitinos with zero 
longitudinal 
momentum, and we find
\[
\Delta_{IJ} = \langle \Psi_I(x) \bar{\Psi}_J(y) \rangle \propto (\Gamma_J 
\slash{\partial} \Gamma_I - 7 \delta_{IJ} \slash{\partial} ) \delta( x^+ - 
y^+) 
{1 \over | \vec{x} - \vec{y}|^7}
\]
We now insert this expression into (\ref{vsup}) to determine the leading-order 
terms in the supergravity potential due to gravitino exchange. We find an 
interaction proportional to 
\[
\int dx dy \bar{\hat{S}}^I (x) \{ \Gamma_J \Gamma^K \Gamma_I - 7 \delta_{IJ} 
\Gamma^K \} \partial_K \delta( x^+ - y^+) {1 \over | \vec{x} - \vec{y}|^7} 
\tilde{S}^J(y) \;.
\]
The only terms at order $1/r^7$ come from choosing $\partial_K$ to be a time 
derivative ($K=+$), since the $x^-$ derivative vanishes, while the transverse 
spatial derivatives acting on $1/r^7$ result in an additional power of $1/r$.  
To find the explicit form of these leading-order interactions, we expand the 
currents as distributions (\ref{dist}) and find
\[
V^{\rm super}_{1/r^7} \propto {R^2 \over r^7} \int d \tau \bar{\hat{S}}^I 
\{\Gamma_J 
\Gamma_- \Gamma_I - 7 \delta_{IJ} \Gamma_- \} \dot{\tilde{S}}^J
\]
where there is an overall rational coefficient we have not calculated.
To connect with our Matrix theory potential, we write $S$ in terms of
sixteen-component spinors $q^I$ and $Q^I$ as
\[
S^I = \left( \ba{c} q^I \\ Q^I \ea \right ) \;, \; \; \; \bar{S}^I = 
(S^I)^\dagger \Gamma^0
\]
and choose gamma matrices   
\[
\Gamma^i = \left( \ba{cc} \gamma^i & 0 \\ 0 & -\gamma^i \ea \right) \; \; 
\Gamma^0 = \left( \ba{cc} 0 & -1 \\ 1 & 0 \ea \right) \; \; 
\Gamma^{10} = \left( \ba{cc} 0 & 1 \\ 1 & 0 \ea \right) \; \;
\]
(Note that the convention used in this section for 11D gamma matrices
is different from the convention (\ref{eq:first-gamma}) used elsewhere
for 10D gamma matrices.)
We have then
\[
\Gamma^+ = - \Gamma_- = \left( \ba{cc} 0 & 0 \\ \sqrt{2} & 0 \ea \right) \; \; 
\Gamma^- = - \Gamma_+ = \left( \ba{cc} 0 & -\sqrt{2} \\ 0 & 0 \ea \right) \; 
\;
\]
and we find
\begin{eqnarray}
\label{sup1}
V^{\rm super}_{1/r^7} &\propto&
\frac{R^2}{r^7}  \left(
  \hat{q}^+\; \partial_t\; (7 \tilde{q}^- - \sqrt{2} 
\gamma^i \tilde{Q}^i) + (7 \hat{q}^- - \sqrt{2} \hat{Q}^i \gamma^i)\; 
\partial_t 
\; \tilde{q}^+ -2 \hat{Q}^+\; \partial_t \;\tilde{Q}^+ \right.\\
 & &\left. \hspace{0.4in} + \hat{q}^i \gamma^i\; \gamma^j \;
\partial_t \; \tilde{q}^j -  
9 \hat{q}^i \; \partial_t \; \tilde{q}^i \right)\nonumber
\end{eqnarray}
Before comparing this with the matrix theory potential, we recall a few 
physical 
facts about the supercurrent $S^I(x)$. Firstly, since the gravitino is 
massless 
and has a vector index, the source $S^I$ must obey the constraint
\[
\partial_I S^I(x) = 0
\]
Thus, the zeroeth moment $S^+$, and therefore the charges $q^+$ and
$Q^+$ are conserved in light-front time. These quantities are the
supersymmetry generators of the theory, whose matrix theory
counterparts are known to be \cite{bss}
\begin{eqnarray}
q^+ &= & {c_1 \over R} \tr(\theta) \label{eq:known-super}\\
Q^+ &= & {c_2 \over R} \tr((\dot{X}^i \gamma_i+ {1 \over 2}F_{ij}
\gamma^{ij})\theta) \nonumber
\end{eqnarray}
One may easily check that $\dot{q}^+ = \dot{Q}^+=0$. 
The overall numerical coefficient multiplying the currents in this section 
depends on the conventions chosen to define the current $S$, so we will not 
keep careful track of it. However, we will be able to fix the normalization on 
the other components of the current using the constants $c_1$ and $c_2$ 
appearing in these expressions.

Looking back at the potential (\ref{sup1}) we now see that last two
terms on the first line vanish, while the first is a total derivative
and therefore does not correspond to a physical interaction. The
remaining terms involve only the component $q^i$.
 
Turning now to the matrix theory potential, we expect that with the proper 
identification of matrix theory quantities for the supercurrents, the 
expression 
(\ref{sup1}) should reproduce all terms in the decomposition of the $1/r^7$ 
potential with two fermionic traces. It is not difficult to check that using 
the 
matrix theory equations of motion, such terms may be rewritten as (ignoring 
four 
fermion terms) 

\begin{eqnarray*}
V_{\rm mat} &\propto& \str\left(\hat{\theta}\right)\;\; 
\partial_\tau\;\;\str\left(\{ -{1  
\over 4} \gamma^{[ijkl]}F_{ij}F_{kl} + {1 \over 2} F_{ij} F_{ji} - 2i 
\dot{\tilde{X}}_iF_{ij} \gamma^j  +  \dot{\tilde{X}}_i \dot{\tilde{X}}_i 
\}\tilde{\theta} \right)\\
& & \hspace{0.2in} + {2i \over 9}\str\left(\hat{\theta}\{i\dot{\hat{X}_i} + 
F_{ij}\gamma^j\}\right)\;\;(\gamma^i \gamma^k - 9\delta_{ik} )\partial_\tau 
\;\;\str\left(\{i \dot{\tilde{X}}_k + F_{kl}\gamma^l\}\tilde{\theta}\right)\\
& & \hspace{0.2in} + \{\widehat{} \leftrightarrow \widetilde{} \} 
\end{eqnarray*}

This matches exactly with the nonvanishing terms in the supergravity potential 
(\ref{sup1}), provided we identify (in Minkowski notation)
\[
q^i = \frac{c_1}{R} 
\str\left(\{\dot{X_i} - F_{ij} \gamma^j\} \theta\right)
\]
and
\begin{equation}
\label{com}
(7q^- - \sqrt{2} \gamma^i Q^i) = {9 c_1\over 2R} \str\left(\{ -{1 \over 4} 
\gamma^{[ijkl]}F_{ij}F_{kl}  + {1 \over 2} F_{ij} F_{ji} + 2 \dot{X}_iF_{ij} 
\gamma^j  - \dot{X}_i \dot{X}_i \}\theta\right)
\end{equation}
(with overall coefficients which are determined relative to those of
(\ref{eq:known-super}) below.)  Note that the latter definition comes
from identifying a total derivative term in the supergravity potential
with a total derivative in the Matrix theory potential. For more solid
evidence of the this identification, it is necessary to look at higher
order interactions at $1/r^8$.  We also note that $Q^+$ and $Q^-$ do
not appear in the potential at all, while $q^-$ and $Q^i$ appear only
in the combination $(7q^- - \sqrt{2} \gamma^i Q^i)$.  To make explicit
identifications of these terms by comparing interaction potentials,
one would again have to look at terms of order $1/r^8$.

As a check on the $q^i$ components we have identified and to
identify the components $Q^i$ and $q^-$, we can use the conservation
relation $\partial_I S^I$ mentioned above. As for the bosonic
currents, this implies that the spatial moments of $q$ and $Q$ are
related to the time derivative of moments with a $+$ index. In
particular, we expect
\begin{eqnarray*}
q^i &=& \partial_t q^{+(i)}\\
Q^i &=& \partial_t Q^{+(i)} 
\end{eqnarray*}
Though we have not identified the first moments $q^{+(i)}$ or $Q^{+(i)}$ 
through 
an analysis of the potential terms as with the bosonic currents, we expect 
these 
to have terms given by the insertion of $X^i$ into the zeroth moments plus 
possibly other terms with the $i$ index appearing on a gamma matrix. Assuming 
no 
such additional terms are present, we would have
\begin{eqnarray*}
q^i =
\partial_t q^{+(i)} &=& \partial_t {c_1 \over R} \str\left(X_i \theta 
\right)\\
 &=& {c_1 \over R} \str\left(\{ \dot{X}^i - F_{ij} \gamma^j \} \theta \right)
\end{eqnarray*} 
This agrees with our previous expression for $q^i$ and fixes the
normalization relative to $q^+$. For the $Q$
components, again assuming that all terms in $Q^{+(i)}$ are given by
$X^i$ insertions into $Q^+$ we have
\begin{eqnarray*}
Q_i =\partial_t Q^{+(i)}
 &=& \partial_t {c_2 \over R}\; \str\left(X^i\{\dot{X^j} \gamma_j+ {1 
\over 2}F_{jk} \gamma^{jk}\}\theta\right)\\
&=&  \frac{c_2}{2R}  \;\str\left(\{2\dot{X}_i \dot{X^j}
 \gamma_j+\dot{X}_iF_{jk}\gamma^j \gamma^k  
+2\dot{X}^j F_{li}\gamma^j \gamma^l + F_{jk} F_{li} \gamma^j \gamma^l \gamma^i 
\} \theta\right)
\end{eqnarray*}
It is possible that there are additional terms in $Q^i$ arising from
extra terms in the moment $Q^{+(i)}$ that we have not accounted for,
however if this is not the case, we may use our identification in
expression (\ref{com}) to define $q^-$. A careful analysis of the
potential at $1/r^8$ would provide a check on our definitions of $q^-$
and $Q^i$ as well as an identification of the component $Q^-$. Analysis of 
these higher order terms should also fix the coefficient $c_2$ relative to 
$c_1$, not possible using only the $1/r^7$ potential. 

We have now given an interpretation of all terms at leading and subleading 
orders in the general matrix theory potential between two isolated systems and 
found within this potential all interactions expected from the exchange of a 
single quantum in supergravity.  

\subsection{Summary of supergravity currents}
\label{sec:currents-summary}

We summarize here our results for the matrix form of the supergravity
currents.  The integrated stress-energy tensor, membrane current and
5-brane current are given in matrix theory by
\begin{eqnarray}
T^{++} &=& {1 \over R}\str\left(\identity\right)\nonumber\\
T^{+i} &=& {1 \over R}\str\left(\dot{X_i}\right)\nonumber\\
T^{+-} &=& {1 \over R}\str\left({1 \over 2} \dot{X_i} \dot{X_i} + {1 \over 4} 
F_{ij}  
F_{ij} + {1 \over 2} \theta\gamma^i[X^i,\theta]\right)\nonumber\\
T^{ij} &=& {1 \over R}\str\left( \dot{X_i} \dot{X_j} +  F_{ik}  F_{kj} - {1 
\over 4} 
\theta\gamma^i[X_j,\theta] - {1 \over 4} 
\theta\gamma^j[X_i,\theta]\right)\nonumber\\
T^{-i} &=& {1 \over R} \str\left({1 \over 2}\dot{X_i}\dot{X_j}\dot{X_j} + {1 
\over 
4} \dot{X_i} F_{jk} F_{jk} + F_{ij} F_{jk} \dot{X_k}\right) \nonumber\\ & & - 
{1 \over 
4R} 
\str\left(\theta_\alpha 
\dot{X_k}[X_m,\theta_\beta]\right)\{\gamma^k\delta_{im} 
+\gamma^i\delta_{mk} -2\gamma^m\delta_{ki} \}_{\alpha \beta}\nonumber\\ & & - 
{1 
\over 8R} 
\str\left(\theta_{\alpha} F_{kl}[X_m,\theta_{\beta}]\right)\{ \gamma^{[iklm]} 
+ 
2 \gamma^{[lm]} \delta_{ki} + 4\delta_{ki}\delta_{lm} \}_{\alpha 
\beta}\nonumber\\  & & + 
{i \over 8R} \tr(\theta \gamma^{[ki]} \theta \; \theta \gamma^k 
\theta)\nonumber\\
T^{--} &=& {1 \over 4R} \str\left(F_{ab}F_{bc}F_{cd}F_{da} - {1 \over 4}F_{ab} 
F_{ab} F_{cd} F_{cd}  + {\theta} \Gamma^a \Gamma^b \Gamma^c F_{ab} F_{cd} 
D_a\theta + {\cal O} ({\theta^4})\right)\nonumber\\
J^{+ij} &=& {1 \over 6R} \str\left(F_{ij}\right) \label{eq:currents-all}\\
J^{+-i} &=& {1 \over 6R} \str\left( F_{ij} \dot{X_j} - {1 \over 2} 
\theta[X_i,\theta] 
+ {1 
\over 4} \theta \gamma^{[ki]} [X_k, \theta]\right)\nonumber\\
J^{ijk} &=& {1 \over 6R} \str\left( \dot{X_i} F_{jk} +  \dot{X_j}
F_{ki}  +
\dot{X_k} F_{ij} -{1 \over 4} \theta 
\gamma^{[ijkl]}[X_l,\theta]\right)\nonumber\\
J^{-ij} &=& {1 \over 6R} \str\left(+\dot{X_i} \dot{X_k} F_{kj} - 
\dot{X_j}\dot{X_k} 
F_{ki} - {1 \over 2} \dot{X_k}\dot{X_k} F_{ij} + {1 \over 4}F_{ij} F_{kl} 
F_{kl} 
+ F_{ik} F_{kl} F_{lj}\right)\nonumber\\
& & +{1 \over 24R} \str\left(\theta_\alpha 
\dot{X_k}[X_m,\theta_\beta]\right)\{\gamma^{[kijm]} + 
\gamma^{[jm]} \delta_{ki} - \gamma^{[im]} \delta_{kj} + 2 \delta_{jm} 
\delta_{ki} - 2 \delta_{im} \delta_{kj}\}_{\alpha \beta}\nonumber\\
& & + {1 \over 8} \str\left(\theta_{\alpha} 
F_{kl}[X_m,\theta_{\beta}]\right)\{\gamma^{[jkl]} 
\delta_{mi} - \gamma^{[ikl]} \delta_{mj} + 2 \gamma^{[lij]} \delta_{km} + 2 
\gamma^l \delta_{jk} \delta_{im} - 2 \gamma^l \delta_{ik} 
\delta_{jm}\nonumber\\
& & \hspace{1in} + 2 \gamma^j \delta_{il} \delta_{km} - 2 \gamma^i \delta_{jl} 
\delta_{km}\}_{\alpha \beta}\nonumber\\ & & + {i \over 48R} \str\left(\theta 
\gamma^{[kij]} \theta \; \theta 
\gamma^k \theta - \theta \gamma^{[ij]} \theta \; \theta 
\theta\right)\nonumber\\
M^{+-ijkl} &=& {1 \over 12R} \str\left(F_{ij}F_{kl} +F_{ik}F_{lj} + 
F_{il}F_{jk} + 
\theta \gamma^{[jkl}[X^{i]},\theta]\right)\nonumber\\
M^{-ijklm} & = & { 5 \over 4R} \str\left( \dot{X}_{[i}F_{jk}F_{lm]}  
-{1 \over 3}\theta\dot{X}^{(i}\gamma^{jkl}[X^{m)},\theta] - {1 \over 6} \theta 
F^{(ij}\gamma^{klm)}\gamma^i [X^i,\theta]\right) \nonumber\\
M^{+ijklm} &= &M^{ijklmn}  =  0 \nonumber
\end{eqnarray}
The first moments of these currents are given by
\begin{eqnarray}
T^{IJ (i)} & = & \sym (T^{IJ}; X^i) + T_{\rm add}^{IJ(i)} \nonumber\\
J^{IJK (i)} & = & \sym (J^{IJK}; X^i) + J_{\rm add}^{IJK(i)}\\
M^{IJKLMN (i)} & = & \sym (M^{IJKLMN}; X^i) + M_{\rm add}^{IJKLMN(i)}\nonumber
\end{eqnarray}
where in particular
\begin{eqnarray*}
T_{\rm add}^{+i(j)} &=& {1 \over 8R} \tr(\theta \gamma^{[ij]} \theta)\\
T_{\rm add}^{+-(i)} &=& {1 \over 16R} \tr(-i\theta F_{kl} \gamma^{[kli]} 
\theta + 
2i\theta \dot{X_l} 
\gamma^{[li]} \theta)\\
J_{\rm add}^{+ij(k)} &=& {i \over 48R} \tr(\theta \gamma^{[ijk]} \theta)\\
J_{\rm add}^{+-i(j)} &=& {1 \over 48R} \str\left(i\theta \dot{X_k} 
\gamma^{[kij]} \theta 
+ i \theta 
F_{ik} 
\gamma^{[kj]}\theta\right)\\
M_{\rm add}^{+-ijkl(m)} &=& -{i \over 16R} \str\left(\theta 
F^{[jk}\gamma^{il]m}\theta\right)
\end{eqnarray*}  
are the two-fermion contributions to the first moments with a $+$ index.

The fermionic components of the supercurrent (excluding possible three fermion 
terms) are given by
\begin{eqnarray*}
q^+ &=& {c_1 \over R} \tr(\theta)\\
q^i &=&{c_1 \over R} \str\left(\{ \dot{X}^i - F_{ij} \gamma^j \} \theta 
\right)\\
Q^+ &=& {c_2 \over R} \str\left((\dot{X^j} \gamma_j+ {1 \over 2}F_{ij} 
\gamma^{ij})\theta\right)\\
Q^i &=&  {c_2 \over 2R} \str\left(\{2\dot{X}_i \dot{X^j}
\gamma_j+\dot{X}_iF_{jk}\gamma^j  
\gamma^k +2\dot{X}^j F_{li}\gamma^j \gamma^l + F_{jk} F_{li} \gamma^j \gamma^l 
\gamma^i \} \theta\right)
\end{eqnarray*}
As discussed above, the components $q^-$ and $Q^-$ as well as the ratio 
$c_1/c_2$ should follow from an analysis of the $1/r^8$ potential. The 
remaining overall coefficient depends on the conventions used for defining the 
supercurrent.

There is also a current for the IIA 6-brane, corresponding to a
Kaluza-Klein monopole in 11 dimensions.  This current has integrated 
components
\begin{eqnarray}
S^{+ijklmn} & = & \frac{1}{R}  \str\left(F_{[ij} F_{kl} F_{mn]}\right) 
\nonumber\\
S^{+ijklmn(p)} & = & \frac{1}{R}  \str\left(F_{[ij} F_{kl} F_{mn]} X_{p} 
-\theta F_{[kl}F_{mn}\gamma_{pqr]}  \theta\right) \label{eq:6-current2}\\
S^{ijklmnp} & = & 
\frac{7}{R}  \str\left(F_{[ij} F_{kl} F_{mn} \dot{X}_{p]}+ f.t.\right) 
\nonumber\\
S^{ijklmnp(q)} & = & 
\frac{7}{R}  \str\left(F_{[ij} F_{kl} F_{mn} \dot{X}_{p]} X_{q}  
- \theta \, \dot{X}_{[j} F_{kl} F_{mn} \gamma_{pqr]} \theta +{i \over 2} 
\theta 
\,
 F_{[jk} F_{lm}  
F_{np} \gamma_{qr]} \theta
\right) \nonumber
\end{eqnarray}

\section{Conclusions and further directions}
\label{sec:applications}

In this paper we have calculated the most general form of the leading
$1/r^7$ term in the two-body interaction potential between an
arbitrary pair of objects in matrix theory.  We included completely
general bosonic and fermionic background matrices corresponding to
arbitrary non-supersymmetric objects.  We found that the general
interaction potential decomposes into a sum of parts which can be
identified with supergravity interactions between extended objects
arising from the exchange of a single graviton, gravitino or 3-form
quantum.  This correspondence was made by giving precise matrix
expressions for the stress-energy tensor, membrane current, 5-brane
current and fermionic supercurrent components of an arbitrary matrix
theory configuration, generalizing the results of \cite{Dan-Wati-2}
for purely bosonic backgrounds.

We calculated the general form of the subleading $1/r^8$ terms in the
interaction potential to quadratic order in the fermion backgrounds.
We found that beginning at this order a new set of terms appear in the
potential corresponding to membrane-5-brane and 0-brane-6-brane
interactions.  These terms are analogous to the effective potential a
probe electron would see in the long-range electromagnetic field of an
object with nonzero magnetic multipole moments.  We also derived
explicit expressions for some of the fermionic contributions to the
first moments of the gravity currents, generalizing the result for the
spin contribution to angular momentum in \cite{Kraus-spin}.

Although we only explicitly discussed the $1/r^7$ and $1/r^8$ terms in
the effective potential in this paper, the terms we calculated here
correspond to an infinite series of higher-moment interaction terms
which appear at all orders in $1/r$.  As discussed in
\cite{Mark-Wati,Dan-Wati-2}, for every interaction term in the $1/r^7$
interaction potential, an infinite number of other terms can be
produced by symmetrizing with  an arbitrary number $k$ of factors of
$r \cdot K$ and including an appropriate combinatorial coefficient.
In fact, most of the terms we discussed here formally vanish for
finite size objects unless some higher moments are taken.  For
example, the ``electric'' membrane-membrane interaction vanishes out
to order $1/r^9$ for finite $N$ configurations, since a finite size
object cannot have net membrane charge, but can have a dipole moment
of membrane charge.  

In addition to the higher-moment terms which arise in the one-loop
potential from factors of $r \cdot K$ there will be terms of arbitrary
order in the fermions which contribute to higher moments, generalizing
the results in \cite{Kraus-spin,mss-2,McArthur,bhp,psw} and the
general results to quadratic order in this paper.  It would be
nice to have some systematic understanding of the complete set of
higher-order  fermion terms.

The fact that there is an infinite family of terms in the matrix
theory effective potential, including terms at all orders in $1/r$,
which agree with supergravity even for finite $N$ is rather
remarkable.  It suggests that it should be possible to extend to
arbitrary backgrounds the nonrenormalization theorems of
\cite{pss,pss2,Lowe-constraints}.  At present there seems
to be very little understanding of why such nonrenormalization
theorems may exist.  This is clearly a crucial issue to understand
better, however, as these nonrenormalization theorems lie at the heart
of matrix theory's ability to reproduce 11-dimensional gravitational
physics.

This work connects to a wide range of questions which are currently of
interest in the study of matrix theory.  We conclude this paper with a
brief discussion of some ways in which this work connects to other
ideas and some directions for further study.  In particular, we
discuss the connection of these results to supersymmetric nonabelian
Born-Infeld theory, the extension of these results to higher
dimensions, the extension of this analysis to the $N$-body problem in
matrix theory, and a proposal for using the gravity currents defined
here (and their generalizations) to formulate matrix theory in a
general background geometry.

\subsection{Supersymmetric nonabelian Born-Infeld theory}

It has been suggested by a number of authors
\cite{Chepelev-Tseytlin2,Esko-Per2,Balasubramanian-gl,Chepelev-Tseytlin3}
that the infinite series of terms in the 2-body matrix theory
potential of the form $F^{2k}/r^{7k -7}$ should naturally take the
form of a nonabelian Born-Infeld action.  In the case of two
gravitons, the Born-Infeld theory is abelian and this is precisely the
form of the effective potential expected from supergravity
\cite{bbpt}.  Similar arguments have been made in other situations
where the theory is truly nonabelian.  The symmetrized trace form of
the $F^4/r^7$ term (\ref{V7}) agrees with the suggestion by Tseytlin
\cite{Tseytlin} that the ordering problems inherent in defining a
nonabelian Born-Infeld action should be resolved by symmetrizing the
trace over all orderings of $F$'s at each order (in fact, (\ref{V7})
was derived in \cite{Metsaev-Tseytlin} from string theory as part of
the nonabelian Born-Infeld action).  If it is true that the 2-body
potential in matrix theory can be expressed in a nonabelian
Born-Infeld form, it should follow that the complete set of terms at
order $1/r^7$ should be a supersymmetric completion of the $F^4/r^7$
bosonic term.  Although the exact form of the corrections to the
tree-level supersymmetry transformations are not known, it was shown
in \cite{brs} and \cite{Metsaev-Rahmanov} that the supersymmetric {\em
abelian} Born-Infeld action is quite constrained at orders $F^2
\bar{\theta} D \theta$ and $\bar{\theta} D \theta \bar{\theta} D
\theta$.  Comparing our results to these two papers provides a further
check on our expressions for the two-fermion and four-fermion parts of
the $1/r^7$ interaction.  Comparing to \cite{brs}, equation (7) we
find perfect agreement.  Setting $c_1 = c_3 = 0$ and $c_2 = -2$ in
that equation through a field redefinition, and integrating by parts
to put the derivative on the fermion field, we find that their
expression agrees perfectly with the abelianization of the two-fermion
terms in (\ref{on7}).  (Note that our normalization of the fermion
fields differs from theirs by a factor of 2).  For the four-fermion
terms, we recall that our action (\ref{eq:4L}) contained $D=10$
covariant terms as well as some non-covariant terms. We expect that
the abelian version of our covariant terms should obey the constraints
found in \cite{Metsaev-Rahmanov}, and we find in fact that the $D=10$
covariant terms in our action reduce exactly to the four-fermion term
in equation (12) of \cite{Metsaev-Rahmanov}. Our action contains one
additional non-covariant term in the abelian case, the first term in
the fourth line of (\ref{eq:4L}), but this part of the result would
seem to be specific to $D=1$.

\subsection{Effective action in other dimensions}

In performing our calculation of the effective action, we found that
the form of the effective action in the $0+0$ dimensional theory was
almost identical to our $0+1$ dimensional result.  For bosonic and two
fermion terms, we found that up to an overall constant at each order,
the $0+0$ dimensional action is obtained simply by ignoring the terms
with insertions of $D_0^2$ and replacing $\vec{K}^2$ with $K^2$ in the
other non-covariant terms.  This suggests that for other dimensions,
the answer would also have an analogous form, generalizing the $D_0^2$
insertions we have found in the $(0 + 1)$-dimensional case to
insertions of a covariant derivative squared in higher dimensions, and
in the other non-covariant terms interpreting $\vec{K}^2$ to be
$K_iK_i$ with $i \ge D-1$.  It is not immediately clear how to
generalize the non-covariant pieces in the four-fermion terms,
however,
since
these do not appear to be simply related to any covariant terms.  It
may be that there is an alternate way of writing these terms that would
make their generalization more transparent.
  
We note also that the results we have found are not really restricted
in validity to the case of a block-diagonal background. If we were to
compute the effective action in the usual sense with a general ($N
\times N$) background, we would use an action formally identical to
(\ref{squad}), omitting the factors of $r$ and interpreting $Y$ and
$K$ as matrices in the adjoint of $U(N)$.  The calculation would then
be the same as what we have done here, except that we would have only
massless propagators and therefore we would have some divergent
coefficients in place of powers of $r$. Thus, our result may
alternately be interpreted as the full effective action with $r$
acting as a regulator.  In the $(0 + 1)$-dimensional theory, this
allows us to interpret the one-loop potential as the linearized
gravity self-interaction of a single matrix theory configuration.
In higher-dimensional theories, the result can be interpreted
as an effective action for excitations of the internal gauge fields as
well as transverse scalars
in the $U(N)$  gauge theory on the world-volume of a system of
parallel D-branes.

The generalization of our results for the fermionic terms in the
effective potential might be of particular interest for the case
$D=4$, where the gauge theory in question is believed to correspond to
IIB string theory on $AdS_5 \times S^5$ \cite{Maldacena-conjecture}.
In this theory it is known that the supersymmetric one-loop effective
potential at order $F^4$ is not renormalized \cite{Dine-Seiberg,Lowe-vu}.  To
our knowledge, the fermionic terms in this potential have not
been written explicitly.  The above discussion gives an explicit
conjecture for the $F^2 \bar{\theta} D \theta$ terms, although further
work is needed to find the terms quartic in fermion backgrounds.
   
\subsection{$n$-body interactions} 

As discussed in the introduction, there has been a great deal of work
done on the problem of understanding how $n$-graviton interactions in
supergravity can be reproduced by matrix theory calculations.
In fact, there has been an extended controversy regarding
the question of whether finite $N$ matrix theory calculations
can correctly reproduce the classical supergravity interactions
between an arbitrary number of gravitons.  It is now generally agreed
that the first nontrivial term in the effective supergravity potential
of a 3-graviton system can be reproduced by a 2-loop matrix theory
calculation.  This issue was discussed in 
\cite{Dine-Rajaraman,ffi,Mark-Wati-2} and was conclusively
decided by the complete 2-loop calculation of Okawa and Yoneya in
\cite{Okawa-Yoneya,Okawa-Yoneya-2}.  Some terms describing
$n$-graviton interactions were shown to be correctly reproduced by
matrix theory in \cite{deg2}.  In the same paper, however, it was
suggested that some terms appear in the 3-loop matrix theory
description of a 4-body interaction which do not correspond to terms
in the supergravity theory.
It has been suggested that 4-body interactions may not be correctly
reproduced from matrix theory based on other discrepancies found when
matrix theory is considered on general non-flat backgrounds; we
discuss the issue of general backgrounds further in the following subsection.

In general, the $(n-1)$st-order nonlinear effects of classical gravity
can be seen  most easily in an interacting system of $n$ bodies.  Clearly,
achieving a better understanding of whether and how finite $N$ matrix
theory reproduces classical supergravity interactions between more
than two objects is a crucial issue.  Resolving this issue will help
answer the questions of whether matrix theory is correct, and if it is
correct whether it is possible to reproduce supergravity effects to an
arbitrary degree of precision by choosing large but finite $N$, or
whether there is some unexpected subtlety in the large $N$ limit
related to effects which appear as the size of a graviton bound state
exceeds the scale of its separation from other objects in the theory.

While comparing amplitudes and potentials for $n$-graviton scattering
is an interesting test of matrix theory, in general the resulting
interactions are fairly constrained by the Galilean invariance of the
light-front theory, and give essentially only some numerical checks
for coefficients in a finite number of terms.  (Although the structure
found in \cite{Okawa-Yoneya} is indeed quite nontrivial.)
Furthermore, single gravitons are BPS objects which preserve much of
the supersymmetry of the theory, so a skeptic may argue that the
agreement between matrix theory and supergravity for interactions
between such systems is easily protected by supersymmetry and does not
constitute a strong check of the theory.
It would be a significantly more powerful check of the theory if it
were possible to generalize the structural arguments used in this
paper and in \cite{Dan-Wati-2} to interactions between more than two
objects.  Even showing that the leading nonlinear term in a general
3-body interaction is correctly reproduced by supergravity using the
definitions we have given of the stress-energy tensor and other
supergravity currents would indicate that matrix theory completely
encodes the 3-supergraviton vertex of the light-front theory.  It is
conceivable that this alone would be sufficient to give an inductive
proof that matrix theory reproduces supergravity at all orders.

The results of this paper are an important intermediate step in the
calculation of the minimal general 3-body interaction potential of
matrix theory.  Note that the backgrounds corresponding to $n$-body
interactions are a special case of the background we have chosen. We
did not assume anything about the two systems we have studied, so we
could take either of our systems to contain two or more widely
separated objects by choosing $\hat{X}$ or $\tilde{X}$ to be block
diagonal.  For example, if we take $\tilde{X}$ to itself be a
block-diagonal matrix representing two matrix theory objects separated
by a distance $\rho$ which is large but still much less than $r$ then
we can calculate the terms in the effective potential at order $1/(r^7
\rho^7)$ by first integrating out the off-diagonal fields
corresponding to strings stretching between $\hat{X}$ and $\tilde{X}$
and then integrating out further fields which are off-diagonal in
$\tilde{X}$.  These terms should correspond to the leading nonlinear
terms in the gravitational effective potential of the corresponding
3-body system.  Such effects can be understood from one point of view
as the effective action for a body $\hat{X}$ considered as a probe
moving in the long-range gravitational field produced by $\tilde{X}$,
where the terms of order $1/(r^7 \rho^7)$ arise from the contribution
to the long-range gravitational field which is quadratic in the
sources encoded by $\tilde{X}$.  This type of argument was used to
show that certain terms in the 3-graviton potential are correctly
reproduced by matrix theory in \cite{Mark-Wati-2}.  Furthermore, the
problematic four graviton term discussed by Dine, Echols, and Gray in
\cite{deg2} is obtained in just this way by performing successive loop
integrals, so it should also appear in our general framework.

There is one important subtlety here which must be considered in carrying
out the calculation at higher-loop order.  The $K$ appearing in our
result for the effective action is really a background field.  Looking
back at our definitions of the fields (\ref{fields}), we see that it is
really off-diagonal fields in $\hat{Z}$ and $\tilde{Z}$, (rather than off
diagonal fields in the background $K$) that we would integrate out to get
an $n$-body interaction.  In the action, these $Z$ matrices appear in
almost the same way as the background field matrices $X$, but they are
absent from the gauge fixing term defining background field gauge.  Thus,
if we start with our expression for the effective action and integrate
out further components of the fields $K$, we have essentially chosen a
gauge different than the conventional background-field gauge for the
3-body system.  As we have seen, with the standard choice of
background-field gauge the one-loop matrix theory potential correctly
reproduces supergravity interactions in a certain gauge.  It may be
rather nontrivial to extract the gravitational physics of the 3-body
system if we choose an unorthodox gauge for the matrix theory
calculation.  (Issues along these lines were responsible for a
discrepancy between calculations at the 2-loop level in
\cite{ffi,Mark-Wati-2}).  These issues will probably need to be resolved
to find a complete correspondence between the minimal 3-body interactions
of matrix theory and supergravity.

\subsection{Matrix theory in a general background} 

As a final application of our results, we discuss matrix theory in a
general background.  We would like to generalize the matrix theory
action to one which includes a general supergravity background given
by a metric tensor, 3-form field, and gravitino field which together
satisfy the equations of motion of 11D supergravity.  This issue has
been discussed previously in
\cite{dos,Douglas-curved,Douglas-curved-2,dko,Seiberg-DLCQ,Douglas-Ooguri,Lif-back}.
In \cite{Seiberg-DLCQ} it was argued that light-front M-theory on an
arbitrary compact manifold should be reproduced by the low-energy
0-brane action on the same compact manifold; no explicit description
of this low-energy theory was given, however.  In \cite{dko} an
explicit prescription was given for the first few terms of a matrix
theory action on a general K\"ahler 3-fold which agreed with a general
set of axioms proposed in \cite{Douglas-curved-2}.  In \cite{dos} and
\cite{Douglas-Ooguri}, however, it was argued that no finite $N$
matrix theory action could correctly reproduce physics on a large K3
surface.  We  propose here an explicit formulation of matrix theory in
an arbitrary background geometry.  We leave an in-depth analysis of
this proposal to further work and only sketch the basic structure of
the theory here.

If we assume that matrix theory is a correct description of M-theory
around a flat background,  then there is a large class of curved backgrounds
for which we know it is possible to construct a matrix theory action
for $N \times N$ matrices.  This is the class of backgrounds which can
be produced as long-range fields produced by some other supergravity matter
configuration with a known description in matrix theory.  Imagine that
a background metric $g_{IJ}= \eta_{IJ} + h_{IJ}$, a 3-form field $A_{IJK}$ and
a gravitino field $\psi_I$ of light-front compactified 11-dimensional
supergravity can be produced by a matter configuration described in
matrix theory by matrices $\tilde{X}^a$.  Then the matrix theory action
describing $N \times N$ matrices $X^a$
in this background should be
precisely the effective action found by considering the block-diagonal
matrix configuration
\[
{\bf X}^a = \left[\begin{array}{cc}
{X}^a & 0\\
0 & \tilde{X}^a
\end{array}\right]
\]
(and a similar fermion configuration) and integrating out the
off-diagonal fields as well as fluctuations around the background
$\tilde{X}$.

{}From the correspondence found in this paper and in \cite{Dan-Wati-2},
we know that for
weak background fields, the first few terms in an expansion of this
derived action in the background metric are given by
\begin{eqnarray}
S_{\rm eff} & = &  S_{{\rm matrix}}  +\int dx \; T^{IJ} (x) h_{IJ} (x) 
+ \cdots\label{eq:first}\\
&=  &  S_{{\rm matrix}}  +\int
dx^+\{T^{IJ} h_{IJ}(0) + T^{IJ(i)}
\partial_i h_{IJ}(0)  + \cdots \} + \cdots
\nonumber
\end{eqnarray}
where $T^{IJ (\cdots)}$ are the moments of the matrix theory
stress-energy tensor, and there are analogous terms for the coupling of the
membrane, 5-brane and fermionic components of the supercurrent to
$A_{IJK}$ and $S_I$.  If the standard formulation of matrix theory in
a flat background is correct, the absence of corrections to the
long-range $1/r^7$ potential around an arbitrary matrix theory object
up to at least order $1/r^{11}$ implies that this formulation must be
correct at least up to terms of order $\partial^4 h$ and $h^2$.
As we have derived it, this formulation of the effective action is
only valid for certain background geometries which can be produced by
well-defined matrix theory configurations.
It is natural, however, to suppose that this result can be generalized
to an arbitrary background.  Thus,
we propose that up to nonlinear terms in the background, the general
form of the matrix theory action in an arbitrary but weak background
is given by
\begin{eqnarray}
S_{\rm weak}  & = & \int d \tau \;
\sum_{n = 0}^{\infty}  \sum_{i_1, \ldots, i_n}\frac{1}{n!} 
\left(
T^{IJ (i_1 \cdots i_n)}  \partial_{i_1} \cdots \partial_{i_n}  h_{IJ}
+J^{IJK (i_1 \cdots
i_n)}
 \partial_{i_1} \cdots \partial_{i_n} A_{IJK} 
\right. \label{eq:general-background}\\
& &\hspace{1.2in} \left.
+M^{IJKLMN (i_1
\cdots i_n)}  \partial_{i_1} \cdots \partial_{i_n}  A^D_{IJKLMN} 
+ {\rm fermion \; terms} \right) \nonumber
\end{eqnarray}
Let us make several comments about this action.  First, this
formulation is only appropriate for backgrounds with no explicit $x^-$
dependence, as we do not understand how to encode higher modes in the
compact direction in the components of the supergravity currents.
Second, note that the coupling to $A^D$ is free of ambiguity since the
net 5-brane charge must vanish for any finite matrices, so that only
first and higher derivatives of $A^D$ appear in the action.  Third,
note that though we have only provided explicit expressions for the
zeroeth and some of the first moments of $T$, the bosonic terms for
the complete set of moments were given in \cite{Dan-Wati-2}, and we
may in principle generalize the calculation of this paper to
determine all the fermionic contributions from higher order terms in the
one loop matrix theory potential.

The action (\ref{eq:general-background}) has precisely the form we
would expect for the  11-dimensional supergravity action around a
fixed classical background configuration. Typically, to study a
quantum field theory in the presence of classical background fields,
one replaces each field $\phi$ in the action with $\phi_{cl} + \phi'$
where $\phi_{cl}$ is the background field, generally taken to satisfy
the equations of motion. Applying this presciption to supergravity
with a background metric, we may expand the resulting action in powers
of the background field (considering the background metric to be a
small perturbation $h_{IJ}$ away from a flat background). Because of
the $\sqrt{g}$ in the gravity action, this will generate an infinite
series of terms with arbitrary powers of $h$, which we may write as
\begin{equation}
\label{sugen}
S = S_{\rm SUGRA} + \int dx \{h_{IJ}T^{IJ} + h_{IJ}h_{KL} U^{IJKL} + \ldots\}
\end{equation}
To leading order in $h$, this is the same as expression (\ref{sources}), but 
now $h$ is the 
classical field and $T^{IJ}$ is a quantum operator. Assuming that $h$ depends 
only on the transverse coordinates, we may make a Taylor expansion about the 
origin, and write
\begin{equation}
\int dx \;h_{IJ}T^{IJ} = \int dx^+\{h_{IJ}(0) T^{IJ} + \partial_i h_{IJ}(0) 
T^{IJ(i)} + \ldots \}
\end{equation}
where
\[
T^{IJ(i_1...i_n)} \equiv \int dx^- d\vec{x} \; T^{IJ}(x) x^{i_1}...x^{i_n}
\]
By generalizing this argument to include the 3-form and gravitino
fields, we can derive the precise form of the action (\ref{eq:first})
which is linear in the background fields.  Of course, it is desirable
to generalize our proposal to a matrix theory formulation for the
action which includes nonlinear terms in the background fields,
corresponding to the higher order terms in (\ref{sugen}). Thus we need
matrix theory quantities corresponding to moments of the operator $U$
in (\ref{sugen}) and of the operators coupling to higher powers of
$h$.  However, in principle, we should be able to determine these from
an analysis of higher-loop terms in the matrix theory potential for a
two-body interaction. To see this, note that for the term $U_m$ that
couples to $m$ powers of the background $h$, there should be a
tree-level interaction in supergravity given by
\begin{equation}
\label{int}
\int dx_1...dx_m dz \; \hat{T}(x_1)... \hat{T}(x_m) \tilde{U}_m(z)
\; \langle 
h(x_1)...h(x_m) (h(z))^m \rangle 
\end{equation}
This will contain $m$ propagators, and will give leading-order terms at order 
$1/r^{7m}$. On the matrix theory side this term must come from a diagram with 
at least $m$ loops, since we need $m$ separate traces to give the $m$ factors 
of $T$ here. But the $m$ loop contribution in matrix theory is expected to 
vanish at orders less than $1/r^{7m}$ (for agreement with classical 
supergravity), so we see that the interaction (\ref{int}) must appear in the 
leading-order $m$-loop terms. From these terms it should be possible to read 
off $U_m$ as the tilded quantity coupled to $m$ hatted $T$s.

We have now given an explicit formulation of the matrix theory action in an
arbitrary background.  The bosonic terms in the action 
are completely  determined to linear order in the background by the higher
moments of the supergravity currents given in \cite{Dan-Wati-2}.  The
terms containing fermions are determined to quadratic order in the
fermions to first order in the derivatives of the backgrounds by the
results in this paper.  In general, the terms appearing at $m$th order in
the backgrounds can be precisely determined by a $m$-loop matrix
theory calculation.  If matrix theory is correct in a flat space-time
background, then this formulation must be correct, at least for
backgrounds which can be generated by other matrix theory
configurations.  We hypothesize that the proposal is correct in a more
general class of backgrounds; more work is necessary, however, to
determine the validity of this generalization of the proposal.
We conclude this section with a few brief comments on this proposed action.

Our derivation of the general background matrix theory action has been
completely within the context of matrix theory, however in priciple
one should be able to follow the arguments of \cite{Seiberg-DLCQ} to
derive this action as a limit of the action for D0 branes in an
arbitrary type IIA supergravity background. It turns out that the
necessary terms in the D0 brane action are known only for the case
$N=1$. In this case, the matrix theory action derived directly from
the single D0 brane action exactly matches our result here.  Requiring
that such a derivation should work for the case $N>1$, one may use our
result to derive previously unknown leading terms in the action for
multiple D0 branes in an arbitrary type IIA supergrvity
background. These results will be presented in a forthcoming paper
\cite{Mark-Wati-4}.

One other approach which one might take to define matrix theory in a
general background geometry is to follow the original derivation of
matrix theory as a regularized membrane theory, but to include a
general background geometry instead of a flat background as was used
in \cite{Goldstone-Hoppe,dhn}.  The superspace formulation of a
supermembrane theory in a general 11D supergravity background was
given in \cite{bst}.  In principle, it should be possible to simply
apply the matrix regularization procedure to this theory to derive
matrix theory in a general background geometry.  Unfortunately,
however, the connection between superspace fields and component fields
is not well-understood in this theory.  Until recently, in fact, the
explicit expressions for the superspace fields were only known up to
first order in the component fermion fields $\theta$
\cite{Cremmer-Ferrara}.  In \cite{dpp}, this analysis was extended to
quadratic order in $\theta$ with the goal of finding an explicit
formulation of the supermembrane in general backgrounds in terms of
component fields, to which the matrix regulation procedure could be
applied to generate a general background formulation of matrix
theory.  It would be very interesting to carry this work through and
compare with our proposal (\ref{eq:general-background}), (\ref{sugen})
for the general background matrix theory action.

It was pointed out by Douglas in \cite{Douglas-curved-2} that a
natural Ansatz for matrix theory in a curved background metric can be
made by simply covariantizing the action in a flat background.  For
the bosonic terms, this would lead to an action of the form
\begin{equation}
S_{\rm covariant} = \int d \tau \;
{\rm Tr}\; \left( g_{ij} (X) D_\tau X^i D_\tau X^i+
 g_{ij} (X)  g_{kl} (X)[X^i, X^k][X^j, X^l] \right).
\label{eq:Douglas-action}
\end{equation}
Douglas suggested that the correct action for matrix theory in a
general background should be a supersymmetrized version of this action
in which the  ordering ambiguities inherent in the traces are somehow
resolved.  In fact, our proposal (\ref{eq:general-background})
corresponds precisely to this sort of structure.  It is
straightforward to check that the terms which arise in an expansion of
(\ref{eq:Douglas-action}) correspond precisely to the components
described in (\ref{eq:currents}) for the components of the
stress-energy tensor coupling to $h_{IJ}$.  The ordering ambiguities
are resolved by using the explicit formulae we have derived for the
higher moments of the stress tensor.  In a general background, the
theory is no longer in fact supersymmetric, since an arbitrary
background metric can break the supersymmetry.  Nonetheless, the
fermionic contributions to the action arise from the generalizations
of the expressions we have given for the fermionic contributions to
the supergravity currents.  We note also that the argument
which suggested (\ref{eq:Douglas-action}) is not easily generalized to
suggest a form for the couplings of the membrane and 5-brane currents
to a background 3-form field, or for the couplings of the fermionic
current components to a background gravitino field.

In \cite{Douglas-curved-2}, it was proposed that any formulation of
matrix theory in a curved background should satisfy a number of
axioms.  All these axioms are satisfied in a straightforward fashion
by our proposal, except one: this exception is the axiom that states
that a pair of 0-branes at points $x^i$ and $y^i$ should correspond to
diagonal $2 \times 2$ matrices where the masses of the off-diagonal
fields should be equal to the geodesic distance between the points
$x^i$ and $y^i$ in the given background metric.  In \cite{Mark-Wati-4}
we verify this axiom for the linear terms in the background $h$ which
we have described here.
It may be possible
to find situations in which this axiom is violated for our proposal at
order $h^2$ or higher.  We note, however, that in order to demonstrate
the correspondence between linearized supergravity and  matrix theory
interactions it was necessary to choose a gauge for the graviton field
$h_{IJ}$.  The explicit formulae given here and in \cite{Dan-Wati-2}
for the supergravity currents depended explicitly on this choice of
gauge.  Thus, if we use these expressions for the currents in
(\ref{eq:general-background}) we should expect that the background
must be chosen in the same gauge that we have chosen to fix the
graviton propagator.    

Thus, we synopsize the discussion in this section as follows:  we have
given a proposal for the matrix theory action in a general metric,
3-form and gravitino background.  This formulation of the action
requires the explicit computation of matrix expressions for higher
moments of all the supergravity currents.  At linear order in the
background fields, these expressions can all be found from a one-loop
matrix theory computation, although we have only performed the analysis
explicitly to quadratic order in fermion backgrounds so far.  At $m$th
order in the background fields, matrix expressions are needed for
quantities which can be determined from a $m$-loop matrix theory
calculation.  The definitions of the matrix theory currents depend upon
gauge choices for the propagating supergravity fields.  For a given
gauge choice, the theory is only defined for backgrounds compatible
with the gauge condition.  We leave the further study of this proposal
for matrix theory in a general background to further work.  It would
clearly be interesting to understand how the issues raised in
\cite{dos,Douglas-Ooguri} can be understood in the context of the
action we have proposed.  It would also be interesting to consider this
action on a general K\"ahler manifold and to understand the connection
with the results of \cite{dko}.  Finally, it would be interesting to
study this proposal for general backgrounds in the T-dual context of
the matrix string theory formulation of Dijkgraaf, Verlinde and
Verlinde \cite{DVV}.  In that framework, our proposal gives a
description of a fundamental string theory in a general background
which can have spatially dependent Ramond-Ramond fields.

\section*{Acknowledgments}

We would particularly like to thank Per Kraus, who collaborated in the
early stages of this project, for numerous productive discussions.
Thanks also to Lorenzo Cornalba, Michael Dine, Dan Freedman and Dan Kabat
for helpful conversations.
The work of
MVR is supported in part by the Natural Sciences and Engineering
Research Council of Canada (NSERC). The work of WT is supported in part
by the A.\  P.\ Sloan Foundation and in part by the DOE through
contract \#DE-FC02-94ER40818. 

\newpage
\appendix

\section{The calculation in detail}

In this section, we present in detail our calculation of the one-loop matrix 
theory effective action for a background corresponding to widely separated 
systems. The setup was presented in Section \ref{sec:setup} above, along with 
the action and propagators.

We divide the calculation up by the number of fermionic fields appearing in 
the 
result. The leading term with $2n$ fermions comes from choosing $n$ each of 
the 
vertices (\ref{fermbos}) and (\ref{bosferm}). Such a term will have $n$ 
bosonic 
and $n$ fermionic propagators, and thus appear at order $3n-1$. Thus, up to 
order $1/r^7$, where we expect the leading-order interactions, we will have 
terms with zero, two, and four fermions.

\subsection{Bosonic terms}
\label{sec:bosterms}

For this part of the calculation, we set $K_0=0$, and calculate derivative 
terms 
explicitly. We have purely bosonic terms coming from a boson loop, a fermion 
loop, and a ghost loop mentioned before. 

  For $K_0=0$, the bosonic vertices (\ref{bos1}) and (\ref{bos2}) vanish, so 
the 
complete contribution from the bosonic loop may be written as a sum over 
the number of insertions of the bosonic vertex (\ref{bos3}) as
\begin{eqnarray*}
\Gamma^{\rm bosloop}&=&\sum_{n=1}^{\infty}{(-1)^n \over n!} \int \prod d 
\tau_i
   \tr \langle (Y_{a_1}^{\dagger}(\tau_1) M_{a_1 b_1}(\tau_1) 
Y_{b_1}(\tau_1))\cdots(Y_{a_n}^{\dagger}(\tau_n) M_{a_n b_n}(\tau_n) 
Y_{b_n}(\tau_n))\\
&=&\sum_{n=1}^{\infty}{(-1)^n \over n} \int \prod d \tau_i 
\tr\left(M_{a_1 
a_2}(\tau_1)M_{a_2 a_3}(\tau_2) \cdots M_{a_n a_1}(\tau_n)\right)\\ && 
\hspace{2.5 in} \Delta(\tau_1 - \tau_2) \Delta(\tau_2 - \tau_3) \cdots 
\Delta(\tau_n - \tau_1)
\end{eqnarray*}
In order to get the complete set of 
derivative 
and non-derivative terms in the effective action, we now perform a Taylor 
expansion of all functions about a specific time, which we choose to be 
$\tau_1$ 
and perform the remaining $\tau$ integrals. Defining $\sigma_i = \tau_i - 
\tau_1$ and $\tau = \tau_1$ we have
\begin{eqnarray*}
\Gamma^{\rm bosloop}&=& \sum_{n=1}^{\infty}{(-1)^n \over n} \sum_{D_i = 
0}^{\infty}  
 \int d \tau \tr\left(M_{a_1 a_2}(\tau)M_{a_2 a_3}^{(D_2)}(\tau) \cdots M_{a_n 
a_1}^{(D_n)}(\tau)\right)\\ && \hspace{1 in}
\int\prod_{i=1}^n {dk_i \over 2\pi}\int \prod_{i=2}^n ({d\sigma_i 
\sigma_i^{D_i} 
\over D_i!})
{e^{-ik_1\sigma_2} \over k_1^2 +r^2}{e^{ik_2(\sigma_2-\sigma_3)} \over k_2^2 
+r^2} \cdots {e^{ik_n\sigma_n} \over k_n^2 +r^2}\\
&=& \sum_{n=1}^{\infty}{(-1)^n \over n} \sum_{D_i = 0}^{\infty} \prod {1 \over 
D_i!}   \int d \tau \tr\left(M_{a_1 a_2}(\tau)M_{a_2 a_3}^{(D_2)}(\tau) \cdots 
M_{a_n a_1}^{(D_n)}(\tau)\right)\\ && \hspace{1 in}
\int{dk_1 \over 2\pi} \left(\prod_{i=2}^n dk_i \{(-i \partial_{k_i})^{D_i} 
\delta(k_i-k_{i-1})\}\right){1 \over k_1^2 +r^2}\cdots{1 \over k_n^2 +r^2}\\
&=& \sum_{n=1}^{\infty}{(-1)^n \over n} \sum_{D_i = 0}^{\infty} \prod {1 \over 
D_i!}   \int d \tau \tr\left(M_{a_1 a_2}(\tau)M_{a_2 a_3}^{(D_2)}(\tau) \cdots 
M_{a_n a_1}^{(D_n)}(\tau)\right)\\ && \hspace{1 in} {1\over r^{2n+D-1}}\int{dk 
\over 2\pi} {1 \over k^2 +1}(i \partial_k)^{D_2}\{{1 \over k^2+1}\cdots(i 
\partial_k)^{D_n}\{{1 \over k^2+1}\}\cdots\}
\end{eqnarray*}
The $k$ integral is a sum of integrals of the type 
\[
\int{k^n \over (1+k^2)^m}
\]
which are always convergent (for $n<2m-1$ and all cases which may arise here)  
and easily computed. Note that this integral vanishes for odd n, so we will 
only 
have terms with an even number of derivatives.
   In an exactly analogous fashion, the fermion loop contribution can be 
computed to be
\begin{eqnarray*}
\Gamma^{\rm fermiloop}&=&-\sum_{m=1}^{\infty}{(-1)^m \over m} \sum_{D_i = 
0}^{\infty} \prod {1 \over D_i!}   \int d \tau 
\tr\left(K_{i_1}(\tau)K_{i_2}^{(D_2)}(\tau) \cdots 
K_{i_m}^{(D_m)}(\tau)\right)\\ 
&& \hspace{0.4 in} \int{dk \over 2\pi}\tr\left(\gamma^{i_1} {\slash{r} + ik 
\over 
k^2 +r^2}\gamma^{i_2}(i \partial_k)^{D_2}\{{\slash{r} + ik \over k^2 
+r^2}\cdots\gamma^{i_m}(i \partial_k)^{D_n}\{{\slash{r} + ik \over k^2 +r^2}\} 
\}\right)
\end{eqnarray*}
Up to order $1/r^8$we find again that only terms with an even number of 
derivatives contribute, since the other terms either have an odd number of 
$k$'s in the numerator or an odd number (less than nine) of gamma matrices in 
the trace.

Finally, the ghost loop contributes a term 
\begin{eqnarray*}
&& -2\sum_{n=1, D_i = 0}^{\infty}{(-1)^n \over n D_i!} 
\int d \tau \tr\left((K^2+2r \cdot 
K)(\tau)(K^2+2r \cdot K)^{(D_2)}(\tau) \cdots (K^2+2r \cdot 
K)^{(D_n)}(\tau)\right)\\ && \hspace{1 in} {1\over r^{2n+D-1}}\int{dk \over 
2\pi} {1 \over k^2 +1}(i \partial_k)^{D_2}\{{1 \over k^2+1}\cdots(i 
\partial_k)^{D_n}\{{1 \over k^2+1}\}\cdots\}
\end{eqnarray*}
Using these general results, we can explicitly calculate the action at each 
order in $1/r$, (with the help of Maple) and we find that all terms with 
explicit derivatives cancel up to and including order $1/r^7$, leaving exactly 
the action 
\[
\Gamma^{\rm bos}_{1/r^7} = {15 \over 16} \str\left(F_{ab} F_{bc} F_{cd} F_{da} 
- {1 \over 
4} F_{ab} F_{ab} F_{cd} F_{cd}\right)
\] 
calculated in \cite{Dan-Wati} using the quasistatic approximation. Thus, to 
this order, the 
result of our full calculation agrees with the quasistatic calculation. 
However, 
as we shall see below, we expect the appearance of higher derivative terms at 
order $1/r^9$ which would not appear in a quasistatic approximation. 

  We will also be interested in terms at $1/r^8$. At this order, we find two 
contributions. The first is proportional to an insertion of $r \cdot K$ into 
the 
$1/r^7$ action, 
\[
-{105 \over 16}  \str\left(F_{ab} F_{bc} F_{cd} F_{da}(r \cdot K) - {1 \over 
4} F_{ab} F_{ab} F_{cd} F_{cd}(r \cdot K)\right)
\]
 The second contribution contains a nine index 
totally antisymmetric tensor coming from the trace of nine gamma matrices in a 
fermion loop term, and 
is given by
\[
{35r_s\over 
256r^9}\str\left(F_{ij}F_{kl}F_{mn}\dot{K_p}K_q\right)\epsilon^{ijklmnpqs}
\]

\subsection{two-fermion terms}
\label{sec:twoferm}

We now move on to terms with two fermions. These arise from loops with a 
single 
insertion of each of the two boson-fermion vertices (\ref{bosferm}) and 
(\ref{fermbos}) to give two factors of $L$. In addition, we may have an 
arbitrary number of boson-boson vertices and fermion-fermion vertices. In this 
case, we shall proceed with the calculation using a non 
vanishing $K_0$ and ignoring derivative terms. In this case, we have two types 
of bosonic vertices, 
\[ 
-Y_a^{\dagger}M_{ab}Y_b
\] 
which we had before, plus the additional vertex
\[
2iY_a^{\dagger} K_0 \partial_{\tau}Y_a.
\]
The derivative here gives an extra factor of $k$ when acting on a propagator. 
Following the same steps as in the calculation above (but keeping only the 
first 
term in the Taylor expansions) we find the two fermion term to be
\begin{eqnarray}
\Gamma^{\rm LL} &=& \sum_{m,n,p=0}^{\infty}(-1)^{m+n+p}2^p\int d\tau 
\tr(L_{\alpha} K_{a_1}\cdots K_{a_m} L_{\beta} \sum_{K_0 \, insertions} ( 
M_{b_1b_2}\cdots M_{b_n b_{n+1}};K_0^p)) \nonumber \\
&& \hspace{0.4 in} \int {dk \over 2\pi} \left( \gamma^{b_{n+1}} {\slash{r} + 
ik 
\over 
k^2 +r^2}\gamma^{a_1}{\slash{r} + ik \over k^2 +r^2}\cdots 
\gamma^{a_m}{\slash{r} + ik \over k^2 +r^2} \gamma^{b_1}{k^p \over (k^2 
+r^2)^{n+p+1}} \right)_{\alpha \beta}
\label{LL}
\end{eqnarray}
Here, we sum over the ${n+p+1 \choose n}$ distinct ways of ordering the $p\; 
K_0$'s and $m \; M$'s. 

   Simplification of this expression becomes quite difficult even at the 
orders 
we are interested in. Though it is possible to calculate the integrals by 
computer, and perhaps to display all terms with gamma matrices rearranged into 
some standard form, the number of distinct terms calculated blindly in this 
way 
would be on the order of 500 at order $1/r^7$, making it a virtually 
impossible 
task to simplify into a meaningful expression. Thus, at this point, it is 
useful to make a couple of observations which will simplify the calculation 
considerably.
   
 First, since our answer must have the $SO(9)$ rotational symmetry of the 
original action, any $r_i$ must appear in an  $r^2$, an $r \cdot K$, or 
coupled 
to a gamma matrix. For two-fermion terms, all gamma matrices appear together 
in 
a product between the $L$'s, so we may always write such terms with a maximum 
of one $r_i$ coupled to a gamma matrix (since $\slash{r} \slash{r} = r^2$).  
Now, note that in the action (\ref{squad}), $r_i$ and $K_i$ appear only in the 
combination $(r_i + K_i)$ (a trivial consequence of their definitions). Thus, 
a 
transformation 
\begin{equation}
\label{rtrans}
r_i \rightarrow r_i + \Lambda_i
\end{equation}
is exactly equivalent to a transformation 
\begin{equation}
\label{Ktrans} 
K_i \rightarrow K_i + \Lambda_i.
\end{equation}
Applying transformation (\ref{rtrans}) to the effective action we see that for 
each term containing  some number of $r \cdot K$'s and/or an $r_i$ coupled to 
a 
gamma matrix, we would get a single term in which all of these $r$'s, but no 
others, were replaced with $\Lambda$s. Such a term must also appear when we 
apply the equivalent transformation (\ref{Ktrans}) to the original effective 
action, and we see that it can come only from a term in which all $r$'s appear 
only as $r^2$'s. We conclude that we may ignore all terms in our calculation 
containing $r \cdot K$ or a single $r$ coupled to a gamma matrix, restoring 
them at the end by making a transformation (\ref{Ktrans}) on the action, then  
replacing all $\Lambda$s not appearing as $\Lambda^2$ with $r$s, and setting 
$\Lambda$ to zero. The only caveat is that one must calculate to higher orders 
to find the terms of a given order which contain $r \cdot K$ or $r_i$ coupled 
to a gamma matrix, however, we will often be able to deduce the form of the 
relevant higher order terms without doing the full calculation.   

Consider first the terms in (\ref{LL}) which contain only spatial $K$s. For 
these terms, $p=0$ and all the indices $a_i$ are spatial. By our simplifying 
assumption above, we may take $\slash{r}$ to anticommute with any $\gamma^i$ 
whose index is coupled to a $K$, since the anticommutator produces a term 
containing $r \cdot K$ which we ignore. Thus, effectively, we have the 
relation 
\[
{\slash{r} + ik \over k^2 +r^2}\gamma^i{\slash{r} + ik \over k^2 +r^2} = 
-{\gamma^i \over r^2 + k^2}.
\]
Using this, the general form (\ref{LL}) simplifies considerably. For even 
orders 
in $1/r$ (even $m$), we are left with a single $(\slash{r} + ik)$ in the 
numerator, and we ignore all such terms, since they either contain a single 
$r$ 
coupled to a gamma matrix or a vanishing integral of $k/(r^2 +k^2)^n$. Thus we 
may ignore all $K_0$ independent terms at even orders in $1/r$.
   For odd orders in $1/r$ the integral in (\ref{LL}) becomes simply
\[
\int {dk \over 2 \pi} {(-1)^{(m+1)/2} \over (r^2+ k^2)^{n+m/2 +3/2}}
= { (2n+m)!! \over (n + m/2 + 1/2)!2^{(n+m/2 +3/2)}r^{2n+m+2}}
\]      
so the $K_0$ independent terms can be written
\begin{equation}
\Gamma^{\rm LL} = \sum_{l=0}^{\infty} \sum_{m+2n=2l-1}{a_l \over r^{2l+1}}\int 
d\tau 
\tr(L_{\alpha} K_{i_1}\cdots K_{i_m} L_{\beta} M_{b_1b_2}\cdots M_{b_n 
b_{n+1}})(\gamma^{b_n+1}\gamma^{i_1} \cdots \gamma^{i_m}\gamma^{b_1})_{\alpha 
\beta}
\label{K0ind}
\end{equation}
where we have the same coefficient for all terms of a given order, given by
\[
a_l = {(-1)^{l-1} \over 2^{2l+1}}{2l \choose l}
\]
  An analogous (but simpler) chain of reasoning shows that the $0+0$ action 
(\ref{szero})
gives exactly the same expression for the terms involving only spatial $K$s, 
but
with a different $a_l$.  Though we do not necessarily expect this equivalence 
to
hold between the full $0+0$ and $0+1$ dimensional calculations (the $0+0$
dimensional action displays an $SO(10)$ symmetry while the $0+1$ dimensional
calculation does not), it is interesting and helpful to carry out the simpler
$0+0$ dimensional calculation including the $K_0$ terms.  We will see that 
this 
calculation
suggests the eventual form of the $0+1$ dimensional effective action.

  From the action (\ref{szero}), we see that the complete set of two fermion 
terms in the 
$0+0$ dimensional calculation at order $1/r^q$ is proportional to:
\begin{equation}
\label{zerofull}
\sum_{m+2n+2=q} {1 \over r^{2n+2m+3}}\tr(L_{\alpha} K_{a_1} \cdots K_{a_m} 
L_{\beta} M_{b_1b_2}\cdots M_{b_n b_{n+1}})(\gamma^{b_n+1} \slash{r} 
\gamma^{a_1} \slash{r} \cdots \gamma^{a_m} \slash{r} \gamma^{b_1})_{\alpha 
\beta}
\end{equation}
Unfortunately, we can no longer assume that $\slash{r}$ anticommutes with the 
other $\gamma$ matrices, since the indices $a_i$ may now take a value of $0$, 
in which case $\slash{r}$ and $\gamma^{a_i}$ clearly commute. However, 
defining 
$\gamma_a$ (lower index) such that
\[
\gamma_i = \gamma^i, \; \; \;  \gamma_0 = -\gamma^0 = i,
\] 
(index lowered by $\eta$, not $\delta$!) we note that 
\[
\gamma^a \gamma_b + \gamma^b \gamma_a = 2 \delta_{ab}
\]
and
\[
\gamma_a \gamma^b + \gamma_b \gamma^a = 2 \delta_{ab}.
\]
Finally, since $r_0 = 0$, we have $r_a \gamma^a = r_a \gamma_a $, and we find
\begin{eqnarray}
\slash{r} \gamma^{a_1} \slash{r} \cdots \gamma^{a_m} \slash{r} &=& 
\gamma_{b_1} 
\gamma^{a_1} \gamma_{b_2} \cdots \gamma^{a_m} \gamma_{b_{m+1}} r_{b_1} \cdots 
r_{b_{m+1}}\nonumber\\
&=& \begin{array}{ll}
   \gamma_{a_1} \gamma^{a_2} \gamma_{a_3} \cdots \gamma^{a_{m-1}}\gamma_{a_m} 
r^{m+1} & (m \; {\rm odd})\\
   \gamma_{a_1} \gamma^{a_2} \gamma_{a_3} \cdots \gamma^{a_m} \gamma_b r_b 
r^{m} & (m \; {\rm even})
\end{array}
\label{gam}
\end{eqnarray}
We may choose to ignore the even $m$ terms since they have a single $r$ 
coupled 
with a gamma matrix, but in either case, we see that there is an odd number of 
gamma matrices with alternating upper and lower indices. This suggests an 
upgrade to ten dimensional notation as follows.

   First, note that the $32 \times 32$ matrices defined by:
\[
\Gamma^a = \left[
\ba{cc}
0 & \gamma^a \\
\gamma_a & 0
\ea
\right]
\]
satisfy the Clifford algebra 
\[
\{ \Gamma^a, \Gamma^b \} = 2 \delta^{ab}
\]
and thus are true gamma matrices for an $SO(10)$ spinor representation. If we 
define a $32$ component spinor by
\[
{\bf L} = \left(
\ba{c}
0  \\
L
\ea
\right)
\]
(ie satisfying a Weyl condition $\Gamma^{10}{\bf L} = {\bf L}$) we find that
\begin{eqnarray}
\bar{{\bf L}} \Gamma^a {\bf L} &=& iL \gamma^a L\nonumber\\ 
\bar{{\bf L}} \Gamma^a \Gamma^b \Gamma^c {\bf L} &=& iL \gamma^a \gamma^b 
\gamma^c L \nonumber\\
\bar{{\bf L}} \Gamma^a \Gamma^b \Gamma^c \Gamma^d \Gamma^e {\bf L} &=& iL 
\gamma^a \gamma_b \gamma^c \gamma_d \gamma^e L
\label{Gam}
\end{eqnarray}
and so on, while expressions with an even number of $\Gamma$s sandwiched 
between 
$\bar{{\bf L}} = {\bf L}\Gamma^0$ and ${\bf L}$ vanish. Thus, we see that the 
alternating upper and lower indices in (\ref{gam}) may be recovered using an 
ordinary 
product of $D=10$ gamma matrices. From now on we will drop the bold font on 32 
component spinors, as the dimension should be clear from the context.

   We may therefore rewrite the expression (\ref{zerofull}) as
\begin{equation}
\label{fullterm}
\sum_{m+2n+2=q\; odd} {1 \over r^{2n+m+2}}\tr(\bar{L_{\alpha}} K_{a_1} \cdots 
K_{a_m} 
L_{\beta} M_{b_1b_2}\cdots M_{b_n b_{n+1}})(\Gamma^{b_n+1} 
\Gamma^{a_1}  \cdots \Gamma^{a_m} \Gamma^{b_1})_{\alpha \beta}
\end{equation}
where as above, we may ignore the terms at even orders in $1/r$. But this is 
exactly obtained from the $1/r^q$ term in (\ref{K0ind}) by replacing 
$\gamma^a$ 
with $\Gamma^a$ everywhere and the left $L$ with $\bar{L}$. Since 
manipulations 
with $\gamma$ matrices and with $\Gamma$ matrices are formally identical, we 
may 
actually get the full set of terms from the $K_0$ independent terms in this 
way 
regardless of what form the $K_0$ independent terms are in. 
 
   Using the tools just outlined, we would like to write the expression 
(\ref{LL}) for 
the two-fermion terms in a more enlightening form.  In particular, we will see 
that at each order in $1/r$, we can divide 
the
terms into a set which are the dimensional reduction of $D=10$ Lorentz and 
gauge
invariant terms, plus some remaining terms which do not display the full 
$D=10$
symmetry.  These terms must still respect the symmetries of the Matrix theory
action, namely $0+1$ dimensional gauge invariance and SO(9) rotational
invariance in the (formerly) spatial indices. We proceed order by order in 
$1/r$, and in each case we begin by computing the $0+0$ dimensional result and 
then determine the remaining terms in the $0+1$ dimensional action. 

\subsubsection{Order ${1 \over r^3}$}

   Beginning at order $1/r^3$, we find a single term in expression 
(\ref{fullterm}), taking $m=1, n=0$, proportional to 
\[
\tr(\bar{L} K_a \Gamma^b \Gamma^a \Gamma^b L) \propto -4\tr(\bar{L} \Gamma^a 
[K_a, L])
\]
This is the dimensional reduction to zero dimensions of a ten dimensional 
Lorentz and gauge invariant term proportional to 
\[
\tr(\bar{L} \slash{D}L)
\]
Now, from the general $0+1$ dimensional expression (\ref{LL}) we see that at 
$1/r^3$, 
we have terms coming from $(p=0,n=0,m=1)$ and $(p=1,n=0,n=0)$. These are 
easily 
calculated, and combine to give the same expression (up to the overall 
coefficient) as in the 0+0 calculation, 
\begin{eqnarray*}
\Gamma^{0+1}_{1/r^3} &=& i\tr(\bar{L} \Gamma^a [K_a,L])\\ 
&=& i\tr(\bar{L} \Gamma^i [K_i,L])+i\tr(\bar{L} \Gamma^0 [K_0,L])\\
&=& i\tr(\bar{L} \Gamma^i [K_i,L])+\tr(\bar{L} \Gamma^0 D_0L)
\end{eqnarray*}
where in the last line we have restored the derivative term by replacing the 
$K_0$ commutator by a covariant derivative (dictated by $0+1$ dimensional 
gauge 
invariance). We see that this is the dimensional reduction to $0+1$ dimensions 
of the same expression, 
\[
\Gamma_{1/r^3} = \tr(\bar{L} \slash{D}L)
\]
that gave the $0+0$ dimensional action. Hence, we may guess that for any terms 
in the $0+0$ dimensional action which are the dimensional reduction of $D=10$ 
Lorentz and gauge covariant terms, there will be terms in the $0+1$ 
dimensional 
action which are the dimensional reduction of the same terms. The coefficient 
of these is determined by the expression (\ref{K0ind}). 
   The presence of nonvanishing terms in the effective action below order 
$1/r^7$ may seem surprising in light of the supposed correspondence with 
supergravity, whose leading-order interactions are at $1/r^7$. However, to 
find 
a potential between two physical systems, we need to require that the matrix 
theory equations of motion be satisfied by the background, and we shall see 
that 
this results in cancellation for all terms below order $1/r^7$. 

\subsubsection{Order ${1 \over r^5}$}

At order $1/r^5$, the $0+0$ dimensional expression (\ref{fullterm}) gives 
terms 
proportional to 

\[
{-3 \over 16} \left(\tr(\bar{L}_\alpha K_{a_1} K_{a_2} K_{a_3} L_{\beta}) 
(\Gamma^{b} \Gamma^{a_1} \Gamma^{a_2} \Gamma^{a_3} \Gamma^{b})_{\alpha \beta} 
+ 
\tr(\bar{L}_\alpha K_{a} L_{\beta} M_{b_1 b_2}) (\Gamma^{b_2} \Gamma^{a} 
\Gamma^{b_1})_{\alpha \beta}\right)
\]
We have chosen coefficients using (\ref{K0ind}) so that $K_0$ independent 
terms 
agree with those in the $D=0+1$ calculation.

Manipulating these expressions, we find terms which are the dimensional 
reduction of the $D=10$ covariant action          
\begin{eqnarray}
\Gamma^{\rm cov}_{1/r^5} &=& 
         {3 \over 8}i \; \tr(\bar{L}\; \Gamma^b \; D_a F_{ab}\, L)\nonumber\\
         & & +{3 \over 16}i \; \tr(\bar{L}\; \Gamma^{[ab]} \; F_{ab}\, 
\slash{D} 
L)  
         -{3 \over 16}i \; \tr(\bar{L} \; \Gamma^{[ab]} \; \slash{D} L \, 
F_{ab})
\label{cov5}
\end{eqnarray}
plus additional terms
\begin{equation}
{3 \over 4}i\left(\tr(\bar{L} K^2 [\slash{K}, L]) +\tr(\bar{L} [\slash{K}, L] 
K^2)\right)
\label{sy}
\end{equation}
At this point, it is convenient to define an operation 
$\sym (\tr(A_1...A_n);B_1;...,B_n)$ to be the average of all possible 
different insertions of the operators $B_i$ between elements of the trace. For 
example, 
\begin{eqnarray*}
\sym (\tr(\bar{L} \; \slash{D} L);\vec{K}^2, D_0^2)
   &=& {1 \over 6} (\tr(D_0^2 \bar{L}\;\vec{K}^2\; \slash{D} L) + 
\tr(\bar{L}\;D_0^2\vec{K}^2\; \slash{D} L) + \tr(\bar{L}\;\vec{K}^2\; 
D_0^2\slash{D} L)\\
   & & +\tr(D_0^2\bar{L}\; \slash{D} L \; \vec{K}^2) + \tr(\bar{L}\; 
D_0^2\slash{D} 
L \; \vec{K}^2) + \tr(\bar{L}\; \slash{D} L \;D_0^2\vec{K}^2))
\end{eqnarray*}
Note that any commutators in the trace are treated as single element.

   Using this definition, the eqpression (\ref{sy}) may be written
\[
 {3 \over 2}i\; \sym (Tr(\bar{L} [\slash{K}, L]) ; K^2)
\]   
Thus, we see that the non-covariant term is proportional to the average of all 
possible insertions of $K^2$ into the covariant $1/r^3$ action. Now, $K^2$ is 
not gauge invariant 
in 
the $D=0+1$ sense since it contains $K_0^2$, so we must have additional terms 
in 
the $D=0+1$ action which restore gauge invariance. We know, however, that the 
$D=0+1$ and $D=0+0$ actions contain the same $K_0$ independent terms, so we 
expect a term 
\begin{equation}
{3 \over 2}i\; \sym(Tr(\bar{L} \slash{D} L) ; \vec{K}^2)
\label{vect5}
\end{equation}
in the $D=0+1$ action plus additional terms whose sum with
\begin{equation}
\label{zero5}
{3 \over 2}i\; \sym(Tr(\bar{L} \slash{D} L) ; K_0^2)
\end{equation}
is gauge invariant. Using (\ref{LL}) we may calculate all terms in the $D=0+1$ 
action 
at $1/r^5$, and subtracting off a set of terms corresponding to (\ref{cov5}), 
(\ref{vect5}) and (\ref{zero5}), it is not hard to show that the remaining 
terms 
are 
\begin{equation}
\label{rem5}
-{1 \over 2}i\;\sym(Tr(\bar{L} \slash{D} L) ; K_0, K_0)
\end{equation}
As desired, these combine with the term (\ref{zero5}) to give a $0+1$ 
dimensional gauge invariant term 
\[
{1 \over 4}i\tr(\bar{L} [K_0, [K_0,\slash{D} L] )
\]
which we write (reinstating the covariant derivatives) as
\[
{1 \over 4}i\; \sym(Tr(\bar{L} \slash{D} L ); D_0^2),
\]
an average over the insertions of the covariant derivative squared into the 
covariant action at the previous order.
 
\subsubsection{Order $1 \over r^7$}

At order $1/r^7$, the picture is not significantly different. The $D=0+0$ 
calculation gives terms which are the dimensional reduction of the $D=10$ 
covariant action
\begin{eqnarray*}
\Gamma^{\rm cov}_{1/r^7} 
&=& -{5 \over 64} \tr(\bar{L}\; \Gamma^e \Gamma^a \Gamma^b \Gamma^d \; F_{ab}
                                        \,  \slash{D} L\, F_{de} )\\ 
& & - {5 \over 16} \tr(\bar{L} \; \Gamma^a \Gamma^c \; F_{ab}\, 
F_{bc}\,\slash{D} 
L) 
- {5 \over 16} \tr(\bar{L} \; \Gamma^c \Gamma^a \; \slash{D} L \,F_{ab} 
\,F_{bc} 
)\\
& & + {5 \over 32} \tr(\bar{L} \; \Gamma^d \Gamma^b \Gamma^c \;  D_a
         F_{ab} \, L \,F_{cd} )\\
& & - {5 \over 16} \tr(\bar{L}\; \Gamma^c \; D_a F_{ab}\, L\, F_{cd} )      
- {5 \over 16} \tr(\bar{L} \; \Gamma^c \; F_{cb}\, D_a F_{ab}\, L)\\ 
& & + {15 \over 16} \str\left(\bar{L} \; \Gamma^b \Gamma^c \Gamma^d \; 
F_{ab}\, 
F_{cd}\, D_a L\right) 
\end{eqnarray*}
plus additional terms proportional to all possible insertions of $K^2$ into 
the 
covariant terms at $1/r^5$, and all insertions of two $K^2$ terms into the 
$1/r^3$ action. The $D=0+1$ action contains additional terms proportional to 
the 
sum of all possible insertions of two $K_0$'s into terms in the $1/r^5$ 
action, 
and as in the $1/r^5$ case these combine with terms containing insertions of 
$K_0^2$ to give terms which correspond to all possible insertions of $D_0^2$ 
into the $1/r^5$ terms. We will write the full $1/r^7$ action with 
coefficients 
below.  

\subsubsection{Even orders, $r \cdot K$ terms}

Recalling the prescription above for restoring terms with $r_i$ not appearing 
in $r^2$, we can take the terms calculated so far, make a transformation $K_i 
\rightarrow K_i + \Lambda_i$, replace $\Lambda$ with $r$ anywhere it is 
coupled 
to $K$ or a gamma matrix, and set $\Lambda$ to zero. Noting that any covariant 
term is invariant under such a transformation, we see that the only new terms 
will be those obtained by replacing any number of $\vec{K}^2$s in a given term 
by $2r \cdot K$. 

   We may also have $r \cdot K$ terms coming by our prescription from higher 
order terms. However, it is easy to show that all $r \cdot K$ terms are simply 
insertions of $r \cdot K$ into lower order terms: By the equivalence of 
transformations  (\ref{rtrans}) and (\ref{Ktrans}), for any term containing $r 
\cdot K$ we must have a lower order term given by the removal of $r \cdot K$, 
since (\ref{Ktrans}) produces a term of this form, and (\ref{rtrans}) leaves 
the 
form of all terms unchanged. 
 
   The only other possible term we may have neglected would be a term at even 
order containing a single $r_i$ coupled to a gamma matrix. However, such a 
term 
would have to come from a term at one higher order with a gamma matrix coupled 
to a $K_i$ not appearing in a commutator. Since no such terms appear at orders 
up to $1 \over r^7$, the only odd order terms appearing up to order $1/r^6$ 
will 
be given by insertions of odd numbers of $r \cdot K$s into terms calculated 
above.   
   
\subsubsection{Order $1 \over r^8$}

In the applications below, it will be useful to determine some of the terms at 
order $1 \over r^8$. Using the $0+0$ dimensional action, we find that apart 
from 
terms containing $r \cdot K$, the two-fermion terms at $1/r^8$ are given by 
the 
$D=10$ gauge invariant expression 
\begin{eqnarray}
\Gamma_{1/r^8} &=& -{35 \over 256r^9}r_g \str\left(\bar{L}\, F_{ab} F_{cd} 
F_{ef} 
\Gamma^{[abcdefg]} L\right) -{105 \over 32r^9}r_e \str\left(\bar{L}\, F_{ab} 
F_{bc} F_{cd} 
\Gamma^{[ade]} L\right)\nonumber\\
 & & -{105 \over 128r^9}r_e \str\left(\bar{L}\, F_{ab} F_{ab} F_{cd} 
\Gamma^{[cde]} 
L\right)
\label{cov8-2}
\end{eqnarray}
which should also be correct for $0+1$ dimensions. Note that in the 
prescription 
above, these terms would have come from non-covariant terms at $1/r^9$ with a 
gamma matrix coupled to a single free $K_i$ not appearing in a commutator. No 
such terms appeared at lower odd orders, thus we did not have any terms of the 
form (\ref{cov8}) at lower even orders.

\subsubsection{off-shell action to order $1/r^7$}

Collecting the various terms calculated above, we now write down the complete 
off-shell action to order $1/r^7$. The terms which come from the dimensional 
reduction of a $D=10$ Lorentz and gauge invariant action are given by
\begin{eqnarray*}
\Gamma^{\rm cov}_{1/r^3} &=& \tr(\bar{L} \slash{D} L)\\
\Gamma^{\rm cov}_{1/r^5} &=& 
         {3 \over 8}i \; \tr(\bar{L}\; \Gamma^b \; D_a F_{ab}\, L)\\
         & & +{3 \over 16}i \; \tr(\bar{L}\; \Gamma^{[ab]} \; F_{ab}\, 
\slash{D} 
L)  
         -{3 \over 16}i \; \tr(\bar{L} \; \Gamma^{[ab]} \; \slash{D} L \, 
F_{ab})\\ 
\Gamma^{\rm cov}_{1/r^7} 
&=& -{5 \over 64} \tr(\bar{L}\; \Gamma^e \Gamma^a \Gamma^b \Gamma^d \; F_{ab}
                                        \,  \slash{D} L\, F_{de} )\\ 
& & - {5 \over 16} \tr(\bar{L} \; \Gamma^a \Gamma^c \; F_{ab}\, 
F_{bc}\,\slash{D} 
L 
) 
- {5 \over 16} \tr(\bar{L} \; \Gamma^c \Gamma^a \; \slash{D} L \,F_{ab} 
\,F_{bc} 
)\\
& & + {5 \over 32} \tr(\bar{L} \; \Gamma^d \Gamma^b \Gamma^c \;  D_a
         F_{ab} \, L \,F_{cd} )\\
& & - {5 \over 16} \tr(\bar{L}\; \Gamma^c \; D_a F_{ab}\, L\, F_{cd} )      
- {5 \over 16} \tr(\bar{L} \; \Gamma^c \; F_{cb}\, D_a F_{ab}\, L)\\ 
& & + {15 \over 16} \str\left((\bar{L} \; \Gamma^b \Gamma^c \Gamma^d \; 
F_{ab}\, 
F_{cd}\, 
D_a L\right) 
\end{eqnarray*}    
All other terms up to order $1/r^7$ come from insertions of $r \cdot K$, 
$\vec{K}^2$, or $D_0^2$. Using the symmetrization operation defined above, we 
can now write the complete set of two-fermion terms in the $D=0+1$ off-shell 
action as
\begin{eqnarray*}
\Gamma_{1/r^3} &=& \Gamma^{\rm cov}_{1/r^3}\\
\Gamma_{1/r^4} &=& -3\;\sym (\Gamma^{\rm cov}_{1/r^3};r \cdot K)\\    
\Gamma_{1/r^5} &=& \Gamma^{\rm cov}_{1/r^5}\\
             & & + {1 \over 4} \sym (\Gamma^{\rm cov}_{1/r^3}; D_0^2)
             -{3 \over 2} \sym (\Gamma^{\rm cov}_{1/r^3};\vec{K}^2) +
              +{15 \over 2}\sym (\Gamma^{\rm cov}_{1/r^3};r \cdot K, r \cdot 
K)\\ 
\Gamma_{1/r^6} &=& -5 \;\sym (\Gamma^{\rm cov}_{1/r^5}; r \cdot K)\\
               & & -{15 \over 8}  \sym (\Gamma^{\rm cov}_{1/r^3}; D_0^2,r 
\cdot K)
             +{15 \over 2} \sym (\Gamma^{\rm cov}_{1/r^3};\vec{K}^2,r \cdot K) 
\\     
\Gamma_{1/r^7} &=& \Gamma^{\rm cov}_{1/r^7}\\ 
       & & + {5 \over 8}\sym (\Gamma^{\rm cov}_{1/r^5}; D_0^2 )
       - {5 \over 2}\sym (\Gamma^{\rm cov}_{1/r^5}; \vec{K}^2) +
          {35 \over 2}\sym (\Gamma^{\rm cov}_{1/r^5}; r \cdot K, r \cdot K)\\ 
       & & + {15 \over 8} \sym (\Gamma^{\rm cov}_{1/r^3}; \vec{K}^2, 
\vec{K}^2)
        -{15 \over 16} \sym (\Gamma^{\rm cov}_{1/r^3}; D_0^2, \vec{K}^2)
        +{1 \over 16} \sym (\Gamma^{\rm cov}_{1/r^3}; D_0^2,D_0^2)\\
       & & -{105 \over 4} \sym (\Gamma^{\rm cov}_{1/r^3}; \vec{K}^2, r \cdot 
K, r 
\cdot K)
        +{105 \over 48} \sym (\Gamma^{\rm cov}_{1/r^3}; D_0^2, r \cdot K, r 
\cdot 
K)\\
        & & +{315 \over 8} \sym (\Gamma^{\rm cov}_{1/r^3}; r \cdot K, r
\cdot K,r \cdot K, r \cdot K)
\end{eqnarray*}   

We will see that things simplify greatly when we restrict to a background that 
satisfies the matrix theory equations of motion. In our $D=10$ notation, these 
equations of motion read   
\begin{equation}
\label{fermeom2}
\slash{D} L = 0
\end{equation}
and
\begin{equation}
\label{boseom2}
D_aF_{ab} = i\bar{L}\Gamma^bL
\end{equation}

Examining the covariant terms above, we see that apart from the final term in 
the $1/r^7$ action, all terms listed contain either $\slash{D} L$ or 
$D_aF_{ab}$, and so these terms either cancel directly or may be canceled by 
the four-fermion terms which we will soon calculate.     

\subsection{four-fermion terms}
\label{sec:fourferm}

 The remaining terms at orders up to $1/r^7$ contain four fermions.  These
arise from a loop with two each of the vertices  
(\ref{fermbos}) and (\ref{bosferm}) plus insertions of boson-boson and 
fermion-fermion vertices.
In this part of the calculation, we will calculate using derivatives and
setting $K_0$ to zero.  Also, we will return to our 16 component spinor
notation and $16 \times 16$ gamma matrices.  The leading-order four-fermion
terms are at $1/r^5$. 

The calculations below rely heavily on Fierz rearrangement identities for the 
$16\times 16$ symmetric $\gamma$ matrices. A treatment of these is given in 
Appendix B.

\subsubsection{Order ${1 \over r^5}$}

At order $1/r^5$ we have a single term, arising from two vertices each of the 
forms (\ref{bosferm}) and (\ref{fermbos})  without any derivatives. This is 
given by
\begin{eqnarray}
\Gamma^{LLLL}_{1/r^5} &=&  {1 \over 2}\int d\tau \int {dk \over 2 \pi} \tr(L 
\gamma^a {\slash{r} + ik \over k^2 +r^2} \gamma^b L {1 \over k^2 + r^2} L 
\gamma^b {\slash{r} + ik \over k^2 +r^2} \gamma^a L{1 \over k^2 + 
r^2})\nonumber\\
&=& {5 \over 64r^7} \int d\tau \tr(L \gamma^a \slash{r} \gamma^b L \; L 
\gamma^b 
\slash{r} \gamma^a L) - {1 \over 64r^5} \int d\tau \tr(L \gamma^a \gamma^b L 
\; 
L 
\gamma^b \gamma^a L)
\label{fourL5}
\end{eqnarray}

Generally we would like to manipulate the gamma matrices in our expression 
into
sums of antisymmetric products in order to apply the Fierz identities from 
Appendix B. For the trace in the first term, we have:
\begin{eqnarray*}
& &\tr(L \gamma^a \slash{r} \gamma^b L \; L \gamma^b \slash{r} \gamma^a L)\\
&=&r_ir_j\{\tr(L\gamma^{[kli]}L \; L \gamma^{[lkj]}L) + 2\tr(L\gamma^{[ki]}L 
\; L \gamma^{[kj]}L) + 8\tr(L\gamma^iL \; L \gamma^jL)\\
 & & \hspace{0.5in} +2\tr(L\gamma^kL \; L \gamma^kL))\delta_{ij}-2\tr(LL \; 
LL)\delta_{ij} \}\\
 &=&r_ir_j\{4\tr(L\gamma^kL \; L \gamma^kL))\delta_{ij}-4\tr(LL \; 
LL)\delta_{ij} 
\} 
\end{eqnarray*}
where we have used Fierz identities (\ref{2fi1},\ref{2fi9}) from
Appendix B. 
Note that after application of the Fierz identity, the two factors of $r$ 
become contracted with one another. The second trace in (\ref{fourL5}) becomes
\begin{eqnarray*}
& &\tr(L \gamma^a \gamma^b L \; L \gamma^b \gamma^a L)\\
&=&\tr(L\gamma^{[kl]}L \; L \gamma^{[lk]}L) - 2\tr(L\gamma^kL 
\; L \gamma^kL) + 10\tr(LL \; L L)\\
&=&-4\tr(L\gamma^k L \; L \gamma^kL))+4\tr(LL \; LL)\delta_{ij} \} 
\end{eqnarray*}
using Fierz identity (\ref{0fi1}). Thus, we find that the four 
fermion contribution at $1/r^5$ is given by 
\begin{eqnarray}
\Gamma^{LLLL}_{1/r^5} &=& {3 \over 8} \int d\tau (\tr(L \gamma^i L \; L 
\gamma^i 
L) - \tr(LL \; LL))\nonumber\\
&=& -{3 \over 8} \tr(\bar{L} \Gamma^a L \; \bar{L} \Gamma^a L)  \label{df5}
\end{eqnarray}
Now, looking back at the result for the two fermion calculation, we see that 
the only term up to order $1/r^5$ which does not vanish by the equations of 
motion is
\[
{3 \over 8} i \; \tr(\bar{L} \Gamma^b D_aF_{ab} L)
\]
Bringing the last $L$ to the front by the cyclic property of the trace and 
combining this with the four-fermion term we have just calculated, we have
\begin{equation}
\label{combo}
-{3 \over 8} i \; \tr(\bar{L} \Gamma^b L \; (D_aF_{ab} -i\bar{L} \Gamma^a L))
\end{equation}
which clearly vanishes on-shell by the bosonic equation of motion 
(\ref{boseom2}). Thus, we have shown so far that up to order $1 \over r^5$ all 
terms in the one loop matrix theory effective action vanish for background 
fields satisfying the equations of motion.

\subsubsection{Order ${1 \over r^6}$}

At order $1/r^6$, we have four-fermion terms coming from taking the four mixed 
vertices above, plus either a single fermion vertex or a single derivative (ie 
using the first order term in the Taylor expansion of one of the background 
$L$ 
fields). In either case, apart from $r^2$ factors in the denominator, each 
non-vanishing term contains exactly one $r_i$ coupled to a $K_i$ or a 
$\gamma^i$, so we may obtain such terms from the $1/r^7$ terms using the 
transformation (\ref{Ktrans}), as described above. We will return to these 
terms in section (\ref{sec:sixterms}) and show that the four-fermion terms at 
$1/r^6$ exactly 
cancel the two-fermion terms by the matrix theory equations of motion.

\subsubsection{Order ${1 \over r^7}$}

The contributions to the four-fermion terms at order $1/r^7$ are of four 
types.
Firstly, we have a term with two each of the mixed vertices (\ref{fermbos}) 
and 
(\ref{bosferm}) plus a single insertion of the bosonic vertex (\ref{bos3}),  
\begin{equation}
\label{LLKKLL}
\Gamma^{LLMLL} =  \int d\tau \int {dk \over 2 \pi} \tr(L \gamma^a (\slash{r} + 
ik) \gamma^b L M_{bc} L \gamma^c (\slash{r} + ik) \gamma^a L){1 \over (k^2 + 
r^2)^5}
\end{equation}
Next, we have terms with the four mixed vertices plus two insertions of the 
fermion fermion vertex, either between the same pair 
of $L$'s, or between different pairs of $L$'s. The first is given by
\begin{equation}
\label{LKKLLL}
\Gamma^{LKKLLL} =  -\int d\tau \int {dk \over 2 \pi} \tr(L \gamma^a (\slash{r} 
+ 
ik) \slash{K} (\slash{r} + ik) \slash{K} (\slash{r} + ik) \gamma^b L \; L 
\gamma^b (\slash{r} + ik) \gamma^a L){1 \over (k^2 + r^2)^6}
\end{equation}
while the second contribution is
\begin{equation}
\label{LKLLKL}
\Gamma^{LKLLKL} =  -{1 \over 2} \int d\tau \int {dk \over 2 \pi} \tr(L 
\gamma^a 
(\slash{r} + ik) \slash{K} (\slash{r} + ik) \gamma^b L \; L \gamma^b 
(\slash{r} 
+ ik) \slash{K} (\slash{r} + ik) \gamma^a L){1 \over (k^2 + r^2)^6}.
\end{equation}
The remaining terms involve Taylor expansions of the background fields (see 
the 
calculation of the boson loop).  
 For the first of these we take two each of the mixed vertices (\ref{fermbos}) 
and (\ref{bosferm}), as in the $1/r^5$ term, but use the second order term in 
the Taylor expansion of the product of background fields. This term will then 
have two derivatives, either on the same $L$ or on two different $L$s. 
Following steps analogous to the boson loop calculation above, we find that 
such terms are given by
\begin{eqnarray}
\Gamma^{dd}&=& \sum_{d_1+d_2+d_3 =2} \int {dk \over 2 \pi} \tr( L 
\gamma^a {\slash{r} + ik \over k^2 +r^2} \gamma^b L^{(d_1)}\nonumber\\ & & 
\hspace{1in}{(i\partial_k)^{d_1} \over d_1!}\{{1 \over k^2 + r^2} L^{(d_2)} 
\gamma^b {(i\partial_k)^{d_2} \over d_2!}\{{\slash{r} + ik \over k^2 +r^2} 
\gamma^a L^{(d_3)}{(i\partial_k)^{d_3} \over d_3!}\{{1 \over k^2 + 
r^2})\}\}\} \label{twoder}
\end{eqnarray}  
Finally, we have a term coming from taking the four mixed vertices plus a 
single fermionic vertex and using the first order term in the Taylor expansion 
(one derivative on one of the background fields). This term may be written 
\begin{eqnarray}
\Gamma^{Kd}&=& -\sum_{d_1+d_2+d_3+d_4 =1} \int {dk \over 2 \pi} \tr( L 
\gamma^a {\slash{r} + ik \over k^2 +r^2} \slash{K}^{(d_1)} 
(i\partial_k)^{d_1}\{{\slash{r} + ik \over k^2 +r^2}\gamma^b 
L^{(d_2)}\label{oneder}\\ & & \hspace{1in}(i\partial_k)^{d_2} \{{1 \over k^2 + 
r^2} L^{(d_3)} 
\gamma^b (i\partial_k)^{d_3} \{{\slash{r} + ik \over k^2 +r^2} \gamma^a 
L^{(d_4)}(i\partial_k)^{d_4} \{{1 \over k^2 + 
r^2})\}\}\} \nonumber
\end{eqnarray} 

  Before simplifying the expressions above, we note the following. Based on 
the 
calculation at $1 \over r^5$, we may expect to get four-fermion terms which 
pair with two-fermion terms containing $D_aF_{ab}$ to cancel them by the 
bosonic equations of motion (\ref{boseom2}). These terms are given by:
\begin{eqnarray}
\Gamma^{LLLL}_{eom} &=& {5 \over 16} \tr(\bar{L} \Gamma^a \vec{K}^2 L \; 
\bar{L} \Gamma^a L) + 
{5 \over 8} \tr(\bar{L} \Gamma^a L \vec{K}^2 \bar{L} \Gamma^a L)\nonumber\\
& & -{5 \over 32} \tr(\bar{L} \Gamma^a L \; D_0^2(\bar{L}) \Gamma^a L) 
-{5 \over 32} \tr(\bar{L} \Gamma^a L \; \bar{L} \Gamma^a (D_0^2L))
-{5 \over 32} \tr(\bar{L} \Gamma^a L \; (D_0\bar{L}) \Gamma^a 
(D_0L))\nonumber\\
& & +{5 \over 32} \tr(\bar{L} \Gamma^c \Gamma^b \Gamma^d F_{cd} L \; \bar{L} 
\Gamma^b L)-{5 \over 8} \tr(\bar{L} \Gamma^a L F_{ab} \bar{L} \Gamma^b L) 
\label{dfterms}
\end{eqnarray}
We will find it convenient to start with these terms and subtract off an 
equivalent set of terms in the remaining calculation. We now calculate the 
remaining terms, proceeding by the number of derivatives appearing.  

{\bf Two derivative terms}

We begin by calculating the terms containing two derivatives. These are 
exactly 
the terms in expression (\ref{twoder}), which we may rewrite as
\begin{eqnarray}
& &\sum_{d_1+d_2+d_3 =2}\tr(L \gamma^a \slash{r} \gamma^b L^{(d_1)} \; 
L^{(d_2)} 
\gamma^b 
\slash{r} \gamma^a L^{(d_3)}) \label{der2r} \\
& & \hspace{0.5in} \int {dk \over 2 \pi} {1 \over k^2 +r^2}   
{(i\partial_k)^{d_1} \over d_1!}\{{1 \over k^2 + r^2} L^{(d_2)} 
{(i\partial_k)^{d_2} \over d_2!}\{{1 \over k^2 +r^2} \gamma^a 
L^{(d_3)}{(i\partial_k)^{d_3} \over d_3!}{1 \over k^2 + r^2}\}\}\nonumber\\
& & - \sum_{d_1+d_2+d_3 =2} \tr(L \gamma^a \gamma^b L^{(d_1)} \; L^{(d_2)} 
\gamma^b \gamma^a L^{(d_3)})\} \label{der2nor}\\
& & \hspace{0.5in}\int {dk \over 2 \pi} {k \over k^2 +r^2}   
{(i\partial_k)^{d_1} \over d_1!}\{{1 \over k^2 + r^2} L^{(d_2)} 
{(i\partial_k)^{d_2} \over d_2!}\{{k \over k^2 +r^2} \gamma^a 
L^{(d_3)}{(i\partial_k)^{d_3} \over d_3!}{1 \over k^2 + r^2}\}\}
\nonumber 
\end{eqnarray}
Here, we have separated terms with zero $k$s and two $k$s in the numerator 
(the 
integral of terms choosing a single $k$ in the numerator vanish). 
Starting with the terms (\ref{der2r}), we can evaluate the integral, arrange 
the gamma matrices into sums of antisymmetric products and perform integration 
by parts to rearrange the derivatives, giving the following expression:
\begin{eqnarray*}
& &{7r_ir_j \over 256r^9} 
\left(2\tr(L*L\;\ddot{L}*L)+2\tr(L*\ddot{L}\;L*L)+\tr(L*\dot{L}\;\dot{L}*L) + 
\tr(\dot{L}*\dot{L}\;L*L)\right)\\
& &\hspace{0.5in}\{\gamma^{[kli]}\otimes \gamma^{[lkj]}+2\gamma^{[ki]}\otimes  
\gamma^{[kj]}+8\gamma^i\otimes \gamma^j+2\gamma^k\otimes 
\gamma^k\delta_{ij}-2\identity \otimes \identity \delta_{ij}\}
\end{eqnarray*}
Here, we have introduced the notation
\[
\tr(A*B\;C*D)\{\gamma \otimes \tilde{\gamma} \} = \tr(A \gamma B \; C 
\tilde{\gamma} D)
\]
Now, using the Fierz identities (\ref{2fi1}, \ref{2fi9}), this reduces to                                                                     
\begin{equation}
\label{d2r}
{7 \over 64r^7}\left( 2\tr(L \gamma^a L\; L \gamma^a \ddot{L}) + 2\tr(L 
\gamma^a 
L\; \ddot{L} \gamma^a L) + \tr(L \gamma^a \dot{L}\; \dot{L} \gamma^a L)
 +\tr(\dot{L} \gamma^a \dot{L}\; L \gamma^a L) \right)
\end{equation} 
We now move to the remaining terms (\ref{der2nor}). After integration,
gamma 
matrix algebra, and differentiation by parts, these become:
\begin{eqnarray*}
& &{-1 \over 256r^7} 
\left(6\tr(L*L\;\ddot{L}*L)+6\tr(L*\ddot{L}\;L*L)+13\tr(L*\dot{L}\;\dot{L}*L) 
+ 
3\tr(\dot{L}*\dot{L}\;L*L)\right)\\
& &\hspace{0.5in}\{\gamma^{[kl]}\otimes \gamma^{[lk]}-2\gamma^k\otimes  
\gamma^k+10 (\identity \otimes \identity) \}
\end{eqnarray*}   
Using the Fierz identity (\ref{0fi1})  we may write this as
\begin{eqnarray*}
& &{3 \over 64r^7}\left( 2\tr(L \gamma^a L\; L \gamma^a \ddot{L}) + 2\tr(L 
\gamma^a L\; \ddot{L} \gamma^a L) + \tr(L \gamma^a \dot{L}\; \dot{L} \gamma^a 
L)
 +\tr(\dot{L} \gamma^a \dot{L}\; L \gamma^a L) \right)\\
& & -{5 \over 128} \left(\tr(L \gamma^{[kl]} \dot{L}\; \dot{L} \gamma^{[lk]} 
L) 
-2 \tr(L \gamma^k \dot{L}\; \dot{L} \gamma^k L)+ 10\tr(L  \dot{L}\; \dot{L}  
L) 
\right)
\end{eqnarray*}  
Combining these terms with those in (\ref{d2r}) and subtracting the two 
derivative terms in (\ref{dfterms}), we find the remaining two derivative term 
to be 
\begin{eqnarray}
\Gamma^{\dot{L}\dot{L}}&=&{5 \over 32r^7}\left( \tr(L \gamma^a \dot{L}\; L 
\gamma^a \dot{L}) + \tr(\dot{L} \gamma^a L\; \dot{L} \gamma^a L) + 
2\tr(\dot{L} 
\gamma^a \dot{L}\; L \gamma^a L) \right)\label{cov2}\\
& & -{5 \over 128r^7} \left(\tr(L \gamma^{[kl]} \dot{L}\; \dot{L} 
\gamma^{[lk]} 
L) +2 \tr(L \gamma^k \dot{L}\; \dot{L} \gamma^k L)+ 6\tr(L  \dot{L}\; \dot{L}  
L) \right)\nonumber
\end{eqnarray} 
Here, we have rewritten the terms with two derivatives on a single $L$ using 
integration by parts to leave only terms with derivatives on separate $L$s. We 
note here that the last line is equivalent by Fierz identity (\ref{0fi1}) to 
\begin{equation}
\label{LLalt}
{5 \over 128r^7} \left(\tr(\dot{L} \gamma^{[kl]} \dot{L} \; L \gamma^{[lk]} L) 
+2 \tr(\dot{L} \gamma^k \dot{L}\; L \gamma^k L)+ 6\tr(\dot{L}  \dot{L}\; L  L) 
\right)
\end{equation}
however, we shall use the form given above.

{\bf One derivative terms}

Terms with a single derivative come either from (\ref{oneder}) or from terms 
in 
(\ref{LLKKLL}) in which the indices on $M$ are taken to be $M_{0i}$ or 
$M_{i0}$. We begin with the latter terms. Performing the $k$ integration in 
(\ref{LLKKLL}), we get
\[
-{35 \over 256} \tr(L \gamma^a \slash{r} \gamma^b L M_{bc} L \gamma^c 
\slash{r} 
\gamma^a L) + {5 \over 256}\tr(L \gamma^a \gamma^b L M_bc L \gamma^c  \gamma^a 
L)
\] 
Keeping only the terms with a derivative ($\dot{K}$), and rearranging the 
gamma 
matrices into sums of antisymmetric products, we have:
\begin{eqnarray}
& & { 35 \over 256} {r_m r_p \over r^9}\tr(L*L\dot{K_i} L*L)
 \{2 \gamma^{[kmi]} \otimes \gamma^{[kp]} -2\gamma^{[km]} \otimes 
\gamma^{[kpi]} + 4\gamma^m \otimes \gamma^{[pi]}\nonumber\\
 & & \hspace{2in} +4 \gamma^{[mi]} \otimes \gamma^p  -2 \delta_{mp} (\gamma^i 
\otimes \identity - \identity \otimes \gamma^i) \} \label{ldkl1} \\
& & +{ 5 \over 256r^7} \tr(L*L\dot{K_i} L*L) \{2 \gamma^{[ki]} \otimes 
\gamma^k 
+ 
2 \gamma^k \otimes \gamma^{[ki]} - 4 \gamma^i \otimes \identity +4 \identity 
\otimes \gamma^i \} \label{ldkl2}
\end{eqnarray}

Moving to expression (\ref{oneder}) we can compute the $k$ integrals to obtain
\begin{eqnarray}
& &{5 \over 256} \left( \tr(\dot{L} \gamma^a \slash{K} \gamma^b L\; L \gamma^b 
\gamma^a L) - \tr(L \gamma^a \slash{K} \gamma^b \dot{L}\; L \gamma^b \gamma^a 
L)\right.\nonumber\\ & & \left. \hspace{1.2in}-3\tr(L \gamma^a \slash{K} 
\gamma^b L\; 
\dot{L} \gamma^b \gamma^a L)+3\tr(L \gamma^a \slash{K} \gamma^b L\; L \gamma^b 
\gamma^a \dot{L})\right) \nonumber\\
& & +{35 \over 256} \tr(L \gamma^a \slash{r} \dot{\slash{K}} \gamma^b L\; L 
\gamma^b \slash{r} \gamma^a L) \label{dldk}              
\end{eqnarray}
Here, we have also used an integration by parts to obtain the left-right 
antisymmetric form of the terms in the first line. From the $\dot{K}$ term in 
the last line, we now arrange the gamma matrices into sums of antisymmetric 
products, to get
\begin{equation}
 { 35 \over 256} {r_m r_p \over r^9} \tr(L* \dot{K_i}L \; L*L)
 \{- \gamma^{[klmi]} \otimes \gamma^{[lkp]} -4\gamma^{[mi]} \otimes \gamma^p + 
2\gamma^m \otimes \gamma^{[pi]}- 2 \delta_{mp} \gamma^{[ki]} \otimes \gamma^k 
\} 
\end{equation} 

At this point, we consider all terms containing $\dot{K}$ together, and use a 
Fierz identity, (\ref{3fi}) to reduce these to a form in which all factors of 
$r$ 
appear 
as $r^2$. The $\dot{K}$ terms then become 
\begin{eqnarray*}
& &{5 \over 16} \left( \tr(L \gamma^{[ki]}  L \dot{K_i} L \gamma^k L) +\tr(L 
\gamma^k  L \dot{K_i} L \gamma^{[ki]} L) -2\tr(L \gamma^i  L \dot{K_i} L L) 
+2\tr(L L \dot{K_i} L \gamma^i L) \right)\\
& & +{35 \over 128} \left(\tr(L \gamma^k \dot{K_i} L \;L \gamma^{[ki]} L) - 
\tr(L 
\gamma^{[ki]} \dot{K_i} L \; L \gamma^k L) \right)
\end{eqnarray*}
Finally, subtracting those terms in (\ref{dfterms}) with a $\dot{K}$, we are 
left with
\begin{eqnarray}
\label{kdot}
\Gamma^{\dot{K}} &=& {5 \over 16} \left( \tr(L \gamma^k \dot{K_i} L \; L 
 \gamma^{[ki]} L) +\tr(L \gamma^{[ki]}  L \dot{K_i} L \gamma^k L) + \tr(L 
\gamma^k 
 L \dot{K_i} L \gamma^{[ki]} L) \right)\\
 & & + {5 \over 128} \left( \tr(L \gamma^{[ki]} [\dot{K_i} , L] \; L \gamma^k 
L) 
+ 
\tr(L \gamma^{[ki]}  L  [\dot{K_i} ,L] \gamma^k L) \right)\nonumber
\end{eqnarray}

Going back to the $\dot{L}$ terms in (\ref{dldk}), we rearrange the gamma 
matrices into sums of antisymmetric products and obtain 
\begin{eqnarray}
\label{ldot}
\Gamma^{\dot{L}} &=& {5 \over 256} \left(\tr(L* K_i \dot{L} \; 
L * L)- \tr(\dot{L} * K_i L \; L * L) 
+3\tr(L * K_i L \; \dot{L} *    L)-3 \tr(L *  K_i 
L \; L *    \dot{L})\right)\nonumber\\
& &\hspace{1in} \{\gamma^{[kli]}\otimes \gamma^{[lk]} + 6\gamma^i\otimes 
\identity + 2\identity \otimes \gamma^i \} 
\end{eqnarray}
We will use this form, however, we note that by Fierz identities (\ref{1fi1}, 
\ref{1fi2}), this is equivalent to
\begin{eqnarray}
\label{ldotalt}
& & {5 \over 512} \left(\tr(L *  \dot{L} K_i L *  
L)-\tr(L *   L K_i \dot{L} *    L) +3 
\tr(\dot{L} *   L K_i L *    L) -3 \tr(L *   L K_i 
L *    \dot{L})\right) \nonumber\\
& &\hspace{0.4in} \{\gamma^{[kli]}\otimes \gamma^{[lk]} + \gamma^{[kl]}\otimes 
\gamma^{[lki]}+ 4\gamma^{[ki]}\otimes \gamma^k -4\gamma^k \otimes 
\gamma^{[ki]}+ 8\gamma^i\otimes \identity + 8\identity \otimes \gamma^i 
\}
\end{eqnarray}

{\bf Zero derivative terms}

  Finally, we calculate the terms with no derivatives. These come from the 
term 
with a boson-boson vertex, which we found above to be
\begin{equation}
\label{A}
-{35 \over 256} \tr(L \gamma^a \slash{r} \gamma^b L M_{bc} L \gamma^c 
\slash{r} 
\gamma^a L) + {5 \over 256}\tr(L \gamma^a \gamma^b L M_{bc} L \gamma^c  
\gamma^a 
L)
\end{equation}  
plus the two terms with a pair of fermion-fermion vertices. Performing the $k$ 
integrals in these,  (\ref{LKKLLL}) becomes
\begin{equation}
\label{B}
-{35 \over 256} \tr(L \gamma^a \slash{r} \slash{K} \slash{K} \gamma^b L\; L 
\gamma^b \slash{r} \gamma^a L) + {5 \over 256}\tr(L \gamma^a \slash{K} 
\slash{K} 
\gamma^b L\; L \gamma^b  \gamma^a L)
\end{equation}
while (\ref{LKLLKL}) gives simply  
\begin{equation}
\label{C}
{5 \over 64} \tr(L \gamma^a \slash{K} \gamma^b L\; L \gamma^b \slash{K} 
\gamma^a 
L)\; .
\end{equation}
Rewriting this last term in our standard form with antisymmetric products of 
gamma matrices, we get
\begin{equation}
\label{LKLtemp}
{5 \over 64}\tr(L*K_iL\;L*K_jL) \{\gamma^{[kli]} \otimes \gamma^{[lkj]} 
+2\gamma^{[ki]} \otimes \gamma^{[kj]}+6\gamma^i \otimes \gamma^j +2\gamma^j 
\otimes \gamma^i + 2 \gamma^k \otimes \gamma^k \delta_{ij} -2 \identity 
\otimes 
\identity \delta_{ij}\}
\end{equation} 
We will return to this expression later.
 
Since terms in expressions (\ref{A}) and (\ref{B}) may be related by Fierz 
identities, we will deal with them together. We first note that since
\[
M_{ab} = \vec{K}^2\delta_{ab} +2iF_{ab}
\]
and
\[
\slash{K} \slash{K} = \vec{K}^2 -{i \over 2} F_{ij}\gamma^i \gamma^j
\]
that terms in (\ref{A}) and (\ref{B}) split naturally into terms involving $F$ 
and terms involving $\vec{K}^2$. After some gamma matrix algebra, the 
$\vec{K}^2$ 
terms become 
\begin{eqnarray*}
& &{5 \over 256r^7}(\tr(L * L \vec{K}^2 L * L) + \tr(L * \vec{K}^2L \; L * L) 
) 
\\
& & \hspace{1in} \{ \gamma^{[kl]} \otimes \gamma^{[lk]} -2\gamma^k \otimes 
\gamma^k +10 \identity \otimes \identity \}\\
& & -{35 \over 256} {r_ir_j \over r^9}(\tr(L * L \vec{K}^2 L * L) + \tr(L * 
\vec{K}^2L \; L * L) )\\
& &\hspace{1in}\{ \gamma^{[kli]} \otimes \gamma^{[lkj]} + 2\gamma^{[ki]} 
\otimes \gamma^{[kj]} +8\gamma^i \otimes \gamma^j +2 \delta_{ij} ( \gamma^k 
\otimes \gamma^k - \identity \otimes \identity ) \}
\end{eqnarray*} 

We may now apply Fierz identities (\ref{2fi1}, \ref{2fi9} \ref{0fi1})
to reduce  all of these terms to simply
\[
-{5 \over 8}\left(\tr(L \gamma^a \vec{K}^2 L L \gamma^a L + \tr(L \gamma^a L 
\vec{K}^2 L \gamma^a L) \right)
\]
Subtracting off the terms containing $\vec{K}^2$ in (\ref{dfterms}), we are 
left with only
\begin{equation}
\Gamma^{K^2} = -{5 \over 16} \tr(L \gamma^a L \; L \gamma^a \vec{K}^2 L)
\label{ksquared}
\end{equation}

At this point, we note that (\ref{ksquared}) and (\ref{LKLtemp}) may be 
combined to give a term 
\[
\Gamma^{\rm cov} = {5 \over 32} \{ \tr(L \gamma^{a} [K_i,L] \; L \gamma^{a} 
[K_i,L]) 
+ \tr(K_i,L] \gamma^{a} L \; [K_i,L] \gamma^{a} L) + 2 \tr(L \gamma^{a} L \; 
[K_i,L] \gamma^{a} [K_i,L]) \}
\]
which combines naturally with (\ref{cov2}) to yield a $D=10$ covariant term, 
plus a remaining term
\begin{eqnarray*}
\Gamma^{LKL} &=& {5 \over 64}\tr(L*K_iL\;L*K_jL) \{\gamma^{[kli]} \otimes 
\gamma^{[lkj]} 
+2\gamma^{[ki]} \otimes \gamma^{[kj]}+6\gamma^i \otimes \gamma^j +2\gamma^j 
\otimes \gamma^i\\
& & \hspace{2.2in} - 2 \gamma^k \otimes \gamma^k \delta_{ij} +2 \identity 
\otimes 
\identity \delta_{ij}\}
\end{eqnarray*}
By the Fierz identities (\ref{2fi1}, \ref{2fi7}), we see that this has an 
equivalent form,
\begin{eqnarray*}
& &{5 \over 64}\tr(L*LK_iL*LK_j) \{ -{1 \over 2}\gamma^{[kli]} \otimes 
\gamma^{[lkj]} -{1 \over 2} \gamma^{[klj]} \otimes \gamma^{[lki]}  
-\gamma^{[ki]} \otimes \gamma^{[kj]}-\gamma^{[kj]} \otimes \gamma^{[ki]}\\
& & \hspace{1.1in} -4\gamma^i \otimes \gamma^j - 4\gamma^j \otimes \gamma^i  
-4 
\gamma^{[kij]} \otimes \gamma^k +4\gamma^{[ij]} \otimes \identity+ 2 \gamma^k 
\otimes \gamma^k \delta_{ij} -2 \identity \otimes \identity \delta_{ij}\}
\end{eqnarray*}

We now return to the terms in (\ref{A}) and (\ref{B}) invloving $F_{ij}$. From 
(\ref{A}), we have $F$ terms given by
\begin{eqnarray*}
& &-{35i r_m r_p \over 256r^9} \tr(L * LF_{ij} L*L)\\
& &\hspace{0.5in} \{2 \gamma^{[kmi]} \otimes \gamma^{[kpj]} + 2\gamma^{[mij]} 
\otimes \gamma^p + 2 \gamma^m  \otimes \gamma^{[pij]} -2 \gamma^{[mi]} \otimes 
\gamma^{[pj]} -2 \gamma^i \otimes \gamma^j \delta_{mp} \}\\
& & -{5i \over 256r^7} \tr(L * LF_{ij} L*L)\{-2 \gamma^{[ki]} \otimes 
\gamma^{[kj]} 
- 2\gamma^{[ij]} \otimes \identity -2 \identity  \otimes \gamma^{[ij]} -2 
\gamma^i \otimes \gamma^j \}\\
\end{eqnarray*}

while from (\ref{B}), the corresponding terms are
\begin{eqnarray*}
& &-{35i r_m r_p \over 256r^9} \tr(L * F_{ij} L\; L*L) \{- {1 \over 2} 
\gamma^{[klmij]} \otimes \gamma^{[lkp]} -\gamma^{[kmij]} \otimes \gamma^{[kp]} 
- 2\gamma^{[mij]} \otimes \gamma^p \\
& &\hspace{2.4in}+  \gamma^m  \otimes \gamma^{[pij]}- 
\delta_{mp}(\gamma^{[kij]} \otimes \gamma^k - \gamma^{[ij]} \otimes \identity 
) 
\}\\
& & -{5i \over 256r^7} \tr(L * F_{ij}L\; L*L)\{{1 \over 2} \gamma^{[klij]} 
\otimes 
\gamma^{[lk]} - \gamma^{[kij]} \otimes \gamma^k - \identity  \otimes 
\gamma^{[ij]} +3 \gamma^{[ij]} \otimes \identity \}\\
\end{eqnarray*}
Considering together all terms from these expressions in the $35/256$ 
brackets, 
we may use Fierz identity (\ref{4fi}) to reduce these to a form with all 
factors of 
$r_i$ appearing as $r^2$. Simplifying the remaining terms in the $5 /256$ 
brackets using Fierz identities (\ref{2fi4}, \ref{2fi6}), we find that these 
reduce to a set of terms 
proportional to the $35/256$ terms, and combining all $F$ terms, we get
\begin{eqnarray*}
& &-{5i \over 32}\tr(L * F_{ij} L\; L*L)\{\gamma^k \otimes \gamma^{[kij]} - 
\gamma^{[kij]} \otimes \gamma^k + \identity \otimes \gamma^{[ij]} 
-\gamma^{[ij]} \otimes \identity \}\\
& & -{5i \over 32}\tr(L * L F_{ij} L*L)\{\gamma^k \otimes \gamma^{[kij]} + 
\gamma^{[kij]} \otimes \gamma^k - \identity \otimes \gamma^{[ij]} 
-\gamma^{[ij]} \otimes \identity -4 \gamma^i \otimes \gamma^j \}
\end{eqnarray*}

Finally, we note that the only terms from (\ref{dfterms}) that we have not yet 
accounted for contain $F_{ij}$, and subtracting these off now, we are left 
with
\begin{eqnarray}
\label{Fterms}
\Gamma^{F} &=& -{5i \over 32} \{ \tr(L \gamma^{k} F_{ij} L \; L \gamma^{[kij]} 
L)+\tr(L \gamma^{[kij]} L F_{ij} L \gamma^{k} L)+\tr(L \gamma^{k} L F_{ij} L 
\gamma^{[kij]} L)\nonumber  \\ 
& & \hspace{0.5in} - \tr(L F_{ij} L \; L \gamma^{[ij]} L)-\tr(L \gamma^{[ij]} 
L 
F_{ij} L  L)+\tr(L  L F_{ij} L \gamma^{[ij]} L) \} \; .
\end{eqnarray}
These terms combine with the $\dot{K}$ terms (\ref{kdot}) into a $D=10$ 
covariant expression which we write below.

\subsubsection{Summary of four-fermion terms}  

We now review the four-fermion terms calculated above. Firstly, we
have a set of terms (\ref{dfterms}) which can be obtained by replacing
$D_aF_{ab}$ with $D_aF_{ab} - \bar{L} \Gamma^a L$ everywhere it
appears in the two fermion action, and thus cancel the corresponding
two-fermion terms when the background fields satisfy the equations of
motion. Collecting all the remaining terms, we define

\begin{eqnarray}
\Gamma^{4L}_{1/r^7} &=& \Gamma^{\dot{L}\dot{L}} + \Gamma^{\dot{L}} + 
\Gamma^{\dot{K}}+\Gamma^{\rm cov}+\Gamma^{F} + \Gamma^{LKL}\nonumber\\
&=& {5 \over 32} \{ \tr(\bar{L} \Gamma^{a} D_bL \; \bar{L} \Gamma^{a} D_bL) + 
\tr(D_b\bar{L} \Gamma^{a} L \; D_b\bar{L} \Gamma^{a} L) + 2 \tr(\bar{L} 
\Gamma^{a} 
L \; D_b\bar{L} \Gamma^{a} D_bL) \}\label{4LA} \\ \nonumber \\ \nonumber \\
& & + {5i \over 32} \{ \tr(\bar{L} \Gamma^{c} F_{ab} L \; \bar{L} 
\Gamma^{[cab]} 
L)+\tr(\bar{L} \Gamma^{[cab]} L F_{ab} \bar{L} \Gamma^{c} L)+\tr(\bar{L} 
\Gamma^{c} \bar{L} F_{ab} L \Gamma^{[cab]} L)\label{4LB}\nonumber  \\  
\nonumber\\
& & + {5 \over 128} \{ \tr(L \gamma^{[ki]} [\dot{K_i} , L] \; L \gamma^k L) + 
\tr(L 
\gamma^{[ki]}  , L \;[\dot{K_i} L] \gamma^k L) \} 
\label{4LC}\\ \nonumber \\  
& & - {5 \over 128} \{ \tr(L \gamma^{[kl]} \dot{L} \; 
\dot{L} \gamma^{[lk]} L) + 2\tr(L \gamma^{k} \dot{L} \; \dot{L} \gamma^{k} L) 
+ 
6\tr(L  \dot{L} \; \dot{L}  L) \}\label{4LD}\\ \nonumber\\ 
& & + { 5 \over 256} \{ \tr(L \gamma^{[kli]} K_i \dot{L} \; L \gamma^{[lk]} 
L)+6\tr(L \slash{K} \dot{L} \; L L)+2\tr(L K_i \dot{L} \; L \gamma^{i} 
L)\nonumber 
 \\
& & \hspace{0.5in} - \tr(\dot{L} \gamma^{[kli]} K_i L \; L \gamma^{[lk]} 
L)-6\tr(\dot{L} \slash{K} L \; L L)-2\tr(\dot{L} K_i L \; L \gamma^{i} L) 
\}\nonumber\\
& & + { 15 \over 256} \{ \tr(L \gamma^{[kli]} K_i L \; \dot{L} \gamma^{[lk]} 
L)+6\tr(L \slash{K} L \; \dot{L} L)+2\tr(L K_i L \; \dot{L} \gamma^{i} 
L)\nonumber 
 \\
& & \hspace{0.5in}- \tr(L \gamma^{[kli]} K_i L \; L \gamma^{[lk]} 
\dot{L})-6\tr(L 
\slash{K} L \; L \dot{L})-2\tr(L K_i L \; L \gamma^{i} \dot{L}) 
\}\label{4LE}\\ 
\nonumber
\\& & + { 5 \over 64} \{ \tr(L \gamma^{[kli]} K_i L \; L \gamma^{[lkj]} K_j  
L)+2\tr(L \gamma^{[ki]} K_i L \; L \gamma^{[kj]} K_j  L)+6\tr(L \gamma^{i} K_i 
L 
\; L \gamma^{j} K_j  L)\nonumber \\
& & \hspace{0.5in} +2\tr(L \gamma^{j} K_i L \; L \gamma^{i} K_j  L)-2\tr(L 
\gamma^{i} K_j L \; L \gamma^{i} K_j  L)+2\tr(L K_i L \; L K_i L) 
\}\label{4LF}
\end{eqnarray}

One might hope that these terms might all be combined into a $D=10$ covariant 
expression, however, this does not appear to be the case. In particular, one 
can show that there is no way to rewrite the above expression in a form with 
all $K$'s appearing in commutators. 

\subsubsection{Order $1/r^6$ terms}
\label{sec:sixterms}

To complete the demonstration of the cancellation of the on-shell action below 
order $1/r^7$ we now calculate the terms at order $1/r^6$ using the 
transformation (\ref{Ktrans}) on the $1/r^7$ terms, as described above. From 
the terms (\ref{dfterms}) we find $ r \cdot K$ terms given by
\begin{equation}
{5 \over 8} \tr(\bar{L} \Gamma^a r \cdot K L \; 
\bar{L} \Gamma^a L) + {5 \over 4} \tr(\bar{L} \Gamma^a L r \cdot K \bar{L} 
\Gamma^a L)
\label{df6}
\end{equation}
These are exactly the terms obtained by replacing $D_aF_{ab}$ with $D_aF_{ab} 
- 
i\bar{L} \Gamma^b L$ in the two fermion $1/r^6$ terms, providing cancellation 
on-shell. We have additional terms at $1/r^6$ from applying our transformation 
to (\ref{4LE}) and (\ref{4LF}), however by Fierz identities 
(\ref{1fi1}),(\ref{1fi2}), (\ref{2fi1}), and (\ref{2fi7}), these may be 
written 
in a form 
with $\slash{D}L$ in each term, and therefore vanish on-shell.

This completes our calculation.

\newpage

\section{Fierz Identities}

In manipulating four-fermion terms in our answer, it is necessary to make use 
of 
various Fierz rearrangement identities which shall now be derived. All 
identities are based on the fact that $1,\gamma^i, \gamma^{[ij]}, 
\gamma^{[ijk]},$ and $\gamma^{[ijkl]}$ $i,j,k,l \in \{1...9\}$ form a complete 
basis of the $16 \times 16$ matrices. This may be expressed by the 
completeness 
relation:
\begin{equation}
\label{fierz}
\delta_{\alpha \beta} \delta_{\gamma \delta} = \sum_{n=0}^4 {1 \over 16n!} 
\gamma^{[a_1...a_n]}_{\beta \delta} \gamma^{[a_n...a_1]}_{\gamma \alpha}
\end{equation}
or equivalently:
\[
M = \sum_{n=0}^4 {1 \over 16n!} \tr(M \gamma^{[a_1...a_n]}) 
\gamma^{[a_n...a_1]}
\]
for an arbitrary matrix $M$. Note in particular that
\begin{equation}
\label{anti}
\gamma^{[a_1...a_{9-n}]} = {1 \over n!} \epsilon^{(a_1...a_{9-n} b_1...b_n)} 
\gamma^{[b_n...b_1]}
\end{equation}

We will derive a large class of Fierz identities as follows. Consider a term
\[
f_n \equiv \tr(A \gamma_1 \gamma^{[a_1...a_n]} \gamma_2 B \; C \gamma_3 
\gamma^{[a_n...a_1]} \gamma_4 D)
\]
where $A, B, C,$ and $D$ are $N \times N$ matrices with a single fermionic 
index 
and $\gamma_i$ are arbitrary products of gamma matrices whose spatial indices 
are not contracted with other gamma matrices. Then we may write a general 
non-abelian four-fermion term as a sum of such terms, since we may always 
rearrange the individual gamma matrices to get a sum of terms in which gamma 
matrices with contracted indices appear together in antisymmetrized products 
as 
in $f_n$. By relation (\ref{anti}), we need only consider $n \le 4$. The terms 
$f_n$ are certainly not all independent (even without appealing to Fierz 
identities), however, choosing this form will be useful for our derivation. 

   Now we will apply our general Fierz relation (\ref{fierz}). We have:
\begin{eqnarray*}
f_n &=& \tr((A \gamma_1)_{\alpha} \delta_{\alpha \beta} (\gamma^{[a_1...a_n]} 
\gamma_2 B)_{\beta} \; (C \gamma_3 \gamma^{[a_n...a_1]})_{\gamma} 
\delta_{\gamma 
\delta} (\gamma_4 D)_{\delta})\\
&=& -\sum_{m=0}^4 {1 \over 16m!}\tr(B \gamma_2^T \gamma^{[a_n...a_1]} 
\gamma^{[b_1...b_m]} \gamma^{[a_1...a_n]} \gamma_3^T C \; D \gamma_4^T 
\gamma^{[b_m...b_1]} \gamma_1^T A)
\end{eqnarray*}
where the minus sign is due to an odd rearrangement of fermionic matrices in 
the 
trace. To simplify this, we may use the following identities, which can be 
easily checked:
\[
\gamma^{[a_n...a_1]} \gamma^{[b_1...b_m]} \gamma^{[a_1...a_n]}  = c_{n,m} 
\gamma^{[b_1...b_m]}
\]
where
\begin{eqnarray*}
c_{0,m} &=& 1\\
c_{1,m} &=&  (-1)^m(9-2m)\\
c_{2,m} &=& \{(9-2m)^2 - 9\}\\
c_{1,m} &=& \{(-1)^m(9-2m)^3 - 25(-1)^m(9-2m)\}\\
c_{1,m} &=& \{(9-2m)^4 -46(9-2m)^2 +189\}
\end{eqnarray*}
Then we have:
\[
f_n = -\sum_{m=0}^4{c_{n,m} \over 16m!}g_m
\]
or simply
\begin{equation}
\label{fmg}
\vec{f} = -{\bf M} \vec{g}
\end{equation}
where we have defined
\[
g_n \equiv \tr(B \gamma_2^T \gamma^{[a_1...a_n]} \gamma_3^T C \; D \gamma_4^T 
\gamma^{[a_n...a_1]} \gamma_1^T A)
\]
and 
\[
{\bf M}_{nm} = {c_{n,m} \over 16m!} 
= \left[ \ba{ccccc} 
{1 \over 16} & {1 \over 16} & {1 \over 32}& {1 \over 96} & {1 \over 384} \\
 {9 \over 16} & {-7 \over 16} & {5 \over 32} & {-1 \over 32} & {1 \over 384} 
\\
{9 \over 2} & {5 \over 2} & {1 \over 2} & {0} & {-1 \over 48} \\
 {63 \over 2} & {-21 \over 2} & {0} & {1 \over 2} & {-1 \over 16} \\
 {189} & {21} & {-21 \over 2} & {-3 \over 2} & {3 \over 8} \ea  \right]   
\]
Note that $g_n$ and $f_n$ have the same form, and more importantly, that we 
may 
follow the same steps to show that
\begin{equation}
\label{gmf}
\vec{g} = -{\bf M} \vec{f}
\end{equation}
Now, (\ref{fmg}) and (\ref{gmf}) are equivalent to the relations 
\begin{eqnarray*}
(\vec{f} + \vec{g}) &=& -{\bf M} (\vec{f} + \vec{g})\\
(\vec{f} - \vec{g}) &=&  {\bf M} (\vec{f} - \vec{g})
\end{eqnarray*}
or
\begin{eqnarray*}
(\identity + {\bf M}) (\vec{f} + \vec{g}) &=& 0\\
(\identity - {\bf M}) (\vec{f} - \vec{g}) &=& 0
\end{eqnarray*}
The matrix $(\identity + {\bf M})$ has rank $3$, and the space of its rows is 
spanned by the vectors:
\begin{eqnarray*}
\vec{v}_4 &=& (1,0,0,24,216)\\
\vec{v}_3 &=& (0,1,0,-6,30)\\
\vec{v}_2 &=& (0,0,1,2,6)\\
\end{eqnarray*}
The matrix $(\identity - {\bf M})$ has rank $2$, and the space of its rows is 
spanned by the vectors:
\begin{eqnarray*}
\vec{u}_4 &=& (1,0,24,-120,-216)\\
\vec{u}_3 &=& (0,1,-3,36,-36)\\
\end{eqnarray*}
Thus, we have five basic relations, given by
\begin{equation}
\label{vrels}
\vec{v}_i \cdot (\vec{f} + \vec{g}) 
\end{equation}
for $(i=2,3,4)$ and 
\begin{equation}
\label{urels}
\vec{u}_j \cdot (\vec{f} + \vec{g})
\end{equation}
for $(j = 3,4)$ which should generate all of our Fierz identities. 

The identities we have derived provide relations among terms of the form 
$f_n$, 
but we may write the set of terms of this form in terms of a smaller set of 
terms of a standard form, for example, terms of the form:
\[
\tr(A \gamma^{[a_1...a_n b_1...b_k]}  B \; C \gamma^{[a_n...a_1 c_1...c_m]} D)
\]
Using the the relation (\ref{anti}) we may further restrict our basis to terms 
with $n+k \le 4$ and $n+m \le 4$. With these transformations, each of the 
relations above reduces to a Fierz identity relating elements of this smaller 
basis.

We will mainly be interested in relations involving terms with a small number 
of free indices. Here, we will derive the complete set of relations involving 
zero, one or two free indices, plus particular relations for three and four 
free 
indices that are needed in our calculation.

\subsection{Zero free indices}

For the case of zero free indices, we have $\gamma_i=1$, and all 
of the terms $f_n$ are already in a standard form. Thus, in this case, we 
simply have the five relations (\ref{vrels}, \ref{urels}) above. Explicitly, 
defining
\begin{eqnarray*}
C_n &=& \tr(A \gamma^{[a_1...a_n]} B C \gamma^{[a_n...a_1]} D)\\ 
D_n &=& \tr(B \gamma^{[a_1...a_n]} C D \gamma^{[a_n...a_1]} A)
\end{eqnarray*}
we have
\begin{eqnarray}
(C_2+D_2) + 2(C_1+D_1) + 6(C_0 + D_0) &=& 0
\label{0fi1}\\
(C_3+D_3) - 6(C_1+D_1) + 30(C_0 + D_0) &=& 0
\label{0fi2}\\
(C_4+D_4) + 24(C_1+D_1) + 216(C_0 + D_0) &=& 0
\label{0fi3}\\
(C_3-D_3) - 3(C_2-D_2) + 36(C_1+D_1) - 36(C_0 + D_0) &=& 0
\label{0fi4}\\
(C_4-D_4) + 24(C_2-D_2) - 120(C_1+D_1) - 216(C_0 + D_0) &=& 0
\label{0fi5}
\end{eqnarray}
 As an example, for the case $A=B=C=D=L$, we see that $C_n = D_n$, so the last 
two relations are trivial, and  and we have three independent Fierz 
identities. 
Explicitly, the relation (\ref{0fi1}) is
\[
\tr(L\gamma^{[ij]}L\;L\gamma^{[ji]}L) +2\tr(L\gamma^{i}L\;L\gamma^{i}L) + 
6\tr(LL\;LL)  = 0
\]
Note that except in this case of equal $A, B, C,$ and $D$, the Fierz 
identities 
will always relate terms with one ordering of the matrices to terms with 
another 
ordering.

\subsection{One free index}

  For the case of one free index, there are four possible choices for $f_n$, 
corresponding to taking three of the matrices $\{\gamma_1, \gamma_2, \gamma_3, 
\gamma_4 \}$ equal to the identity and the fourth equal to $\gamma^i$. For 
each 
of these choices, we have five relations (\ref{vrels}, \ref{urels}), however, 
not all twenty of these 
relations are independent. 
  In order to rewrite the Fierz identities in terms of our standard basis with 
only antisymmetrized products of four or less gamma matrices, we can use the 
relations
\begin{equation}
\label{otids}
\gamma^i\gamma^{[a_1...a_n]}\otimes \gamma^{[a_n...a_1]}= 
(-1)^n\gamma^{[a_1...a_ni]}\otimes 
\gamma^{[a_n...a_1]}+n\gamma^{[a_2...a_n]}\otimes \gamma^{[a_n...a_2i]}
\end{equation}
and its mirror image
\[
\gamma^{[a_1...a_n]}\gamma^i\otimes \gamma^{[a_n...a_1]}= 
\gamma^{[a_1...a_ni]}\otimes 
\gamma^{[a_n...a_1]}+(-1)^{n-1}n\gamma^{[a_2...a_n]}\otimes 
\gamma^{[a_n...a_2i]}
\]  
By (\ref{anti}), we note also that
\begin{equation}
\label{extra1}
\gamma^{[klmni]} \otimes \gamma^{[nmlk]} = \gamma^{[klmn]} \otimes 
\gamma^{[nmlki]} = {1 \over 4!}\epsilon^{(wxyzklmni)} \gamma^{[zyxw]} \otimes 
\gamma^{[klmn]}
\end{equation}
so we see that certain relations in the set we are considering (those 
corresponding to $v_4$ and $u_4$) will actually relate terms in our standard 
basis containing one free index to terms containing eight free indices, via 
the 
nine-index epsilon tensor.

We now define  
\begin{eqnarray*}
E_n &=& \tr(A \gamma^{[a_1...a_ni]} B C \gamma^{[a_n...a_1]} D)\\ 
\bar{E_n} &=& \tr(A \gamma^{[a_1...a_n]} B C \gamma^{[a_n...a_1i]} D)\\ 
F_n &=& \tr(B \gamma^{[a_1...a_ni]} C D \gamma^{[a_n...a_1]} A)\\
\bar{F_n} &=& \tr(B \gamma^{[a_1...a_n]} C D \gamma^{[a_n...a_1i]} A)
\end{eqnarray*}
where the bar denotes swapping the two sets of gamma matrices in a term. Then 
writing the basic relations (\ref{vrels}) for each of our four choices of 
$f_n$ 
give:
\begin{eqnarray*}
v^n_i((-1)^n E_n^+ + nE_{n-1}^+ +F_{n}^+ + (-1)^{n-1}nF_{n-1}^+) &=&0\\
v^n_i((-1)^n E_n^- - nE_{n-1}^- -F_{n}^- + (-1)^{n-1}nF_{n-1}^-) &=&0\\
v^n_i( E_n^+ + (-1)^{n-1}nE_{n-1}^+ + (-1)^nF_{n}^+ + nF_{n-1}^+) &=&0\\
v^n_i( E_n^- + (-1)^n nE_{n-1}^- + (-1)^nF_{n}^- - nF_{n-1}^-) &=&0
\end{eqnarray*}
Here we have written the relations in linear combinations involving only 
$E_n^+ 
\equiv E_n + \bar{E_n}$ and $E_n^- \equiv E_n - \bar{E_n}$. From 
(\ref{urels}), 
we get a set of relations of identical form with $v$ replaced by $u$ and the 
signs on all $B$ terms reversed. 

   It remains only to write these twenty relations explicitly for each $v$ and 
$u$, and determine a linearly independent set. This is easily done, and we 
find 
the following nine independent Fierz identities:
\begin{eqnarray}
(E_2^+ + F_2^+) + 8(E_0^+ + F_0^+) &=& 0
\label{1fi1}\\
E_2^- + 4E_0^- - 4F_1^- &=& 0
\label{1fi2}\\
F_2^- + 4F_0^- + 4E_1^- &=& 0
\label{1fi3}\\
(E_3^+ + F_3^+) + 42(E_1^+ + F_1^+) &=& 0
\label{1fi4}\\     
(E_3^+ - F_3^+) - 6(E_1^+ - F_1^+) &=& 0
\label{1fi5}\\
E_3^- + 6E_1^- + 48F_0^- &=& 0
\label{1fi6}\\
F_3^- + 6F_1^- + 48E_0^- &=& 0
\label{1fi7}\\
P + Q + 120(E_0^+ + F_0^+) &=& 0
\label{1fi8}\\      
P - Q + 12(E_2^+ - F_2^+) -168(E_0^+ - F_0^+) &=& 0
\label{1fi9}
\end{eqnarray}
In the last two identities, we have defined
\[
P \equiv E_4 = \bar{E_4} = {1 \over 4!}\epsilon^{(wxyzklmni)} \tr( A 
\gamma^{[wxyz]} B C \gamma^{[klmn]} D)
\]
and
\[
Q \equiv F_4 = \bar{F_4} = {1 \over 4!}\epsilon^{(wxyzklmni)} \tr( B 
\gamma^{[wxyz]} C D \gamma^{[klmn]} A)
\]

\subsection{Two free indices}

Here we proceed exactly as in the preceding section. In this case, there are 
sixteen independent ways to choose $f_n$ to have two free indices on gamma 
matrices. For each of these, we get Fierz identities from (\ref{vrels}) and 
(\ref{urels}). We can use identities analogous to (\ref{otids}) to rewrite 
these 
in terms of our standard basis. The relevant basis elements in this case are
\begin{eqnarray*}
A_n &=& \tr(A \gamma^{[a_1...a_nij]} B C \gamma^{[a_n...a_1]} D)\\ 
a_n &=& \tr(A \gamma^{[a_1...a_ni]} B C \gamma^{[a_n...a_1j]} D)\\
B_n &=& \tr(B \gamma^{[a_1...a_nij]} C D \gamma^{[a_n...a_1]} A)\\ 
B_n &=& \tr(B \gamma^{[a_1...a_ni]} C D \gamma^{[a_n...a_1j]} A)\\ 
c_n &=& C_n \delta_{ij}\\
d_n &=& D_n \delta_{ij}
\end{eqnarray*}
plus the corresponding quantities $\bar{A_n}, \bar{a_n},\bar{B_n}$ and 
$\bar{b_n}$ with gamma matrices swapped. Using these definitions, the Fierz 
identities relating quantities with two free indices are then given by
\[
v_n((-1)^na_n^+ + (-1)^{n-1} n(n-1) a_{n-2}^+ +2nb_{n-1}^+ + 2nc_{n-1} + 
2(-1)^nd_n) = 0\\ 
\]
plus all other relations derived from this by one or more of the 
transformations
\[
\ba{c}
 (A \leftrightarrow B, \;a \leftrightarrow b,\; c \leftrightarrow d)\\
 (a \leftrightarrow A, \; b \leftrightarrow -B,\; c \rightarrow 0,\; d 
\rightarrow 0)\\
 (a_n^+ \rightarrow (-1)^na_n^-, \; b_n^+ \rightarrow (-1)^nB_n^-, \; c_n 
\rightarrow 0, \; d_n \rightarrow 0)\\
 (v \rightarrow u, \; B \rightarrow -B, \; b \rightarrow -b, \; d \rightarrow 
-d)
\ea
\] 
Here, we write explicitly only the relations corresponding to $v_2$, which we 
find to be
\begin{eqnarray}
0 &=& a_2^+ - 2a_1^+ + 4a_0^+ + 4b_1^+ +4b_0^+ + 4c_1 +4c_0 +2d_2 -4d_1 +12d_0 
\label{2fi1}\\
0 &=&b_2^+ - 2b_1^+ + 4b_0^+ + 4a_1^+ +4a_0^+ + 4d_1 +4d_0 +2c_2 -4c_1 +12c_0 
\label{2fi2}\\
0 &=&A_2^+ - 2A_1^+ + 4A_0^+ - 4B_1^+ -4B_0^+
\label{2fi3}\\
0 &=&B_2^+ - 2B_1^+ + 4B_0^+ - 4A_1^+ -4A_0^+
\label{2fi4}\\
0 &=&A_2^- + 2A_1^- + 4A_0^- + 4b_1^- -4b_0^-
\label{2fi5}\\
0 &=&B_2^- + 2B_1^- + 4B_0^- - 4a_1^- +4a_0^-
\label{2fi6}\\
0 &=&a_2^- + 2a_1^- + 4a_0^- - 4B_1^- +4B_0^-
\label{2fi7}\\
0 &=&b_2^- + 2b_1^- + 4b_0^- + 4A_1^- -4A_0^-
\label{2fi8}\\
0 &=&(c_2+d_2) +2(c_1+d_1)+6(c_0+d_0)\label{2fi9} .
\end{eqnarray}
The last of these is simply relation (\ref{0fi1}) multiplied by $\delta_{ij}$.    
 
\subsection{Three free indices}

We will need only a single Fierz identity with three free indices, in order to 
eliminate terms at $1/r^7$ with two $\slash{r}$s and a $\dot{K_i}$. By a 
judicious choice of $f_n$, we can generate the required identity. Taking
\[
f_n = r_mr_p\tr(L \gamma^m \gamma^{[a_1...a_n]} \gamma^i \dot{K_i} L \; L 
\gamma^{[a_n...a_1]} \gamma^p L) + r_mr_p\tr(L \gamma^i \gamma^{[a_1...a_n]} 
\gamma^m \dot{K_i} L \; L \gamma^p \gamma^{[a_n...a_1]}  L),
\]
we have
\[
g_n = r_mr_p\tr(L \gamma^n \gamma^{[a_1...a_n]} \gamma^m  L \dot{K_i} L 
\gamma^i
\gamma^{[a_n...a_1]}  L) + r_mr_p\tr(L  \gamma^{[a_1...a_n]} \gamma^i L 
\dot{K_i} 
L \gamma^m \gamma^{[a_n...a_1]} \gamma^n L),
\]
and the required relation is $v_2^n(f_n+g_n) = 0$. This becomes (scaling by 
1/2
and ignoring $r \cdot K$ terms), 
\begin{eqnarray}
0 &=& r_mr_p\tr(L* \dot{K_i} L \; L*L)\nonumber \\ 
& & \hspace{0.2in}\{ \gamma^{[klmi]} \otimes  \gamma^{[lkp]} + 4 
\gamma^{[mi]} \otimes  \gamma^p - 2 \gamma^m \otimes  \gamma^{[pi]} 
 + 2\delta_{mp}\gamma^{[ki]} \otimes \gamma^k \nonumber \\
 & & \hspace{0.5in} - 2\delta_{mp} (  \gamma^{[ki]} \otimes \gamma^k - 
2\gamma^k \otimes \gamma^{[ki]})\}\nonumber\\
& & +r_mr_p\tr(L*L \dot{K_i} L*L)\nonumber \\ 
& & \hspace{0.2in} \{-2\gamma^{[kmi]} \otimes  \gamma^{[kp]} + 2 \gamma^{[km]} 
\otimes  \gamma^{[kpi]} - 4\gamma^{[mi]} \otimes  \gamma^p -4 \gamma^m \otimes  
\gamma^{[pi]} \nonumber \\ 
 & & \hspace{0.5in} + \delta_{mp} ( 2 \gamma^{[ki]} \otimes \gamma^k + 
2\gamma^k \otimes \gamma^{[ki]}  - 2 \gamma^i \otimes \identity + 2 \identity 
\otimes \gamma^i )\}
 \label{3fi}
 \end{eqnarray}
where we have used identity (\ref{1fi3}) to put the $\delta_{mp}$ terms in a 
simpler form.

\subsection{Four free indices}

Finally, we will need a single Fierz identity with four free indices, in order 
to eliminate terms at $1/r^7$ with two $\slash{r}$s and an $F_{ij}$. The 
required identity is given by $v_2^n(f_n+g_n) = 0$, with
\[
f_n = r_mr_p\tr(L \gamma^i \gamma^m \gamma^{[a_1...a_n]} \gamma^j F_{ij} L \; 
L 
\gamma^{[a_n...a_1]} \gamma^p L) + r_mr_p\tr(L \gamma^i \gamma^{[a_1...a_n]} 
\gamma^m \gamma^j 
F_{ij} L \; L \gamma^p \gamma^{[a_n...a_1]} L),
\]
and
\[
g_n = r_mr_p\tr(L \gamma^p \gamma^{[a_1...a_n]} \gamma^m \gamma^i  L F_{ij} L 
\gamma^j 
\gamma^{[lk]} L) + r_mr_p\tr(L  \gamma^{[a_1...a_n]} \gamma^i  L F_{ij} L 
\gamma^j 
\gamma^m  \gamma^{[a_n...a_1]} \gamma^p L)
\]
Writing this identity explicitly with all gamma matrices appearing in 
antisymmetric products, we have (scaling by 1/4 and ignoring $r \cdot K$ 
terms), 
\begin{eqnarray}
0 &=& r_mr_p\tr(L*F_{ij}L \; L*L)\nonumber \\ 
& & \hspace{0.2in}\{-{1 \over 2} \gamma^{[klmij]} \otimes  \gamma^{[lkp]} -2 
\gamma^{[mij]} \otimes  \gamma^p + \gamma^m \otimes  \gamma^{[pij]} 
-\gamma^{[kmij]} \otimes  \gamma^{[kp]}\nonumber \\ 
 & & \hspace{1in} + \delta_{mp}(-\gamma^k \otimes \gamma^{[kij]} + \identity 
\otimes \gamma^{[ij]}  ) \} \nonumber \\ 
& & +r_mr_p\tr (L*L F_{ij} L*L)\nonumber \\ 
& & \hspace{0.2in} \{2\gamma^{[lmi]} \otimes  \gamma^{[lpj]} +2 \gamma^{[mij]} 
\otimes  \gamma^p + 2\gamma^m \otimes  \gamma^{[pij]} -2\gamma^{[mi]} \otimes  
\gamma^{[pj]} \nonumber \\ 
 & & \hspace{0.5in} + \delta_{mp} ( 2 \gamma^i \otimes \gamma^j - 
\gamma^{[kij]} \otimes \gamma^k - \gamma^k \otimes \gamma^{[kij]} + 
\gamma^{[ij]} \otimes \identity + \identity \otimes \gamma^{[ij]} \}
 \label{4fi}
 \end{eqnarray}
where we have used identity (\ref{2fi7}) to put the $\delta_{mp}$ terms in a 
simpler form.

\bibliographystyle{plain}


\end{document}